\documentclass[aps,pra,twocolumn,groupedaddress,amssymb,amsfonts,showpacs]{revtex4}
\usepackage{amsbsy,amsmath,bm}

\usepackage{color}
\usepackage[dvips]{graphics,graphicx}
\usepackage{calc,ifthen}

\newlength{\elimdepthdim}
\newlength{\elimheightdim}
\newlength{\elimwidthdim}

\newlength{\strutdepthdim}
\newlength{\strutheightdim}
\newlength{\strutwidthdim}

\def\one{{\mathchoice {\rm 1\mskip-4mu l} {\rm 1\mskip-4mu l} {\rm
1\mskip-4.5mu l} {\rm 1\mskip-5mu l}}}

\newcommand{\ket}[1]{\qvbar{#1}\qrangle}
\newcommand{\bra}[1]{\qlangle{#1}\qvbar}

\newcommand{\proof}{\paragraph*{Proof.}}

\newcommand{\qed}{\hspace*{\fill}\rule{2.5mm}{2.5mm}%
\vspace*{8pt}\par}

\newcounter{herefignum}

\newcommand{\shortqph}[1]{}

\providecommand{\ignore}[1]{}

\def\Z{{\mathbb{Z}}}

\def\proof{{\em Proof:  }}

\def\openone{\leavevmode\hbox{\small1\kern-3.8pt\normalsize1}}
\def\RR{{\rm I\kern-.2emR}}
\def\tr{{\rm tr}\; }

\def\fg{\mathfrak{g}}

\def\proof{{\em Proof:  }}

\def\openone{\leavevmode\hbox{\small1\kern-3.8pt\normalsize1}}
\def\RR{{\rm I\kern-.2emR}}
\def\tr{{\rm tr}\; }

\providecommand{\ignore}[1]{}

\renewcommand{\ket}[1]{| #1 \rangle}
\renewcommand{\bra}[1]{\langle #1 |}
\newcommand{\proj}[1]{\ket{#1}\! \bra{#1}}

\newcommand{\bitem}{\begin{itemize}}
\newcommand{\eitem}{\end{itemize}}
\newcommand{\benum}{\begin{enumerate}}
\newcommand{\eenum}{\end{enumerate}}
\newcommand{\beq}{\begin{equation}}
\newcommand{\eeq}{\end{equation}}
\newcommand{\beqa}{\begin{eqnarray}}
\newcommand{\eeqa}{\end{eqnarray}}
\newcommand{\bi}{{\bf i}}
\newcommand{\bj}{{\bf j}}
\newcommand{\bk}{{\bf k}}
\newcommand{\bl}{{\bf l}}

\newtheorem{definition}{Definition}

\newtheorem{proposition}{Proposition}

\newcommand{\bproof}{\begin{proof}}
\newcommand{\eproof}{\end{proof}}
\newcommand{\bprop}{\begin{proposition}}

\newcommand{\bdef}{\begin{definition}}

\begin{document}


\title{A symmetry principle for Topological Quantum Order}



\author{Zohar Nussinov$^1$ and Gerardo Ortiz$^{2}$}
\affiliation{$^1$Department of Physics, Washington University, St.
Louis, MO 63160, USA}
\affiliation{$^2$Department of Physics, Indiana University, Bloomington,
IN 47405, USA}

\date{\today}

\pacs{05.30.-d, 11.15.-q, 71.10.-w, 71.10.Pm}

\begin{abstract}
We present a unifying framework to study physical systems which exhibit
topological quantum order (TQO). The major guiding principle behind our
approach is that of symmetries and entanglement. To this end, we
introduce the concept of low-dimensional Gauge-Like Symmetries (GLSs),
and the physical conservation laws (including  topological terms,
fractionalization, and the absence of quasi-particle excitations) which
emerge from them. We prove then sufficient conditions for TQO at both
zero and finite temperatures. The physical engine for TQO are
topological defects associated with the restoration of GLSs. These
defects propagate freely through the system and enforce TQO. Our
results are strongest for gapped systems with continuous GLSs. At zero 
temperature, selection rules associated with the GLSs enable us to
systematically construct general states with TQO; these selection rules
do not rely on the existence of a finite gap between the  ground states
to all other excited states. Indices associated  with these symmetries
correspond to different {\it topological} sectors. All currently known
examples of TQO display GLSs. Other systems exhibiting such symmetries
include Hamiltonians depicting orbital-dependent spin-exchange and
Jahn-Teller effects in transition metal orbital compounds, short-range
frustrated Klein spin models, and p+ip superconducting arrays. The
symmetry based framework  discussed herein allows us to go beyond
standard topological field theories and systematically engineer new
physical models with finite temperature TQO (both Abelian and
non-Abelian). Furthermore, we analyze the insufficiency of entanglement
entropy (we introduce $SU(N)$ Klein models on small world networks to
make the argument even sharper), spectral structures, maximal string
correlators, and fractionalization in establishing TQO. We show that
Kitaev's Toric code model and Wen's plaquette model are equivalent and
reduce, by a duality mapping,  to an Ising chain, demonstrating that
despite the spectral gap in these systems  the toric operator
expectation values may vanish once thermal fluctuations are present.
This illustrates the fact that  the quantum states themselves in a
particular (operator language) representation encode TQO and that the
duality mappings, being non-local in the original representation,
disentangle the order. We present a general algorithm  for the
construction of long-range {\it string and brane orders} in general
systems with entangled ground states; this algorithm relies
on general ground states selection rules and becomes of the broadest
applicability in gapped systems in arbitrary dimensions. 
We discuss relations to problems
in graph theory.
\end{abstract}

\maketitle

\section{Introduction}
\label{Section 0}

The emergence of complex phenomena in strongly coupled quantum matter 
may be the result of quite simple guiding principles yet to be
discovered. Nevertheless, we expect that the key players of   {\it
Symmetry} and {\it Topology} are rooted in those principles, and their
cross fertilization may help to unveil those guiding principles or even 
lead to new phenomena. That symmetry is a guiding principle is
demonstrated by  the classical Landau theory of phase transitions
\cite{Landau}.  Two key contributions form the crux of this theory: one
is the association of a {\it local} order parameter ${\cal O}({\bf r})$
to a broken (ordered) symmetry state, thus to (the lack of) symmetry,
and the other is a phenomenological description of the transition in
terms of an analytic functional of ${\cal O}({\bf r})$ (Landau free
energy). Despite its simplicity and success, can all possible phases of
matter and their transitions (even approximately) be described within
Landau's framework?

A new paradigm, Topological Quantum Order (TQO), extends the Landau
symmetry-breaking framework.  At its core, TQO is intuitively
associated with insensitivity to {\it local} perturbations: the order
is {\it topological}.  As such, TQO cannot be described by {\it local}
order parameters. Some examples of systems displaying TQO also display 
ground state (GS) subspace degeneracy  highly sensitivity to boundary
conditions or surface topology. What are the physical principles and
consequences behind such an order? How do we mathematically
characterize it? Although much has been learned about this topic, much
still remains shrouded in mystery, in part, because there is no
universally accepted definition of TQO from which many results may then
be shown to follow. Further interest in TQO  is catalyzed by the
prospect of  fault-tolerant topological quantum computation
\cite{kitaev}.


Notwithstanding the lack of an unambiguous accepted mathematical
definition, several inter-related  concepts are typically invoked in
connection to TQO: symmetry, degeneracy, fractionalization of quantum
numbers, maximal string correlations ({\it non-local} order), among
others. A fundamental question is {\it what} is needed to have a system
exhibiting TQO. The current article aims to show relations between
these {\it different} concepts, by rigorously defining and establishing
the equivalence between some of them and more lax relations amongst
others. Our main contribution is to provide a unifying symmetry
principle to characterize existent physical models and engineer new
systems displaying TQO. Clearly, this principle leads to unique
physical consequences that we will expand on below. Most importantly,
we (i) prove that systems harboring generalized $d$-dimensional (with
$d=0,1,2$) Gauge-Like Symmetries ($d$-GLSs) exhibit TQO; (ii) analyze
the resulting conservation laws and the emergence of topological terms
in the action of theories in high space dimensions (dimensional
reduction); (iii) affirm that the structure of the energy spectrum is
irrelevant for the existence of TQO (the {\it devil} is in the
states);  (iv) establish that, fractionalization, string correlators,
and entanglement entropy are insufficient criteria for TQO; (v) report
on a general algorithm for the construction of string and brane
correlators; (vi) suggest links between TQO and problems in graph
theory. The current work \cite{NOlong} is an  elaborate extension of 
Ref. \cite{NO} in which our central results were stated yet due to the
lack of space were not proven or explained in detail. 

The present work aims to investigate many recurring themes in the study
of  TQO. We start in Section \ref{classop} by classifying the degrees
of locality/non-locality of general operators acting on a physical
Hilbert space, and the possible spectral structures. 

In Section \ref{Section 2}, we define both zero and finite temperature
TQO. Thus far,  most work on TQO focused on the GS properties of
(gapped) quantum systems.  In this work, we formalize a notion of
finite temperature TQO. This notion is based on the influence of
boundary conditions on both local and topological observables. 

Then, in Section \ref{Section 3}, we review the existence of an
$SU(N_g)$ type symmetry (or of subgroups thereof) in a general system
with an $N_g$-fold degenerate GS.  As well known, $SU(N)$ symmetry can
lead to a fractionalization of the basic charge (e.g., the quark charges
which are multiple of $e/3$ in $SU(3)$ or {\it $N$-ality} in the
general  case). Thus degeneracy can lead to fractionalization in much
the same  way as the same groups lead to fractionalization elsewhere. A
similar link was also looked at anew lately by Oshikawa and  coworkers
\cite{oshikawa}.

Next, in Section \ref{Section 4}, we introduce one of our key results.
First, we introduce the concept of GLSs - symmetries that generally lie
midway between local symmetries (standard {\it gauge symmetries}) and
global symmetries. In general, these symmetry operators act on a
$d$-dimensional region. Here, $d=0$ correspond to local gauge
symmetries while $d=D$ operations with $D$ the dimensionality of the
system correspond to global symmetry operations. Intermediate GLSs
allow for $0 \le d \le D$. In Section \ref{SubSection 1}, we review
known examples of TQO and analyze the GLSs which are  present  in each
of these systems. In Section \ref{section12}, we introduce several
other systems which exhibit GLSs. In Section \ref{ABg}, we make links
between these symmetries and Aharonov-Bohm-type unitary 
transformations. 

In Section \ref{consequence}, we study the physical consequences of
having a system endowed with $d$-dimensional GLSs. In particular, 
we discuss the appearance of new conserved Noether and topological
charges in high-dimensional systems as a result of dimensional
reduction. We also explain the conditions to break those symmetries, and
show that well-defined quasi-particle poles are precluded in systems
that display low-dimensional GLSs. Finally, we comment on the relation
between charge fractionalization and GLSs, and the appearance of
low-dimensional topological terms in higher-dimensional theories. 

In Section \ref{Section 5}, we show that there exists an intimate 
relation between GLSs and TQO: GLSs of sufficiently low-dimensionality
$d$ mandate the appearance of TQO. At finite temperature, the proof
relies on a recent extension of Elitzur's theorem.   The results are
strongest for gapped systems which harbor {\em continuous} $d \le 2$
GLSs (Section \ref{gapsection}); here, the  requisite $T=0$ TQO can be
proven very generally. 
 
As we show in Section \ref{t0tqo} in other (also potentially non-gapped)
systems with either continuous and discrete GLSs, at zero temperature,
the results are also a consequence of selection rules. These rely on
selection  rules (such as the Wigner-Eckart theorem) as applied to the
GLS generators. As a particular corollary, these GLSs selection rules
imply (for Hamiltonians which contain finite-range interactions) a
degeneracy of energy levels  in the thermodynamic limit. These
selection rules further enable us to construct special GLSs eigenstates
in which $T=0$ TQO appears. To date, the prominent examples of zero
temperature TQO have GSs which are of that special form.  In Section
\ref{engineer}, we illustrate how these special states indeed appear as
the GSs of the Kitaev and related models. Moreover, we show how to
systematically engineer models that have $T=0$ TQO. 

In Section \ref{Energyspectrumsection}, we expand on the general
equivalence in the spectral structure between TQO systems on the one
hand  and systems with standard local orders on the other  \cite{NO}.
This affirms that there is no spectral signature of TQO. In this
Section, we prove that although Kitaev's model exhibits TQO, at any
finite  temperature the Toric code operators have vanishing expectation
values. This occurs notwithstanding the fact that the Kitaev model has
a finite gap between the ground and all other excited states. A
spectral gap is not sufficient to ensure that topological quantities
do  not have a vanishing expectation value at finite temperature. The
crux of this demonstration is that the Kitaev model (being a sum of
commuting operators) can be mapped onto the one-dimensional ($D=1$)
Ising model. Similar considerations apply to Wen's plaquette model
(which as we show is  none other than a rotated version of Kitaev's
model). These results may have important implications for Topological
Quantum Computation.

In Section \ref{eent}, we examine the entanglement entropy. We will
show that the quantitative  identification of TQO as a deviation from
an area law form recently suggested by several authors  differs from
the original definition of TQO as a robust {\it topological} order. At
best, all states in which TQO appears in  the entanglement entropy
should be a subset of  all robust {\it topological} orders which are
insensitive to all quasi-local perturbations. To make our arguments
sharper we introduce a new class of $SU(N)$ Klein models on a small
worlds network. 

In Section \ref{graphsection}, we make analogies between the
characterization of TQO orders and that of identifying the topology of
a graph. In the Appendix, we carry this analogy further and construct
wavefunctions on graphs for which TQO appears that we dubbed {\it
Gauge-Graph Wavefunctions}.

In Section \ref{section20}, we discuss one-dimensional  {\it string}
and higher-dimensional {\it brane correlations}. Such orders appear in
the $S$=1 AKLT chains as well as in other electronic systems (e.g.,
doped Hubbard chains). We show a general algorithm for the construction
of such correlations. Our approach is detailed for general gapped
systems (in any dimension) as well as in general entangled systems with
GS selection rules. In many cases, this construct leads to
the identification of a gauge-like structure. This structure should not
be confused with the GLSs.  It is not a symmetry of the system - rather
it corresponds to a transformation between the GSs of the systems to a
uniform factorizable state.  We show that systems (e.g. the AKLT spin
chain) can display string orders yet not exhibit  TQO. Moreover, we show
that the hidden {\it non-local} string order in the AKLT problem is equal to
the expectation value of a {\it local} nematic-type order.

In Section \ref{topology_not_enough}, we illustrate that systems can
have a degeneracy which depends on the topology of the manifold on
which they are embedded yet not be topologically ordered in the sense
of robustness to local perturbations.  Historically, such a dependence 
of the GS degeneracy on the  topology of the manifold on which the
system is embedded sparked much of the study  of TQO.

We conclude in Section \ref{section21}, with a summary of our results.
Technical but quite important details (some of them novel) have been
relegated to the various Appendices. These include Sections on
degenerate perturbation theory, generalized Elitzur's theorem, the
Peierls' problem and derivation of connecting symmetry operators,
relation between the quantum dimer and Kitaev's models, spin $S=1$
chains and the derivation of string operators as {\it partial
polarizers}.

\section{Classification of local and non-local operators}
\label{classop}

In the bulk of this work, we  will focus on quantum lattice systems
(and their continuum extension)  which have $N_s= \prod_{\mu=1}^{D}
L_{\mu}$  sites, with $L_\mu$ the number of sites along each space
direction $\mu$, and $D$ the  dimensionality of the lattice $\Lambda$. 
Unless stated otherwise, we will assume that $L_{\mu} = L$ for all
directions $\mu$. Associated to each lattice site (or mode, or bond,
etc.) ${\bf i} \in \Z^{N_s}$ there is a Hilbert space ${\cal H}_{\bf
i}$ of finite dimension ${\cal D}$. The Hilbert space is the tensor
product of the local state spaces, ${\cal H} = \bigotimes_{\bf i} {\cal
H}_{\bf i}$, in the case of distinguishable subsystems, or a proper
subspace in the case of indistinguishable ones. On a few occasions,  we
will briefly introduce extensions of our analysis to several network 
and graph systems. 

Statements about local order, TQO, fractionalization, entanglement,
etc., are made {\it relative} to the particular decomposition used to
describe the physical system. Typically, the most natural {\it local
language} \cite{GJW} is physically motivated: This is the essence of
local Quantum Field Theory.  For the present purposes we will classify
operators according to the following:

(a) {\bf Local}: $\hat{\cal O}_\bi$ (or finite linear combinations).

(b) {\bf Quasi-local}: $\prod_{\bi \in \Z^{M_s}}\hat{\cal O}_\bi$ (or
finite linear combinations), where $M_s$ represents a {\it finite}
(non-extensive) integer number (even in the thermodynamic limit).

(c) {\bf Non-local}: $\prod_{\bi \in \Z^{M_s}}\hat{\cal O}_\bi$ (or
finite linear combinations), where an extensive number $M_{s} \leq
N_s$  of lattice sites are involved (e.g., $M_s=L_1$ in a
$D$-dimensional lattice).

(d) {\bf Global}: $\sum_{\bi \in \Z^{N_s}}\hat{\cal O}_\bi$, or the
extensive sum of quasi-locals or non-locals.

(e) {\bf Quasi-global}: $\sum_{\bi \in \Z^{M_s}}\hat{\cal O}_\bi$, or the
extensive sum of quasi-locals or non-locals. $M_{s} \leq N_s$  is an
extensive integer number.

This classification characterizes either the product (cases (a)-(c)) or
the sum of local operators (cases (d)-(e)). For physical systems not
represented by an ${\cal H}$ with a tensor product structure, there is
no natural subsystem decomposition. In general one needs the concept of
{\it generalized entanglement} \cite{BKOV} to understand the nature of
the correlations of their quantum states. In particular, as we will see
below, the nature of entanglement in systems which harbor TQO is of a
highly non-local character (with respect to the {\it local language}),
as opposed to systems with {\it local} order parameters. 

The dynamics of the physical system will be governed by a generic
Hamiltonian $H$  whose form is constrained, in terms of the local
language \cite{GJW}, to linear combinations of (polynomial in $N_s$)
quasi-local operators. 

\begin{figure}
\centerline{\includegraphics[width=0.85\columnwidth]{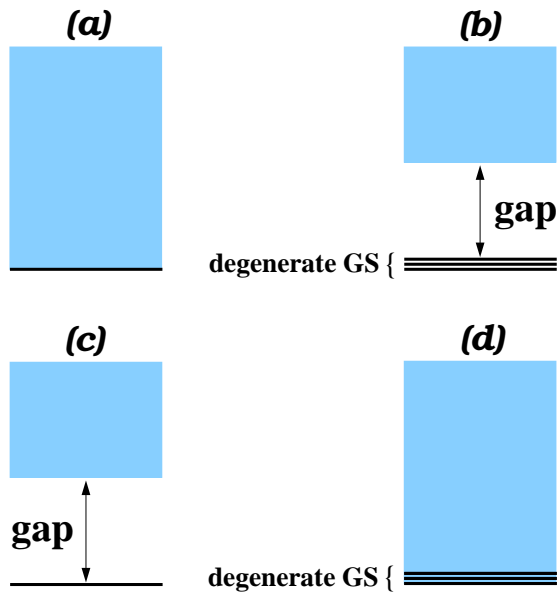}}
\caption{Schematic representation of potential energy spectra. Case
$(a)$  may represent a Fermi liquid, case $(b)$ a Fractional Quantum
Hall liquid, case $(c)$ a band insulator, and case $(d)$.}
\label{fig1}
\end{figure}

To put the subject of TQO in perspective let us start by tabulating all
of the possible low-energy spectra and discuss their relationship to 
different states of matter. Figure \ref{fig1} shows a schematics of the
possible low-energy spectra which is self-explanatory. Basically, the
spectrum may be gapless or gapped and the GS Hilbert subspace may be
degenerate or not. The main case of interest for the present manuscript
is case $(b)$. As we will see, however, in order to have a TQO state it
is not at all mandatory to have an energy spectrum like $(b)$ - symmetry
alone can potentially mandate that gapless systems also exhibit TQO. The
Lieb-Schulz-Mattis theorem pertaining to one dimension \cite{LSM} and
a  recent extension by Hastings \cite{Hastings} in higher dimensions
preclude, in the thermodynamic limit, rather general (locally
interacting)  $S=1/2$ systems from having energy spectra of type $(c)$
shown in Fig. \ref{fig1}.

\section{What is Topological Quantum Order (TQO)?}
\label{Section 2}

To determine what is needed for TQO, we start by defining it. Various
discussions in the literature emphasize the characteristics between
only a pair of states while others aim to link TQO to the general form
of the GS sector. To date, all of the treatments  seen in the
literature focused on zero temperature ($T=0$) TQO. We will discuss
$T=0$ TQO  below and then define its finite-$T$ extension.  In later
Sections, we provide results on finite-$T$ TQO.

\subsection{Zero temperature TQO}
\label{zerotempthing}

To allow  for greater flexibility and precision, we now define {\it
rank-$n$} TQO. Given a set of $n$ orthonormal GSs $\{|g_{\alpha}
\rangle\}_{\alpha=1,\cdots, n}$, with $1< n \le N_{g}$  where $N_{g}$
is the total number of GSs of a given Hamiltonian,  rank-$n$ TQO exists
iff for any bounded operator $V$ with compact support (i.e.  any  {\it
quasi-local operator} $V$),
\begin{eqnarray}
\langle g_{\alpha} | V | g_{\beta} \rangle = v \ \delta_{\alpha \beta} +
c_{\alpha \beta} ,
\label{def.}
\end{eqnarray}
where $v$ is a constant and $c_{\alpha \beta}$ is a correction that 
vanishes in the thermodynamic limit. The physical motivation for this
definition lies in the irrelevance of finite order terms in degenerate
perturbation theory for any quasi-local operator $V$. (In Appendix
\ref{app1}, we review the rudiments of degenerate perturbation theory.)
We will now demonstrate a simple result concerning the $T=0$ definition
of Eq.~(\ref{def.}) and then turn to the finite-$T$  extension.

{\bf Lemma.}
Given a set of orthonormal GSs $\{|g_{\alpha} \rangle\}$ which span an
$n$-dimensional space,  condition (\ref{def.}) holds iff for {\it any}
GS  in the same space, $|g \rangle = \sum_\alpha a_{\alpha} |g_{\alpha}
\rangle$  ($\sum_{\alpha} |a_{\alpha}|^{2} =1$), we have that
\begin{eqnarray}
\langle g | V| g \rangle = v + c_g ,
\label{general_cond}
\end{eqnarray}
with $c_g$ a correction that vanishes  in the thermodynamic limit. 

Proof:
Let us define the operator $V_{0} = P_0 VP_0$ where $P_0$ is the
projection operator  onto the GS basis.  Condition (\ref{general_cond})
then implies that  $\langle g | V_{0} | g \rangle = v + c_g$. If this
holds true for any  $|g \rangle$, it is also true for an orthonormal
set $\{|\tilde{g}_{\alpha} \rangle\}$ that diagonalizes $V_{0}$, i.e.
$V_0 |\tilde{g}_{\alpha} \rangle =
(v+\tilde{c}_\alpha)|\tilde{g}_{\alpha} \rangle$. In this basis
$V_{0}=\sum_\alpha (v+\tilde{c}_\alpha) \proj{\tilde{g}_{\alpha}}$.
Then, given another set of orthonormal GSs  [generally, linked by a
unitary transformation to the eigenstates of $V_{0}$]
$\ket{g_\alpha}=\sum_\beta a^\alpha_\beta \ket{\tilde{g}_\beta}$
\begin{eqnarray}
\langle g_{\alpha} | V | g_{\beta}  \rangle =  \langle g_{\alpha} |
V_{0} | g_{\beta} \rangle  = v \ \delta_{\alpha \beta} + c_{\alpha\beta},
\label{final+equiv}
\end{eqnarray}
with vanishing $c_{\alpha\beta}\!\!= \!\!\sum_\gamma \tilde{c}_\gamma
(a^\alpha_\gamma)^* a^\beta_\gamma$. Similarly,  (\ref{def.}) 
trivially implies (\ref{general_cond}): $\langle g | V| g \rangle
=\sum_{\alpha, \beta} (a_\alpha)^* a_\beta \langle g_{\alpha} | V |
g_{\beta} \rangle = v + c_g$, with $c_g=\sum_{\alpha, \beta}
(a_\alpha)^* a_\beta c_{\alpha\beta}$.  \qed

This simple Lemma naturally lends itself to a finite-$T$ generalization
which we turn to next.

In passing, we note that a similar but not as easily generalizable
relation  to finite-$T$ follows from the off-diagonal portion
of Eq.~(\ref{def.}): Equation~(\ref{def.}) is satisfied  if and only if
$\langle g_{\alpha}|V| g_{\beta} \rangle=c_{\alpha\beta}$ for any two
orthogonal states $|g_{\alpha} \rangle$ and $|g_{\beta} \rangle$ which
lie in the GS manifold.

\subsection{Finite temperature TQO}

Although the definition of TQO is cleanest for the $T=0$ problem, in
all physical problems of relevance, we must consider the effect of
finite temperatures. In \cite{NO}, we introduced for the first time, a
finite-$T$ ($T >0$) generalization of TQO. Extending the general $T=0$
condition  for TQO of Eq. (\ref{general_cond}), we have that
\begin{eqnarray}
\langle V \rangle_{\alpha}  \equiv \tr(\rho_{\alpha} V) &=& v 
+ c_{\alpha \alpha}(L) ,
\label{vt} 
\end{eqnarray} 
(independent of $\alpha$) with $c_{\alpha \alpha}$ a correction that
tends to zero in the thermodynamic limit and  $\rho_{\alpha} =
\exp[-H_{\alpha}/(k_{B} T)]$  a density matrix corresponding to the
Hamiltonian $H$ endowed with boundary terms which favor order in the
state  $|g_{\alpha} \rangle$ \cite{explain_field}. These boundary
terms  are exactly of the same form of  perturbations that allow us to
manipulate the system initially and set to be in  one or the other {\it
topological} GSs at $T=0$ for quantum computing purposes. In later
Sections, we will show that the excitation spectrum precisely
associated with  non-local $d=1$ discrete symmetry operators or $d=1,2$
continuous symmetries (which  is dominated by topological defects) 
enables finite-$T$ TQO to appear. We define finite-$T$ TQO to hold iff
Eq.~(\ref{vt}) holds at finite temperatures on those systems which also
exhibit the $T=0$ TQO defined by Eq.~(\ref{def.}).   As will become
apparent in later Sections (see, e.g., Section \ref{Tfragile}),  the
general finite-$T$ robustness condition of Eq. (\ref{vt}) by which we
define TQO  is different from finite expectation values of non-local
{\it topological} operators; the latter often  capture $T=0$ TQO in
many known instances. 

What is the relation between the definition of TQO above and its
sensitivity to  boundary conditions? To avoid the use of spurious
perturbations such as singular projection operators onto a particular
GS, we define the perturbations to correspond to boundary conditions
associated with a particular GS. A system exhibits  finite-$T$ TQO if
it (i) obeys the $T=0$ TQO conditions  of Eq. (\ref{def.}) and  (ii)
satisfies Eq. (\ref{vt})  for  $T>0$ with the expectation value
subscript $\alpha$  corresponding to the constraint of boundary
conditions which favor the state $\alpha$. If the expectation value of
any quasi-local operator $V$ is independent of the boundary conditions
$\alpha$ then the system exhibits TQO: the topological information on
$\alpha$ encoded in the boundary conditions is not accessible (and
protected from) any local  measurement.  In that sense, TQO systems are
similar to disordered systems in which no local order is found due to 
a condensation of topological defects. 

To make the definition of finite-$T$ TQO clear and intuitive, let us 
consider two well-known systems - the $U(1)$ gauge  theory which
exhibits TQO and the Ising ferromagnet in dimensions $D \ge 2$ which
does not exhibit finite-$T$ TQO.   Within the $U(1)$ gauge theory, if
we fix the gauge fields on the boundary of this gauge system, the
Aharonov-Bohm phase  is well specified. This phase (and fields on the
boundary) are unaltered by fluctuations at any place which does not
involve gauge fields  on the boundary of the system. By contrast, for a
ferromagnet, fluctuations inside the system (local perturbations) do
lift the degeneracy between different boundary conditions. Here, the
boundary conditions and the local fields are not independent of one
another. 

To make this distinction more transparent, we may set $V = \chi_{\bi}$
in Eq. (\ref{vt}). Here, $\chi_{\bi}$ is a pointer to a quasi-local
configuration about site $\bi$. That is, $\chi_{\bi} =1$ if that
quasi-local configuration appears about site $\bi$ and $\chi_{\bi} =0$
otherwise. Here, we will  have by Bayes' theorem
\begin{eqnarray}
\langle \chi_{\bi} \rangle_{\alpha} = P(\bi| \alpha) =  \frac{P(\bi
\cap \alpha)}{P(\alpha)}.
\label{Bayes}
\end{eqnarray}
In Eq.~(\ref{Bayes}),  $P(\bi|\alpha)$ is the conditional probability
of finding that local configuration about site $\bi$ given that the
boundary is of the type $\alpha$.  If various boundary configurations
are  equally probable - $P(\alpha) = P(\beta)$ then Eq. (\ref{vt})
implies that, up to exponential corrections, 
\begin{eqnarray}
P(\bi \cap \alpha) = P( \bi \cap \beta).
\label{probeq}
\end{eqnarray}
Equation (\ref{probeq}) implies that the {\it correlations} $P(\bi \cap
\alpha)$ between the local configuration ($\bi$)  and the boundary
conditions ($\alpha$) are the same for all boundary conditions. A
sufficient condition  for Eq. (\ref{probeq}) to hold is that all
correlations are sufficiently short ranged. Here, the only way to
determine anything about the boundary is to perform a non-local
measurement. Degeneracy between different topological sectors  or
boundary conditions cannot be lifted by local perturbations. The system
is robust to local  fluctuations. If we specify some {\it topological}
condition corresponding to a large set of possible boundary conditions
then as for every unique boundary condition, we have $\langle V
\rangle_{\alpha} = \langle V \rangle_{\beta}$ (up to exponential
corrections). Such a  topological sector can correspond to the total
number of vortices in an XY system, Hopf invariants, total domain wall
parity in an Ising chain, etc.  

In their disordered phase, systems often explicitly exhibits a
condensate  of the dual (disordering) fields. The dual  fields (domain
walls/vortices/$\cdots$) are  topological in character - they are a
product of an infinite number  of local operators (i.e. of
${\cal{O}}(L^{d})$ local  operators with $d>0$) which contains fields
all of  the way up to the boundary of the system. Here, the disordering
({\it defect}) fields are given by operators of the form
\begin{eqnarray}
\hat{T}_{\sf defect} = \prod_{\bf r \in \Omega} \hat{O}_{\bf r}
\label{Td}
\end{eqnarray}
with $\hat{O}_{\bf r}$ a quasi-local operator and  $\Omega$ a region
which spans ${\cal{O}}(L^{d})$  sites/bonds, etc. 

\section{Symmetry, degeneracy, and fractionalization}
\label{Section 3} 

As is well known, the presence of a symmetry allows for degeneracies
yet does mandate their presence par-tout. For instance, the trivial
global  $\Z_{2}$ up/down symmetry of the Ising magnet is intimately
tied to the twofold degeneracy of its model. More formally, if the
group of symmetries of the Hamiltonian $H$ is ${\cal G}=\{\fg_i\}$,
i.e. $[H,\fg_i]=0$, and $\ket{\Psi}$ is an eigenstate of $H$ with
eigenvalue $E$, then $\fg_i\ket{\Psi}$ is also an eigenstate with the
same eigenvalue. Symmetries do not mandate the existence of
degeneracies at all energies. It may happen (as it often does) that a
symmetry operator when acting on a certain state does not alter it at
all: the state transforms as a singlet under that
symmetry.

What is just as obvious yet not as often alluded to is that {\em
degeneracies (of any energy level) always imply the existence of
symmetries} (and indeed not the converse). The proof of this assertion
is trivial. In what follows, we focus on the GS manifold.  By a simple
relabeling, the same argument applied to any other energy level.

As the lowest energy subspace of $H$ will play a key role in the
following note, let us assume that the GS subspace, ${\cal
M}_0=\{\ket{g_\alpha} \}$, $\alpha \in {\cal S}_0=[1,N_g]$, is
degenerate with $\langle g_\alpha|g_\beta \rangle=\delta_{\alpha\beta}$
for $\alpha,\beta  \in {\cal S}_0$.  Let us now prove that this implies
the existence of an, at most, $SU(N_g)$ symmetry.  Let us arrange the
eigenstates of $H$ such that the first $N_g$ states are the degenerate
states in question $\{\ket{g_\alpha}\}$. The remaining eigenvectors,
${\cal M}_0^\perp=\{\ket{v_\beta} \}$ with $\beta \in {\cal
S}_0^\perp=N_g+1, \cdots$, form the orthogonal complement with $\langle
v_\alpha|v_\beta \rangle=\delta_{\alpha\beta}$ for $\alpha,\beta  \in
{\cal S}_0^\perp$. Next, we construct a matrix ${\cal A}$  such that 
\begin{eqnarray}
\langle g_\alpha | {\cal A} | g_\beta \rangle &=& \langle g_\alpha |
U_i | g_\beta \rangle ~  (\mbox{if }~ \alpha,\beta \in {\cal S}_0) ,
\nonumber \\ 
\langle v_\alpha | {\cal A} | v_\beta \rangle &=& \delta_{\alpha\beta}~ 
(\mbox{if }~ \alpha,\beta \in   {\cal S}_0^\perp) , \nonumber \\  
\langle g_\alpha | {\cal A} | v_\beta \rangle &=& 0 ~(\mbox{if } ~\alpha
\in {\cal S}_0 ~ \mbox{and} ~   \beta \in  {\cal S}_0^\perp) ,
\end{eqnarray}
with $U_i$ any $SU(N_g)$ operator. It is straightforward to prove that
${\cal A}$ is a unitary matrix, ${\cal A}{\cal A}^{\dagger} = \one$. 
Moreover, in this eigenbasis, we see that $[H,{\cal A}]=0$. Thus,
$H^{\prime} = {\cal A}^\dagger H {\cal A}$  commutes with $H$ and
shares the same set of eigenvalues. This illustrates that
${\cal{A}}$ realizes an $SU(N_{g})$ symmetry \cite{wk}
of the GS manifold.

For pedagogical purposes and clarity, we revisit  the link between
symmetry groups (in our case GS symmetry groups) and fractionalization.
Although some of these ideas have been around, we have not seen the
direct derivation of our central and general approach to
fractionalization: deducing its  existence by an analysis of the  
unitary group which resides on the GS manifold which we just derived.
Most treatments to date either focused on the unitary group of the 
fundamental interactions or on counting arguments of GSs or associated
zero modes. Here, we will show how the unitary group acting within the
GS manifold directly unifies the  two treatments in leads to
fractionalization. 
 
The notion of fractionalization is defined relative to the original
degrees of freedom of the system.  For $N_g$ degenerate GSs, we saw
earlier that an $SU(N_g)$ group residing within the internal space of
the degenerate manifold is a symmetry of the Hamiltonian, $[H,U]=0$. In
many cases, the  $SU(N_g)$ symmetry may splinter into small subgroups
which suffice to characterize physical aspects of the GS manifold. 
Fractional charges are associated with the center of such symmetry
groups. The center of the group is enough to characterize the coupling
of the elements of the group to the $U(1)$ gauge potential (the
electric charge). The center of the group $SU(N_g)$ is the modular
additive group $\Z_{N_g}$ which can be associated with topological
homotopy groups. That $\Z_{N_g}$ is the center is seen by noting that
if any element of the group commutes with all other elements then it
must be a multiple of the  identity. Any such unitary matrix, having
equal diagonal only elements must have equal pure phases ($z$) along
the diagonal. For the matrices to have unit determinant, $z^{N_g} =1$.
This, in turn, immediately leads us to see that the center of the group
is $\Z_{N_g}$. If we complete a $2 \pi$ revolution, then the
Aharonov-Bohm phase for a integer quantized charge  would lead to no
change (the  exponentiated phase factor would be 1). For an $SU(2)$
representation however,  for half-integer spins, the elements of
$SU(2)$ suffer a factor of $(-1)$. This may be seen as an Aharonov-Bohm
phase of a charge $q=1/2$. This leads us to see that in the fundamental
representation $S=1/2$ and  that a general spin $S$ can only be integer
multiples thereof. Similarly, for the lowest  rank faithful
representation of $SU(3)$, the phases are $e^{2 \pi i/3}$. The latter
may be similarly viewed as the Aharonov-Bohm phase of a charge $1/3$
object when taken around a quantized monopole. In quantum
chromodynamics (governed by $SU(3)$), the fractional $1/N$ charge
adduced to $SU(N)$ goes by the name of {\it $N$-ality}.  This is
related to the simpler fractional charge as deduced from the
Su-Schrieffer counting arguments in polyacetylene  \cite{fraction,NS}.
A related approach was recently highlighted by \cite{oshikawa}. It is
important to stress that other (non-fundamental) charges may appear:
our discussion above only leads to a lower bound for the  simplest
always present $SU(N_{g})$ symmetry. Furthermore, in several
non-Abelian systems (various Quantum Hall states and vortices  in p+ip
superconductors) \cite{oksnt} there are symmetry operators which link
GSs of systems which are related by a gauge symmetry - there the
considerations are richer than in the usual  $N$-ality counting
arguments. 

Degeneracies imply the existence of symmetries which effect general
unitary transformations within the degenerate manifold and act as the
identity operator outside it. In this way, GS degeneracies act as exact
GLSs. The existence of such unitary symmetry operators (generally a
subset of $SU(N)$) allows for fractional charge  ($N$-ality in the
$SU(3)$ terminology of quantum chromodynamics) implying  that
degeneracy allows for fractionalization defined by the center of the
symmetry group. The ($m$-)rized Peierls chains \cite{fraction,NS}
constitute a typical example of a system with universal
($m-$independent) symmetry operators,  where fractional charge
quantized in units of  $e^{*} = e/m$ with  $e$ the electronic charge is
known to occur  \cite{fraction}. In the Appendix, we will provide
explicit expressions for the  symmetry operations which link all of the
GSs to one another.  Different Peierls chain GSs break  discrete
symmetries in this $D=1$ system: this leads to a violation of
Eq. (\ref{def.}) in this system with  fractionalization. The Fermi
number $N_{f}$  in the Peierls chain and related Dirac-like theories 
is an integral over spectral functions \cite{NS};  the fractional
portion of $N_{f}$ stems from  soliton contributions and is invariant
under local background deformations \cite{NS}.

\section{Gauge-like symmetry: A definition}
\label{Section 4}

A $d$-GLS of a theory given by $H$ (or action $S$) is a group of
symmetry transformations ${\cal{G}}_d$ such that the minimal non-empty
set of fields $\phi_{\bf i}$ changed by the group operations spans a
$d$-dimensional subset  ${\cal C} \subset \Lambda$.  Here, $\Lambda$
denotes the entire lattice (or space) on which the  theory is defined.
These transformations can be expressed as \cite{BN}:
\begin{equation}
{U}_{{\bl\bk}} = \prod_{{\bf i} \in {\cal C}_{\bl}} {\bf g}_{{\bf
i}{\bk}},
\label{tran}
\end{equation}
where ${\cal C}_{\bl}$ denotes the subregion $\bl$, and $\Lambda=
\bigcup_{\bl} {\cal C}_{\bl}$. (The extension of this definition to
the  continuum is straightforward.) Gauge (local) symmetries correspond
to $d=0$, while in global symmetries the region influenced by the
symmetry operation is $d=D$-dimensional. These symmetries may be
Abelian or non-Abelian. In the next two Sections (Sections
\ref{SubSection 1}, and \ref{section12}), we analyze the  $d$-GLSs
present in various systems in great detail.  These Sections are lengthy
and aimed at being pedagogical. In Section \ref{SubSection 1}, we
review the prominent examples of TQO and identify the $d$-GLSs present
in each of these systems. In Section \ref{section12}, we discuss other
well known physical systems which  exhibit $d$-GLSs and are candidates
for TQO.  After this long list of examples in which $d$-GLSs are
present we turn in the following Sections to the consequences of
$d$-GLSs. After a list of properties relating to low-dimensional
behavior (for low $d$) enforced in high-dimensional systems (Section
\ref{consequence}), we turn to our central result concerning the link
between TQO and $d$-GLSs. In particular, in Section \ref{Section 5}, we
show how exact low-dimensional $d$-GLSs mandate TQO.

\section{Known examples of TQO and their Gauge-Like Symmetries}
\label{SubSection 1}

We now list the most prominent examples of TQO and  identify the
$d$-GLSs present in each of these systems. These symmetries either
appear exactly in the system or appear in some limit in a system which
is adiabatically connected to the system in question. Symmetries do not
need to be exact in order to exert their influence: so long as points
in parameter space may be adiabatically linked to each other, they lie
in the same phase associated  with the same set of symmetries.  At
issue is whether or not such a smooth adiabatic connection exists.  In
what follows, we start with a list of systems in which the $d$-GLSs are
exact (examples (a)-(c)). We then turn to  the Fractional Quantum Hall
Effect (FQHE) (in (d)) where a clever choice of parameters
adiabatically  links the $D=2$ system to a  $D=1$ in which a discrete
$d=1$ symmetry is present. It is important to stress that in any
Quantum Hall system, the two-dimensional magnetic translation group is
an exact symmetry  which links all GSs.  However, a more transparent
understanding  is attained by considering the one-dimensional limit. 
In (e), we comment on the similarities between the symmetries found in
the suggested  non-Abelian states at the core of vortices in (p+ip)
superconductors to the symmetries of the FQHE problem. In both (d) and
(e), the physical picture behind a  one-dimensional ($d=1$) symmetry
which links different GSs (an evolution around a toric cycle) is
intimately linked to suggestions regarding the use of such braiding
operations in quantum computing.

\begin{figure}
\centerline{\includegraphics[width=0.73\columnwidth]{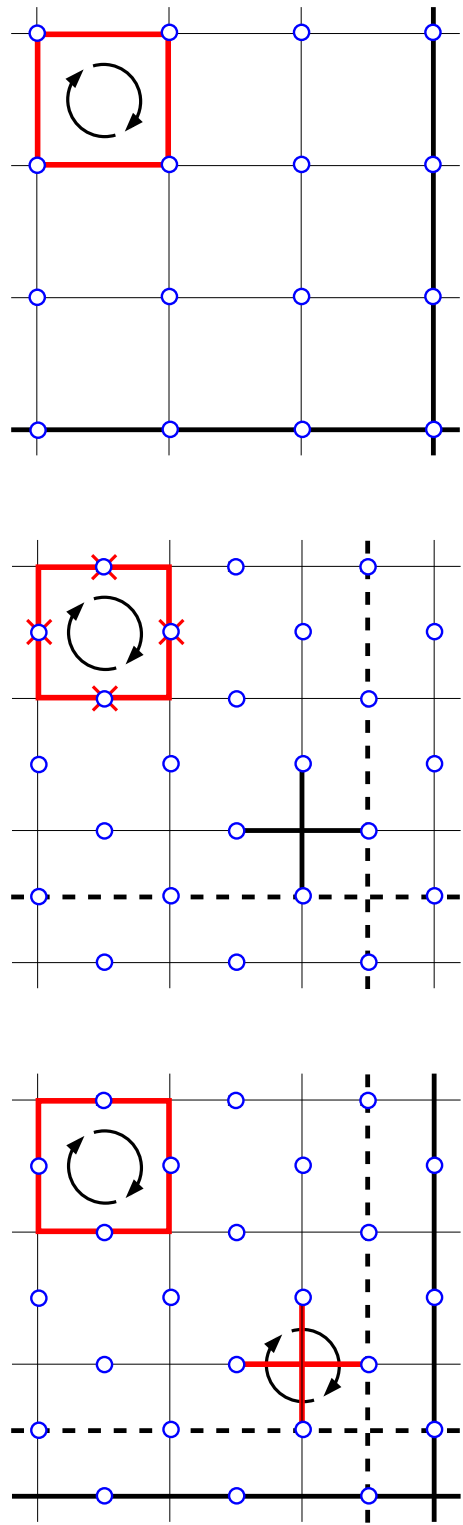}}
\caption{Schematics of the interactions and symmetries involved in
three of the known examples which display TQO. The lower panel
represents Kitaev's Toric-code model, the middle one corresponds to the
$\Z_{2}$ gauge theory and the upper panel corresponds to Wen's
plaquette model. Dark lines (solid or dashed) represent symmetries
while the brighter (red) lines or crosses correspond to the interaction
terms. The hollow circles represent the spin locations. See text.}
\label{fig1_Model}
\end{figure}

(a) In {\em{Kitaev's Toric code model}} on the square lattice, the
Hamiltonian \cite{kitaev}
\begin{eqnarray}
H_{K} = -\sum_{s} A_{s} -\sum_{p} B_{p}
\label{kitaevmodel}
\end{eqnarray}
where 
\begin{eqnarray}
A_{s} = \prod_{ij \in {\sf star}(s)} \sigma_{ij}^{x}, 
~~B_{p} = \prod_{ij \in {\sf boundary}(p)} \sigma_{ij}^{z}.
\label{AB_defn}
\end{eqnarray}
An illustration is provided in the bottom panel of
Fig.~\ref{fig1_Model}. The product defining $A_{s}$ spans all bonds
$ij$ which  have site $s$ as an endpoint (see the cross-shaped object
in Fig. \ref{fig1_Model}).  The plaquette product $B_{p}$ spans all
bonds which lie in the plaquette $p$ (see the plaquette in the bottom
panel of  Fig. \ref{fig1_Model}). In the presence of periodic boundary
conditions, this model possesses two $d=1$ $\Z_{2}$ symmetries which
are given by 
\cite{kitaev}
\begin{eqnarray}
Z_{1,2} = \prod_{ij \in C_{1,2}} \sigma_{ij}^{z}, ~X_{1,2} = \prod_{ij
\in C^{\prime}_{1,2}} \sigma^{x}_{ij},  ~\{X_{i}, Z_{j} \} =
\delta_{ij}.
\label{kit}
\end{eqnarray}
These $d=1$ symmetries are denoted in the lower panel of 
Fig.~\ref{fig1_Model} by the dark solid and dashed  lines. Along the
solid lines a product of the $\sigma^{z}_{ij}$ fields is taken while
along the dashed lines a product  of the $\sigma^{x}_{ij}$ fields is
performed. The contour $C_{1}$ represents the solid horizontal line,
$C_{2}$ denotes the solid vertical line, $C_{1}^{\prime}$ is the dashed
vertical line, and $C_{2}^{\prime}$ is the dashed horizontal line. That
these are, in our definition of Eq.~(\ref{tran}), $d=1$ $\Z_{2}$
symmetries  follows as the regions $\{C_{l}\}$ are lines ($d=1)$ on
which $\Z_{2}$ generators $(\sigma_{ij}^{a=x,z})$ are placed. The {\it
topological} symmetries of Eq. (\ref{kit}) are none other than discrete
$d=1$ symmetries in the general classification of GLSs.

(b) {\em Gauge theories.} 
The Kitaev model of (a) may be viewed  as a $\Z_{2}$ gauge theory in
which usual local gauge symmetries \cite{kogut} have been removed by
hand, leaving bare only  the two additional $d=1$ symmetries that a
$\Z_{2}$ gauge theory exhibits. On a square lattice, the $\Z_{2}$ gauge
theory Hamiltonian reads
\begin{eqnarray}
H_{\Z_{2}} = - K \sum_{\Box} \sigma_{ij}^{z} \sigma_{jk}^{z} \sigma_{kl}^{z} 
\sigma_{li}^{z} - h_{x} \sum_{ij} \sigma_{ij}^{x}.
\label{gz2+}
\end{eqnarray} 
Here, aside from the $d=1$ symmetries $X_{1,2}$ of  Eq. (\ref{kit}),
there are also the standard local $(d=0$) gauge symmetries
$\{A_{s}\}$.  This set of low-dimensional ($d=0,1)$ GLSs suffices to
link all GSs to each other. As a consequence, this gauge theory is an
example of TQO. (Indeed, gauge theories form the most prominent
examples of TQO.)

The middle panel Fig.~\ref{fig1_Model} provides a cartoon of the
interactions of Eq.~(\ref{gz2+}) - the bright (red) plaquette terms and
crosses (on site magnetic field)  as well as the symmetries of the
system (the dark dashed lines representing the the $d=1$ symmetries
$X_{1,2}$ and the dark solid cross representing the $d=0$ symmetries
captured by $\{A_{s}\}$  of Eq.~(\ref{AB_defn})).

(c) {\em{Wen's plaquette model}} \cite{wen_plaq}  is given by 
\begin{eqnarray}
H_{W} = - K \sum_{i} \sigma_{i}^{x} \sigma_{i+ \hat{e}_{x}}^{y}
\sigma_{i + \hat{e}_{x}+\hat{e}_{y}}^{x} \sigma_{i+\hat{e}_{y}}^{y},
\label{hw}
\end{eqnarray}
with all fields lying on the vertices of a square lattice. This system
displays $d=1$ symmetries whose generators 
\begin{eqnarray}
\hat{O}_{P} = \prod_{i \in  ~P} 
\sigma_{i}^{z}.
\label{wensym1}
\end{eqnarray}
With the line $P$ chosen to be any horizontal or vertical line of sites
we find that $[H, \hat{O}_{P}]=0$. Similar symmetry operators can be
extracted for the other known examples of TQO. An illustration of the
interactions and the symmetries is provided by the top panel of
Fig.~\ref{fig1_Model}. Here, the solid dark lines denote the
symmetries  of Eq.~(\ref{wensym1}). The plaquette  in
Fig.~\ref{fig1_Model} denotes the interaction terms in Eq.~(\ref{hw}). 
As in Kitaev's Toric code model (of (a)) where all operators appearing
in the summand of the Hamiltonian commute, all of the operators in the
summand of Eq.~(\ref{hw}) commute with one another. We will later show
(Section \ref{wentemp})  that Wen's model is identically the same as
Kitaev's model when written in a rotated basis. The gauge theory of (b)
in the  absence of an applied field [Eq.~(\ref{gz2+}) with $h_{x}=0$]
also has an energy spectrum which is identical to that of the Kitaev's
Toric model (of (a)) or Wen's plaquette mode (of (c)). As we will show
in  Section \ref{Energyspectrumsection}, all of these systems have an
energy spectrum which is identical to that of the Ising chain.
Schematically, we write
\newline
\newline
Kitaev's model = Wen's model $\leftrightarrow$ $D$=2 Ising gauge.
\newline
\newline

(d) {\em Fractional Quantum Hall States}

Fractional quantum Hall (FQH) states have long been regarded as the
quintessential example of TQO. These states are realized in $D=2$
electronic systems subjected to a strong magnetic field $\vec{B}$
applied perpendicular to the plane where the electrons reside. A minimal
model Hamiltonian (for $N_e$ electrons in a surface of size $[L_{x}
\times L_{y}]$) that captures the FQH physics is 
\begin{equation}
H = \frac{1}{2m} \sum_{i=1}^{N_e}  \vec{\Pi}_i^{2}  + \sum_{i<j}
V(|{\vec{r}_i}- \vec{r}_j|) ,
\label{QHSS}
\end{equation}
with canonical momentum $\vec{\Pi}_i=\vec{p}_{i} - \frac{e}{c}
\vec{A}_{i}$, $\vec{A}_i=\vec{A}(\vec{r}_i)$ the vector potential
corresponding to a uniform magnetic field $\vec{B}=\nabla \wedge
\vec{A}$, and $V(|{\vec{r}_i}- \vec{r}_j|)$ a two-body (Coulomb)
interaction. 

This system has a magnetic translation group (reviewed in the Appendix)
which resides in $d=2$ dimensions. It has long been recognized,
however, \cite{weniu} that the symmetry operators linking the 
different GSs also correspond to operators which move a  quasi-particle
and quasi-hole along a $d=1$ trajectory - a symmetry of a $d=1$
character in the scheme of Eq.~(\ref{tran}).  This symmetry relates to
the magnetic translation group. Recent work \cite{Karl,sdh} allows us
to make this intuition more precise and  cast these symmetries as
explicit $d=1$ symmetry operators which nicely fit in our general
scheme of classifying the symmetries which enable TQO. As shown in 
\cite{Karl,sdh},  the Quantum Hall problem in $D=2$ dimensions is
adiabatically connected to a $D=1$ Peierls-type problem where
the magnetic translation symmetries which we briefly review in the
Appendix  become explicitly discrete symmetries of dimension $d=1$ 
\cite{Karl,sdh}. 

This dimensional reduction makes explicit the one-dimensional $(d=1)$
character of the magnetic translation group operator which is
responsible not only for topological order but also re-explains in
greater depth why the degeneracy of  a Quantum Hall system scales in
the way that it does,  see e.g. \cite{oshikawa,oksnt,weniu}.  This
relies on the correspondence with the discrete $d=1$ symmetries of the
Peierls chain to which we briefly alluded to in the Section
\ref{Section 3} and which we explore in detail in the Appendix.  In an
Abelian Quantum Hall system of  filling fraction $\nu = p/q$ with $p$
and $q$ relatively prime integers which is embedded on a manifold of
genus $g$ (a manifold which  has $g$ {\it handles}) the GS degeneracy
scales as $q^{g}$ as would be expected for a system harboring $q$
independent discrete $d=1$ translation symmetries. This is the
degeneracy of an open chain  of degeneracy $q$ when punctured by
$(g-1)$ holes.  

By contrast, the degeneracy for a system with two independent discrete
symmetries in two dimensions would have a degeneracy which is expected
to scale  as $q^{g} \times q^{g}$. Similar considerations, albeit
slightly more sophisticated, apply in the non-Abelian arena
\cite{oksnt,Karl,sdh}.

(e) {\em Majorana Fermions in Vortex cores of p+ip 
superconductors}

The $d=1$ symmetry operators of  an evolution of a quasi-particle
around a  toric cycle is very similar to that appearing in the Quantum
Hall problem \cite{oksnt}. The considerations outlined above can be
replicated for these systems. These $d=1$  (braiding) symmetries in
non-Abelian states of  FQHE and in (p+ip) have received  much attention
because of their viable use for quantum computation.

\section{Other systems with Gauge-Like Symmetries}
\label{section12}

We now present several other systems which have low-dimensional GLSs.
Following the proofs to be provided  in Section \ref{t0tqo},
many of these systems exhibit TQO.

In the systems of ({\bf{a (i)}}) and ({\bf d}), a low energy {\em
emergent} lower-dimensional GLS appears which  is not sufficient in
leading to finite-$T$  TQO. In all other systems introduced in this
Section, the  $d$-GLSs which appear are exact and  hold over the entire
spectrum. In what follows, we will discuss whether  or not these
systems may exhibit broken symmetries - our conclusions rely on the
theorem of \cite{BN} which  is stated anew in Appendix \ref{app2}. As
we  will explicitly show in Section \ref{Section 5},  if all GSs can be
linked by the exclusive use of low $d$-GLSs  ($d \le 2$ continuous
symmetries or $d \le 1$ discrete symmetries) then finite-$T$ TQO  may
follow .

{\bf{a}}) {\it Orbitals} - In transition metal (TM) systems on cubic
lattices, each TM atom is surrounded by  an octahedral cage of oxygens.
Crystal fields lift the degeneracy of the five 3d orbitals of the TM to
two higher energy $e_{g}$  levels ($|d_{3z^{2}-r^{2}} \rangle$ and $|
d_{x^{2}-y^{2}} \rangle$) and to three lower energy $t_{2g}$ levels ($|
d_{xy} \rangle$, $| d_{xz} \rangle$, and $| d_{yz} \rangle$).  A
super-exchange calculation leads to the  Kugel-Khomskii (KK)
Hamiltonian  \cite{KK,Brink03}
\begin{equation}
\label{1}
H = \sum_{\langle r,r' \rangle} H_{orb}^{r,r'} 
(\vec{S}_{r}\cdot \vec{S}_{r'} + \frac{1}{4}).
\end{equation}
Here, $\vec{S}_{r}$ denotes the spin of the electron at site~$r$  and
$H_{orb}^{r,r'}$ are operators acting on the orbital degrees of
freedom. For TM-atoms arranged in a cubic lattice,
\begin{eqnarray}
\label{orbHam}
H_{orb}^{r,r'} = J(4\hat{\pi}_{r}^\alpha \hat{\pi}_{r'}^\alpha 
-2\hat{\pi}_{r}^\alpha - 2\hat{\pi}_{r'}^\alpha+1),
\end{eqnarray}
where ~$\hat{\pi}_{r}^\alpha$ are orbital pseudo-spins and
$\alpha=x,y,z$ is the direction of the bond~$\langle r, r' \rangle$. 

\begin{figure}[t]
\refstepcounter{obrazek}
\label{figrubick}
\centerline{\includegraphics[width=3.74in]{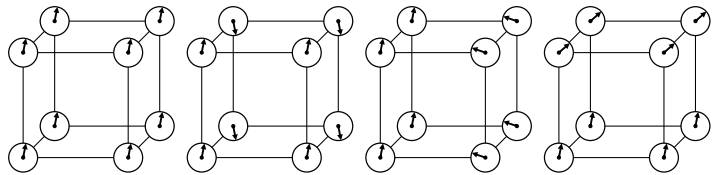}}
\medskip
\caption{From Refs. \cite{NBCv,BCN}. The symmetries of Eq.
(\ref{symorb}) applied on the uniform state (at left).}
\end{figure}

{\bf(i)}  In the ~$e_{g}$ compounds,
\begin{eqnarray}
\hat\pi_{r}^{x,y}=\frac{1}{4}(-\sigma_{r}^z \pm \sqrt{3}\sigma_{r}^x),\qquad
\hat\pi_{r}^z= \frac{1}{2}\sigma_{r}^z.
\label{120_op}
\end{eqnarray}
This also defines the orbital only ``120$^\circ$-Hamiltonian'' which is
given by 
\begin{equation}
H_{orb} = J \sum_{r,r'} \sum_{\alpha=x,y,z} \hat{\pi}_{r}^{\alpha} 
\hat{\pi}_{r+ \hat{e}_{\alpha}}^{\alpha}. 
\label{orb}
\end{equation}
Jahn-Teller effects in $e_{g}$ compounds also lead, on their own, to
orbital interactions of the 120$^\circ$-type \cite{Brink03}. The 
``$120^{\circ}$ model'' model of Eqs. (\ref{120_op}), and (\ref{orb})
displays discrete ($d=2$)  $[\Z_{2}]^{3L}$ GLSs  (corresponding to
planar Rubik's-cube-like reflections about internal spin directions -
Fig. \ref{figrubick}).  Here, there is only an {\it emergent}
symmetry in the GS sector of the classical ($S \to \infty$) rendition
of the problem. In this rendition the spin-1/2 operators $\sigma$ are
replaced by spin-S generators. These symmetry operators  $\hat{O}^{\alpha}$
are \cite{BN,NBCv,BCN}
\begin{eqnarray}
\hat{O}^{\alpha} = \prod_{r \in P_{\alpha}} \hat{\pi}_{r}^{\alpha}.
\label{symorb}
\end{eqnarray}
These operators correspond to a rotation by 180 degrees about the
directions $(- \frac{1}{2}, \pm \frac{\sqrt{3}}{2})$  and $(0,1)$ in
the internal $xz$ spin plane.   Here, $\alpha = x,y,z$ and $P_{\alpha}$
may denote any plane orthogonal to the cubic  $\hat{e}_{\alpha}$ axis.
These  are discrete $d=2$ symmetries which emerge  only within the low
energy sector and may be broken at finite temperature.  Within the
large $S$ limit, and at finite temperatures, the discrete $d=2$
$\Z_{2}$ symmetry is broken by entropic fluctuations \cite{NBCv,BCN}. 

\begin{figure}
\centerline{\includegraphics[width=1.1\columnwidth]{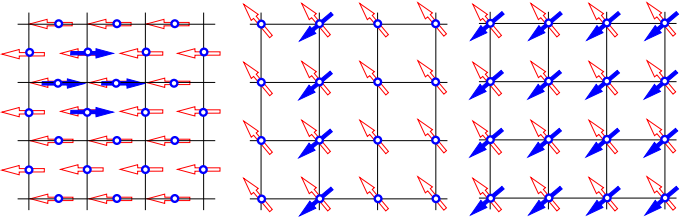}}
\caption{Schematics of the interactions and symmetries involved in the
classical rendition of three $D=2$ examples. The left panel represents
(i) the local ($d=0$) symmetries of the Ising gauge theory [Eq.
(\ref{gz2+})].   The middle panel represents (ii) an orbital compass
model  [Eqs. (\ref{orb}), and (\ref{compass1})] with $d=1$ symmetries;
here the symmetry operations span lines [see Eq. (\ref{symorb}).  The
right panel depicts (iii) an XY model with $d=2$ symmetries; the symmetry
here spans the entire $D=2$ dimensional plane.}
\label{examples}
\end{figure}

{\bf (ii)} In the $t_{2g}$ compounds  (e.g., LaTiO$_3$), we have in
$H_{orb}$ of Eq. (\ref{orb}) 
\begin{eqnarray}
\hat\pi_{r}^\alpha = 
\frac{1}{2} \sigma_{r}^\alpha.
\label{compass1}
\end{eqnarray} 
This is called the \emph{orbital compass} model
\cite{NBCv,BCN,Mishra,NF,Brink03}, The symmetries of this Hamiltonian
are given  by Eqs. (\ref{symorb}), and (\ref{compass1}).  In the $D=3$
model of Eq.~(\ref{orb}),  rotations of individual lower-dimensional
planes  about an axis orthogonal to them leave the system invariant
[see Fig.~\ref{examples}].  It should be noted that here Berry phase
terms lift the $d=1$ $\Z_{2}$ symmetries  present in the (large $S$)
classical orbital compass model  (the $b$ spin component $S_{b} \to (1
- 2 \delta_{ab}) S_{b}$  along a ray parallel to the spatial $a-$th
axis) to a weaker $d=2$ symmetry  (whose lattice version is given in
Eq. (\ref{symorb})). Insofar as their algebra relations are concerned,
in $D=2$ variants  of the orbital compass model of Eq.~(\ref{orb}) (in
which $\alpha =x,z$), where  the planes $P$ become one-dimensional
lines parallel to the coordinate axes, we can regard $O^{P}$  as
$S=1/2$ operators. For example, for a certain given plane $P$
orthogonal to the $x$ axis, $O^{P}$ is set to be $\tau^{x}$ and for a
plane orthogonal to the $z-$ axis, we set $O^{P}$ to be $\tau^{z}$. The
operators $\{\tau_{x}, \tau_{z}\}$ satisfy a spin $S=1/2$ $SU(2)$
algebra. In this regard, these operators  can be viewed as Abelian {\it
anyons}. Here, 
\begin{eqnarray}
\hat{O}_{x} \hat{O}_{y} = \hat{O}_{y} \hat{O}_{x} e^{2i \theta}
\label{qhs}
\end{eqnarray} 
with $\theta  = \pi/4$ ({\it semions}).  A correspondence between the
statistics $(\theta)$ and toric ($d=1$) translation operators in, for
example the Quantum Hall problem is reviewed in \cite{oksnt}. 

The set of $d=1$ symmetries discussed  above does not suffice to link
all of the GSs. In the $d=2$ orbital compass model, an additional 
global discrete symmetry is required. The $D=2$ system displays an
additional independent $\Z_{2}$ reflection symmetry ($\sigma_{x} \to
\sigma_{z}$, $\sigma_{z} \to \sigma_{x})$ - a rotation by $\pi$ about
the symmetric line (the 45 degree line in the  plane), i.e. 
\begin{eqnarray}
\hat{O}_{\sf Reflection} = \prod_{i} \exp[i \frac{\pi \sqrt{2}}{4}
(\sigma_{i}^{x} + \sigma_{i}^{z})].
\label{reflbad}
\end{eqnarray}
This symmetry ($\hat{O}_{\sf Reflection}$) is a manifestation of a
self-duality present in the model. The $D=3$ orbital compass model
displays similar reflection symmetries (permutations). As discrete
$d=2$ Ising  symmetries can be broken, the $D=2$ orbital  compass
systems display finite-$T$ nematic orders \cite{NBCv,BCN,Mishra},
\begin{eqnarray}
Q = \langle \sigma_{i}^{x} \sigma_{i+\hat{e}_{x}}^{x}
- \sigma_{i}^{y} \sigma_{i+\hat{e}_{y}}^{y} \rangle.  
\end{eqnarray}
The additional reflection symmetry of Eq.~(\ref{reflbad}), is needed
in addition the $d=1$ discussed earlier, to link all GSs to each other.
As we will later discuss, such a high-dimensional $(d=2)$  Ising
symmetry can be broken (and indeed is) \cite{NBCv,BCN,Mishra}. Such a
broken symmetry does not make this system a candidate for TQO. This
symmetry is removed in strained variants of the orbital compass model
of Eq.~(\ref{orb}), which are given by
\begin{equation}
H_{orb} = \sum_{r,r'} \sum_{\alpha=x,y,z} J_{\alpha} 
\hat{\pi}_{r}^{\alpha} 
\hat{\pi}_{r+ \hat{e}_{\alpha}}^{\alpha},
\label{orb_strain}
\end{equation}
with all $\{J_{\alpha}\}$ different. It is readily verified that no
$d>1$ symmetries exist for the system of Eq.~(\ref{orb_strain}). 
Orbital systems approximated by Eq. (\ref{orb_strain}) are candidates
for TQO. It should be noted that these symmetries naturally allow for a
high degeneracy of most levels, yet they do not mandate it. For
instance, a state which is an equal amplitude  superposition of all
states in which  the number of sites $i$ on which $\sigma^{z}_{i} = 1$
is  even on every row and on every column is invariant under the $d=1$
symmetries of the $D=2$ orbital compass model. 

To make contact with the previously examined models, we remark that
upon performing standard $D=1$  spin duality transformations along the
diagonals (see \cite{NF}), we can see that the transverse field
$\Z_{2}$  gauge theory of Eq.~(\ref{gz2+}) is dual to an orbital
compass model in which half of the interactions corresponding to one
polarization direction of the spins are missing. Alternatively, if we
define, by reference to Eq. (\ref{gz2+}), 
\begin{eqnarray}
H_{C} \equiv H_{\Z_{2}} - K \sum_{i} C_{i},
\end{eqnarray}
with 
\begin{eqnarray}
C_{i} = \prod_{j} \sigma^{z}_{ij}, 
\end{eqnarray}
a {\it star} term of the same geometry as that of $A_{s}$ in
Eq.~(\ref{AB_defn}) then 
\begin{eqnarray}
\mbox{\sf Spec}\{H_{c}\} = \mbox{\sf Spec}\{H_{orb}[J_{x}= K, J_{y}= h]\}.
\label{specdualt}
\end{eqnarray}
In Eq.~(\ref{specdualt}),  {\sf Spec} denotes the energy spectrum
of the theory.   
 
\newcounter{obrazek}

In the classical (large $S$) limit of the orbital compass model, the
symmetries discussed above mandate that each state is, at least,
$2^{L}$ degenerate \cite{NBCv,BCN}.  As noted by \cite{doucot} and
nicely expanded on by \cite{dorier}, in the quantum problem, the
non-commuting nature of the symmetry operators $\hat{O}_{P_{1}}$ and
$\hat{O}_{P_{2}}$ for two orthogonal planes $P_{1}$ and $P_{2}$
mandates that the GS of the $D=2$-dimensional orbital compass model is,
at least, two fold degenerate. That this lower bound is indeed realized
was verified numerically \cite{doucot,dorier}. We find that this lower
bound on the degeneracy follows, for odd size lattices, even more
generally as a direct consequence of Kramers' degeneracy for the time
reversal invariant Hamiltonian of the orbital compass model. We further
remark here that on a cubic lattice ($D=3$) of size  $L \times L \times
L$ where $L$ is even, all symmetry operators of Eqs. (\ref{symorb}),
and (\ref{compass1}) commute and in principle, a single non-degenerate
GS may be realized. This  state can be separated by exponentially small
energy gaps from all other states in the same topological sector. If,
at least one of the sides of the lattice ($L_{x,y,z}$) is odd then the
GSs are, at least, two fold degenerate. 

A schematic of the $d=1$ symmetries of the orbital compass model  and
its comparison to local and global symmetries is provided in
Fig. \ref{examples}

\begin{figure}[htb]
\vspace*{-0.5cm}
\includegraphics[angle=-90,width=8cm]{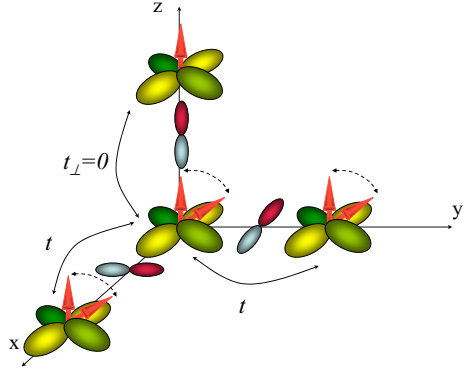}
\vspace*{-0.0cm}
\caption{From Ref. \cite{BN}. The anisotropic hopping amplitudes leading
to  the Kugel-Khomskii (KK) Hamiltonian.  Similar to  \cite{Harris}, 
the four lobed states denote the $3d$ orbitals of a transition metal
while the intermediate small $p$ orbitals are oxygen orbital through
which the super-exchange process  occurs. The dark and bright shades 
denote positive and negative regions of the orbital wave-function. Due
to orthogonality with intermediate  oxygen $p$ states, in  any orbital
state $|\alpha \rangle$ (e.g. $| Z \rangle  \equiv | d_{xy} \rangle $
above), hopping is  forbidden between sites separated along the cubic 
$\alpha$ ($z$ above) axis. The ensuing super-exchange (KK) Hamiltonian
exhibits a $d=2$ $SU(2)$ symmetry corresponding to a uniform rotation
of all spins whose orbital state is $|\alpha \rangle$  in any plane
orthogonal to the cubic direction $\alpha$.}
\label{figu}
\end{figure}

{\bf{b}}) {\it Spins in transition metal compounds}-  Following 
\cite{Harris}, we label the three $t_{2g}$ states $|d_{yz} \rangle,  |
d_{xz} \rangle, | d_{xy} \rangle$  by $|X \rangle, | Y \rangle,$  and
$|Z \rangle$. In the $t_{2g}$ compounds, 
hopping is prohibited via intermediate  oxygen p orbitals between any
two electronic states of orbital flavor $\alpha$ ($\alpha = X, Y$, or
$Z$) along the $\alpha$ axis of the cubic lattice  (see Fig.
\ref{figu}). As a consequence, as noted in \cite{Harris}, a uniform
rotation of all spins, whose electronic orbital state is $|\alpha
\rangle$, in  any given plane ($P$) orthogonal   to the $\alpha$ axis 
$c^{\dagger}_{i \alpha \sigma}  = \sum_{\eta} U^{(P)}_{\sigma, \eta}
d^{\dagger}_{i \alpha \eta}$ with $\sigma, \eta$ the spin  directions,
leaves Eq. (\ref{1}) invariant. The net spin of the electrons of
orbital flavor $|\alpha \rangle$  in any plane orthogonal to the cubic
$\alpha$ axis is conserved; this constitutes the conserved {\em
topological charge} in this case.  Here, we have $d=2$ $SU(2)$
symmetries which are given by \cite{Harris}
\begin{eqnarray} 
\hat{O}_{P;\alpha} \equiv [\exp(i\vec{S}^{\alpha}_{P} \cdot 
\vec{\theta}^{\alpha}_{P})/\hbar], ~ ~ [H, \hat{O}_{P;\alpha}]=0,
\label{symt2g}
\end{eqnarray}
with 
\begin{eqnarray}
\vec{S}^{\alpha}_{P} \equiv \sum_{i \in P} \vec{S}_{i}^{\alpha},
\label{spintopo}
\end{eqnarray}
the sum of all the spins $\vec{S}^{i, \alpha}$ in the orbital state
$\alpha$ in any plane $P$  orthogonal to the direction $\alpha$  (see
Fig. \ref{figu}). The spin of Eq. (\ref{spintopo}) constitutes a
topological charge. There are (3L) conserved topological charges. 
There is a {\em spin continuity} equation in every plane  conjugate to
each of the conserved total planar spin. 

Similar to the orbital compass model, an additional reflection
symmetry  is necessary to link all GSs.  The discrete high-dimensional
reflection symmetry may be broken to lead to nematic type order. The
gauge-like invariance (Eqs. (\ref{vt}), and (\ref{long})) of nematic
type order parameters with respect to the $d=2$ $SU(2)$ symmetries
allows spin nematic order.  If the KK Hamiltonian is the most dominant
spin exchange process, one may predict a spin nematic order which
onsets at temperatures much higher than the currently  measured gauge
non-invariant  magnetization \cite{BN}. In the strained orbital system 
of Eq. (\ref{orb_strain}), the discrete high-dimensional reflection
symmetry is removed. The remaining $d=2$ $SU(2)$ symmetries allow for
TQO.

A $D=2$ rendition of the $t_{2g}$  KK Hamiltonian is defined
by  Eqs. (\ref{orbHam}), and (\ref{compass1})  on the $xz$ plane with
only two of the three $t_{2g}$ orbital flavors ($\alpha =x,z$). In this
system, the spin conservation of Eq. (\ref{spintopo}) holds for each
individual line (a continuous $d=1$ symmetry). 

{\bf c})  Superconducting arrays: A Hamiltonian for superconducting
$(p+ip)$ grains  (e.g. of Sr$_{2}$RuO$_{4}$)  on a square grid, was
recently proposed,  
 \cite{Xu03}
\begin{eqnarray}
H = - K \sum_{\Box} \sigma^{z} \sigma^{z} \sigma^{z} \sigma^{z}
- h \sum_{\bf r} \sigma_{\bf r}^{x}.
\label{XM}
\end{eqnarray}
Here, the four spin product is the product of all spins common to a
given plaquette $\Box$. The spins reside on the vertices on the
plaquette (not on its bonds as gauge fields). These systems  have
$(d=1$ $\Z_{2}$)  symmetries similar to those of the $D=2$
orbital compass model. With $P$ any row or column,
\begin{eqnarray}
\hat{O}_{P} = \prod_{\vec{r} \in  P} \sigma^{x}_{\vec{r}}, 
~~[H, \hat{O}_{P}]=0.
\label{xms}
\end{eqnarray}

In fact, the model of Eq.~(\ref{XM}) can be shown to not only have the
same symmetries but to also have the same spectrum as the orbital
compass model. This is shown by duality transformations \cite{NF} 
which link the two systems,
\newline
\newline
(p+ip) model ~~~$\leftrightarrow$~~~~  $D$=2 orbital compass model.
\newline
\newline 
 
There is a deep link between the system of Eq. (\ref{XM})  and the
$\Z_{2}$ gauge theory of Eq. (\ref{gz2+}).  To see this link, we may
express  Eq. (\ref{gz2+}) in terms of sites which  are located at the
centers of these bonds. Rotating the lattice by 45 degrees, we
immediately obtain the Hamiltonian of Eq. (\ref{XM}) with one
difference: the transformed  Eq. (\ref{gz2+}) will not correspond to
the sum of the products of $\sigma^{z}_{i*}$ around all plaquettes.
Rather, only half of the  plaquettes appear in the rotated Eq.
(\ref{gz2+}). This smaller number of terms allows for a higher or (in
the worst case) an equal number of   symmetry operations which leave
the Hamiltonian of Eq. (\ref{gz2+})  invariant. In the aftermath, all
of the $d=1$ symmetries of  Eq. (\ref{XM}) are symmetries of the
$\Z_{2}$ lattice gauge theory of Eq. (\ref{gz2+}) (as they must be).
The local symmetries of the gauge theory (the $d=0$ operators
$\{A_{s}\}$ of Eq.~(\ref{AB_defn}))  are no longer symmetries of the
system of  the more constrained system of Eq.~(\ref{XM}).  These
residual $d=1$ symmetries are none other than the {\it topological
symmetries} of the  $\Z_{2}$ lattice gauge theory emphasized by Kitaev 
and Wen \cite{kitaev,wenbook}.

{\bf d}) Klein models:
Klein spin models \cite{klein} have an exceptionally high number of 
low-dimensional symmetries.  As an example with the highest degree of
symmetry, we quote the model on the $D=3$ pyrochlore lattice and on its
$D=2$ (checkerboard) incarnation. Here, in both cases (the pyrochlore
and checkerboard lattices), the corresponding short-range  Hamiltonian
reads
\begin{eqnarray}
H&=& \frac{12}{5} J \sum_{\alpha} P_{\alpha}^{S^{tot}= 2} \nonumber
\\ &=& \Big( J \sum_{\langle i j \rangle \alpha}
\vec{S}_{i}^{\alpha} \cdot \vec{S}_{j}^{\alpha} \nonumber
\\ &+& K \Big[ \sum_{\alpha} (\vec{S}_{i}^{\alpha} \cdot \vec{S}_{j}^{\alpha})
(\vec{S}_{k}^{\alpha} \cdot \vec{S}_{l}^{\alpha}) +
(\vec{S}_{i}^{\alpha} \cdot \vec{S}_{l}^{\alpha})
(\vec{S}_{j}^{\alpha} \cdot \vec{S}_{k}^{\alpha}) \nonumber
\\ &+ &(\vec{S}_{i}^{\alpha} \cdot \vec{S}_{k}^{\alpha})
(\vec{S}_{j}^{\alpha} \cdot \vec{S}_{l}^{\alpha}) \Big] \Big),
\label{KleinHamN}
\end{eqnarray}
with $K = 4J/5>0$ and $P_{\alpha}^{S^{tot}=2}$ the projection  operator
onto the state [in each tetrahedron/plaquette  $\alpha \equiv ijkl$ for
the pyrochlore/checkerboard lattices respectively] with maximal total
spin ($S=2$). All GSs in these systems are  superpositions of singlet
coverings \cite{chayes,NBnew}
\begin{eqnarray}
| \psi \rangle &=& \sum_{W} \alpha_{W} \prod_{ij \in W} |S_{ij} 
\rangle, \nonumber
\\ \mbox{with~} |S_{ij} \rangle &=& \frac{1}{\sqrt{2}} (\ket{\!\uparrow
\downarrow} - \ket{\!\downarrow \uparrow}),
\label{GSK}
\end{eqnarray}
and $\alpha_{W}$ arbitrary amplitudes (up to an overall
normalization).  Here, the dimer coverings are labeled by $W$. The 
condition for each dimer covering is that a single  dimer appears in
each tetrahedral unit of the pyrochlore lattice.  These states are
separated from all other excited states by a finite gap \cite{NBnew}.
Here, within the GS manifold the systems maps into the  six vertex
model (see Fig.~  \ref{fig_rules}), the number of dimer state GSs is 
bounded from below by $(3/2)^{N_{s}/2}$. (On the $D=2$ version of the
pyrochlore lattice - the checkerboard lattice - the dimer state GS
degeneracy is exactly equal to $(4/3)^{3 N_{s}/4}$.) As the exponential
in volume degeneracy suggests, these states may be linked to each by
local $(d=0$) symmetry operations. An  illustration of such a process
is shown in Fig. \ref{fig_local}.

\begin{figure}[htb]
\vspace*{-0.5cm}
\includegraphics[angle=0,width=8cm]{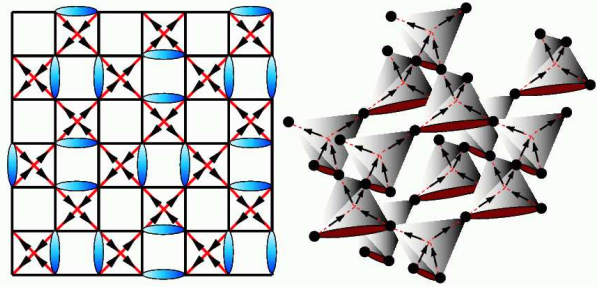}
\vspace*{-0.0cm}
\caption{From Ref. \cite{NBNT}.  All GSs of the Klein model on the
pyrochlore and checkerboard lattices are composed of dimer coverings.
The ovals denote singlet dimer states. The arrows denote the
representation of these dimer states within the six vertex model. On
each plaquette (tetrahedron) the dimer connects the base of the two
incoming arrows. The GSs here are highly regular. Many other GSs exist.
The GSs exhibit an emergent local ($d=0$) GLS. Local ($d=0$) 
variations of dimer coverings product new GSs.}
\label{fig_klein}
\end{figure}

\begin{figure}[htb]
\vspace*{-0.5cm}
\includegraphics[angle=90,width=7cm]{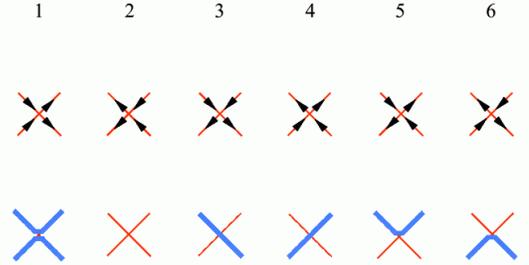}
\vspace*{-0.0cm}
\caption{From Ref. \cite{NBNT}.  Standard representation of the six
vertex states in terms of lines. Every line is composed of links whose
arrows flow to the right in the vertex representation.}
\label{fig_rules}
\end{figure}

\begin{figure}[htb]
\vspace*{-0.5cm}
\includegraphics[angle=90,width=9cm]{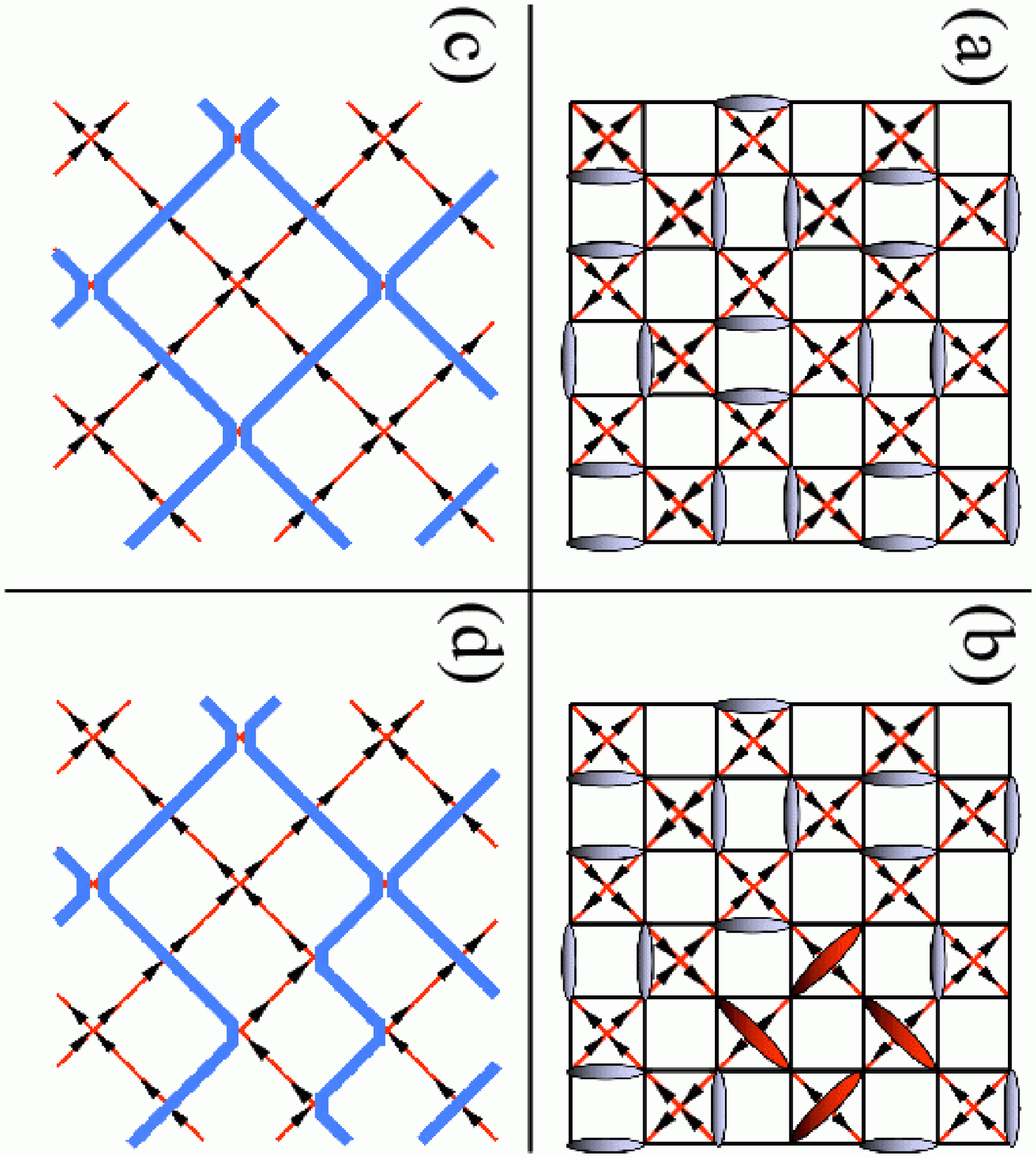}
\vspace*{-0.0cm}
\caption{From Ref. \cite{NBNT}.  Local GLSs in the GS manifold. Left: a
representation of the the six vertex states in a highly  regular
configuration (a) in terms of lines [see Fig. \ref{fig_rules}] (c).
Joining the vertex coverings according to the standard prescription 
allows on to label all GSs. Right: an elementary local (gauge-like)
spin interaction process leading to a new dimer configuration in the GS
basis (b). This corresponds  in the line representation (d) to a flip
of one of the line corners.}
\label{fig_local}
\end{figure} 

We illustrated earlier that the existence of a  GS degeneracy implies
the existence  of symmetry operators which perform unitary 
transformations within the GS basis and which, otherwise, act
trivially. The $d=0$  symmetries within the GS manifold are exact
symmetries of the Hamiltonian. Although (due to the $d=0$ symmetries
that this model displays), the system does not possess local  order, it
does display TQO.  Different sets of states in the GS manifold are
characterized (in the line representation of Fig. \ref{fig_rules}) by
the net number of vertical lines on each horizontal row of lattices
(and conversely). This number is conserved and  corresponds to a {\em
topologically invariant} {\it flux} of lines.  Here, these $d=0$
symmetries (and TQO) appears exactly at the  onset of a deconfined
$T=0$ critical  point \cite{NBNT,senthil}. 

The GSs of the Klein model on a pyrochlore lattice can be linked by
discrete $d=0,1$ symmetry operations. The $d=1$ symmetries link GSs
differing by the net number of lines (the flux). We can remove
the $d=1$ symmetries by  prescribing boundary conditions which fix the
number of lines at a given value and leave only the $d=0$ symmetries.
These lead to TQO at $T=0$. The GS of the  Klein model on the
checkerboard lattice map onto the six vertex model at its  ice point.
Here, the correlations between dimers (and finite sets of dimers) fall
off algebraically (as in a dipole gas). 

{\bf e}) Other systems: 
similar symmetries were found in frustrated spin systems. Ring exchange
Bose metals, in the absence of nearest-neighbor boson hopping, exhibit
$d=1$ symmetries \cite{Arun02}. Continuous {\em sliding symmetries} of 
Hamiltonians (actions) invariant under  arbitrary deformations along a
transverse direction,
\begin{eqnarray}
\phi(x,y) \to 
\phi(x,y) + f(y),
\label{slideq}
\end{eqnarray} 
appear in many systems.  Amongst others, such systems were discovered
in works on Quantum Hall liquid crystalline phases \cite{lawler,rad}, 
a number of models of lipid  bilayers with intercalated DNA
strands \cite{ohern}, and sliding Luttinger liquids \cite{emery2000}. 

\section{Gauge-Like Symmetries and Aharonov-Bohm-type unitary 
transformations}
\label{ABg}

\begin{figure}
\centerline{\includegraphics[width=0.5\columnwidth]{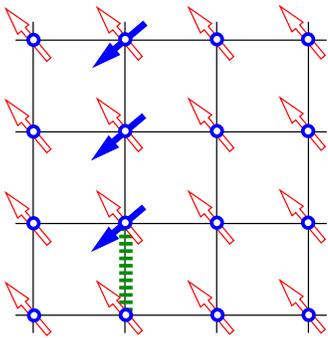}}
\caption{A soliton (topological defect) in an otherwise uniform state of
a general (anisotropic) orbital compass model [see  Eq.
(\ref{orb_strain})] leads to only a finite energy cost. A single
energetic bond (dashed line) indicates that penalty. The energy-entropy
balance associated with such $d=1$ Ising type domain walls is the same
as that in a $D=1$ Ising system. At all finite temperatures, entropic
contributions  overwhelm energy penalties and no  local order is
possible. Order is only manifest in non-local quantities associated
with topological defects. Similar results occur in other systems with
low-dimensional GLSs.}
\label{soliton}
\end{figure}

In this Section we formally write down our group symmetry elements and
topological defects in terms of Aharonov-Bohm-type unitary operators.
In the  continuum limit the group elements of $d=1$ GLSs can be
written  as a path-ordered (${\cal P}$) product
\begin{equation}
U = {\cal P} e^{ i \oint_C \vec{A}  \cdot \vec{ds}} ,
\label{u1}
\end{equation}
where $C$ is a closed path in configuration space and $\vec{A}$ is the
corresponding connection. On the other hand, a defect creating
operator in such a system is 
\begin{eqnarray}
\hat{T}_{{\sf +}} =  {\cal P} e^{ i \int_{C_{+}} \vec{A}  \cdot \vec{ds}},
\end{eqnarray}
where $C_{+}$ is an open contour that only spans  a portion of the
entire closed cycle. On a discrete lattice these expressions are
replaced by  equivalent discrete sums, and $C$ ($C_+$) represents a
closed (open) path on the lattice. For instance, for the orbital
compass model  [Eqs. (\ref{orb}), and (\ref{compass1})],  the $d=1$
group elements can be written in a form which formally looks like  an
Aharonov-Bohm (AB) phase \cite{AB} [Eq. (\ref{symorb})]. For the same
model,  the defect depicted in Fig.~\ref{soliton} is generated by
\begin{eqnarray} 
\hat{T}_{{\sf +}} = e^{i \frac{\pi}{2} \sum_{j \in C_{+}} \sigma^{x}_{j}} .
\end{eqnarray}

The unitary operators of Eqs. (\ref{symorb}), and (\ref{u1})   belong
to a unitary group different from $U(1)$.  As such, Eqs.
(\ref{symorb}), and (\ref{u1}) extend  the AB phases associated with
$U(1)$ generators.   We can interpret $\hat{T}_{{\sf +}}$ as the
creation of a defect-antidefect pair and  the displacement of each
member of the pair to the opposite endpoints of $C_{+}$.  The $d=1$ GLS
operators linking different GSs (the generators  of Eqs.
(\ref{symorb}), and (\ref{u1})) correspond to the displacement  of a
defect-antidefect pair along a toric cycle. Formally, this is similar
to the quasiparticle-quasihole pair operator linking different GSs  in
the FQHE.  {\it Monodromy} and toric {\it topology} are manifest in 
the $d=1$ symmetry operators.  The generator $\vec{A}$ (i.e. the
effective connection $\vec{A}$ that appears in the integral for the
$d=1$ symmetry operators)  is  borne by parallel displacement on spin
(or other) fiber bundles. More complicated topological properties are
associated with $d=2$ symmetries. 

If $\vec{A}$ is also a generator of a local symmetry (i.e. a $d=0$ GLS)
then by local symmetry operations we can deform the contour $C$
continuously. Here, the group element of Eq. (\ref{u1}) will only
depend on the topology of the configuration space and $C$. This is a
particular case of a more general set of GLS generators that we discuss
in the present paper.

\section{Consequences of Gauge-like symmetries}
\label{consequence}

What are the physical consequences of having a system endowed with a 
symmetry group ${\cal{G}}_d$? In the following Sections, we outline a
few of those. First, we remark on the conserved Noether charges  which
are associated with these symmetries. We then give  several toy
examples in which topological indices which are usually associated with
low-dimensional systems make an  appearance in high-dimensional
systems. Next, we discuss a property which will be of key interest to
our forthcoming discussion and proof of  finite-$T$ TQO in several
systems - that regarding the absence  of spontaneous symmetry breaking
(SSB). Then, we discuss an extension of that result which shows that
well-defined quasiparticle poles are precluded in systems which exhibit
GLSs of low dimensionality. We will conclude this Section  with a brief
discussion regarding the appearance of lower-dimensional topological
terms when these symmetries are present.

\subsection{Conserved Noether charges}

Symmetries generally imply the existence  of conservation laws and
topological charges with associated continuity equations. We find that
systems with $d$-GLSs lead to conservation laws within $d$-dimensional
regions. To illustrate, consider the (continuum) Euclidean Lagrangian
density of a complex field  $\vec{\phi}(\vec{x})=(\phi_1(\vec{x}),
\phi_2(\vec{x}),\phi_3(\vec{x}))$ ($\vec{x}=(x_1,x_2,x_3)):$
\begin{eqnarray}
{\cal{L}} = \frac{1}{2} \sum_{\mu} |\partial_{\mu} \phi_{\mu}|^{2} 
+ \frac{1}{2} \sum_{\mu} |\partial_{\tau} \phi_{\mu}|^{2} +W(\phi_\mu),
\label{soft}
\end{eqnarray}
with 
\begin{eqnarray}
W(\phi_\mu)= u(\sum_{\mu} |\phi_{\mu}|^{2} )^{2} - \frac{1}{2}
\sum_{\mu} m^{2}(|\phi_{\mu}|^{2})
\end{eqnarray} 
and $\mu,\nu=1,2,3$. The Lagrangian density ${\cal L}$  displays the
continuous $d=1$ symmetries 
\begin{eqnarray}
\phi_{\mu} \to
e^{i \psi_{\mu}(\{x_{\nu}\}_{\nu\neq\mu})}  \phi_{\mu}.
\label{contsymn}
\end{eqnarray}  
The conserved $d=1$ Noether currents are tensors given by  
\begin{eqnarray}
j_{\mu \nu}= i
[\phi_{\mu}^{*} \partial_{\nu}  \phi_{\mu} - (\partial_{\nu} 
\phi_{\mu}^{*}) \phi_{\mu}],
\end{eqnarray} 
which satisfy $d=1$ conservation laws
\begin{eqnarray}
[\partial_{\nu} j_{\mu \nu} + \partial_{\tau} j_{\mu \tau}] =0 
\end{eqnarray}
(with
no summation over repeated indices implicit). What is special about 
$d$-GLSs is that there is a conservation law for each line associated
with a fixed value of all coordinates $x_{\nu \neq \mu}$ relating to
the $d=1$ charge  
\begin{eqnarray}
Q_{\mu}(\{x_{\nu \neq \mu}\}) = \int dx_{\mu}
~j_{\mu \tau}(\vec{x}).
\label{QN1d}
\end{eqnarray}

\subsection{Topological indices}
\label{tindex}

Noether's theorem stems from the continuous symmetries of the 
Lagrangian. Its application to the toy system of Eq.~(\ref{soft}) with
the  $d=1$ symmetries of Eq.~(\ref{contsymn})  led to the conserved
$d=1$ charges of Eq.~(\ref{QN1d}). By contrast, topological numbers in
field theories  are related to boundary conditions.  Their presence is
dictated by the requirement of a finite energy (or action).

In what follows, we will show how in specific continuum field theories,
topological numbers which  are associated with low-dimensional systems 
(of dimension $d$) can  appear in high-dimensional systems (of
dimension $D>d$). 

(i)  $d=1$ $\Z_{2}$ soliton numbers:

To illustrate the basic premise, consider the theory of
Eq.~(\ref{soft}) on a ($D=2$) square lattice  for a system with two
component fields $\vec{\phi}= (\phi_1,\phi_2)$ and in which the {\it
potential term} is given by an extension to 2+1 of the example in
Section 2.3 of \cite{raj} 
\begin{eqnarray}
W(\phi_1,\phi_2) = \frac{1}{4}(\phi_1^2-1)^2 + \frac{1}{2}m^{2} 
\phi_{2}^{2} \nonumber \\ 
+ \frac{\lambda}{4} \phi_2^4+ \frac{1}{2} \eta \ 
\phi_{2}^{2}(\phi_{1}^2-1) ,
\label{Wraj}
\end{eqnarray}
with $\lambda, \eta, m^{2}>0$. The system given by Eqs.~(\ref{soft}),
and (\ref{Wraj}) displays $d=1$ $\Z_{2}$ symmetries: these are given
by  $( \phi_{1} \to - \phi_{1})$ on each  horizontal line and 
$(\phi_{2} \to -\phi_{2}$) on each vertical line. The minima of $W$ 
occur at 
\begin{eqnarray}
(\phi_1,\phi_2) = (\pm 1,0),
\label{topexam}
\end{eqnarray}
at which $W=0$.
For $\lambda>(\eta-m^2)^2$, the functional $W$ is positive elsewhere.
The requirement of finite energy dictates that at infinity, the field
must be of the form of Eq.~(\ref{topexam}). The form of $W$ enables us
to define a topological charge as that of soliton number in a
$D=1$-dimensional system,
\begin{eqnarray}
Q_{soliton}(x_2)= [\phi_{1}(x_{1}= \infty,x_{2}) - \phi_{1}(x_{1}= 
-\infty,x_{2})].
\label{solitoncharge}
\end{eqnarray}
By virtue of the gradient terms in Eq.~(\ref{soft}), for  finite energy
penalties, we must have a constant topological charge:
$Q_{soliton}(x_{2}) = Q_{soliton}$ for all $x_{2}$.  

Various contending soliton configurations are given as a direct product
of two $D=1$  systems (one for $\phi_1$ along the  $x_1$ axis and one
for $\phi_2$ along the $x_2$ axis). A solution is given by 
\begin{eqnarray}
\phi_1(x_1,x_2) &=& \tanh[m(x_1-c)],\nonumber
\\ \phi_2(x_{1},x_{2})&=& \pm \sqrt{\frac{1 - 2m^2}{\eta}} ~~\mbox{sech} 
[m(x_2-c)],
\label{examtopraj}
\end{eqnarray}
with $c$ an arbitrary real constant. The main feature of
Eq.~(\ref{examtopraj}) is the existence of an effective
$d=1$-dimensional soliton configuration for the field component
$\phi_1$. Such a soliton-like configuration is expected for  $D=1$
$\Z_{2}$ systems.

(ii) Pontryagin indices for $d=2$ $SU(2)$ GLSs:

Next, we consider another toy model in 3+1 dimensions for which 
similar topological indices appear for  continuous $d=2$ $SO(3)$ GLSs.
Consider, at every lattice  site a field composed of an $O(3)$ vector
triad
\begin{eqnarray}
\vec{\phi}(\vec{x}) = (\vec{\phi}_{1}(\vec{x}), 
\vec{\phi}_{2}(\vec{x}), \vec{\phi}_{3}(\vec{x})).
\end{eqnarray}
Here, each $\vec{\phi}_{i}$ $(i=1,2,3)$ is a vector with three components
[and thus $\vec{\phi}$ contains nine components]. 
In the ``hard spin'' limit of $\vec{\phi}_{i}$, 
for each $i=1,2,3$, we will demand the normalization
condition 
\begin{eqnarray}
\vec{\phi}_{i}(\vec{x}) \cdot \vec{\phi}_{i}(\vec{x}) =1.
\label{pp1}
\end{eqnarray}
For a general Lagrangian density of the form
\begin{eqnarray}
{\cal{L}} = \frac{1}{2} \sum_{\mu_{1} = 0,2,3} (\partial_{\mu_{1}}
\vec{\phi}_{1}) \cdot (\partial^{\mu_{1}} \vec{\phi}_{1}) \nonumber
\\ + 
\frac{1}{2} \sum_{\mu_{2} = 0,1,3} (\partial_{\mu_{2}}
\vec{\phi}_{2}) \cdot (\partial^{\mu_{2}} \vec{\phi}_{2}) \nonumber 
\\ + 
\frac{1}{2} \sum_{\mu_{3} = 0,1,2} (\partial_{\mu_{3}}
\vec{\phi}_{3}) \cdot (\partial^{\mu_{3}} \vec{\phi}_{3}) \nonumber 
\\ + W, 
\label{LWd}
\end{eqnarray}
with $W$ a functional of $\{\vec{\phi}_{i}(\vec{x})\}$  which is
invariant under all global rotations of each 
$\vec{\phi}_{i}(\vec{x})$: For example,  for a soft spin implementation
of Eq.~(\ref{pp1}), we may set $W= \sum_{i=1}^{3} (\phi_{i}^{2}(x)
-1)^{2}$. Here, each field $\vec{\phi}_{i}$ exhibits a $d=2$ $SO(3)$
rotation symmetry in each spatial plane $[jk]$ for which $i \neq j$ and
$i \neq k$ ($j \neq k$).  The static finite energy solutions to the 
Lagrangian density of Eq.~(\ref{LWd}), exhibit three topological
indices -  one for each of the three fields
$\{\vec{\phi}_{i}\}_{i=1}^{3}$. These are given by the Pontryagin
indices
\begin{eqnarray}
{\cal Q}_{i} = \frac{1}{8 \pi} \int \epsilon_{a b} \vec{\phi}_{i} \cdot
(\partial_{a} \vec{\phi}_{i} \times \partial_{b} \vec{\phi}_{i})
~dx_{j} dx_{k}.
\label{pointr}
\end{eqnarray}
In Eq.~(\ref{pointr}), $a, b$ span all spatial indices different from
$i$: ($a,b \in \{ j,k\}$). These (integer) topological indices ${\cal
Q}_{i}$ can be related to the homotopy group
\begin{eqnarray}
\Pi_{2}(S^{2}) = \Z.
\label{homotopy2}
\end{eqnarray}
The integers ${\cal Q}_{i}$ denote the wrapping numbers - the number of
times that the spatial two sphere [(the subscript in 
Eq.~(\ref{homotopy2}) defined by the plane $[jk]$ with a unique point
at infinity] can be wrapped onto the the {\it internal} sphere $S^2$
relevant to the  three component fields $\vec{\phi}_{i}$ of unit norm. 
The likes of Eq.~(\ref{homotopy2}) are seen in numerous  $o(3)$ spin
systems in two spatial  or in 1+1 space-time dimensions yet not seen in
three spatial dimensions. What is novel about systems with $d$-GLSs
with $d<D$ is that they allow for defects of lower-dimensional
structure in high dimensions.

Later on, we will return to similar forms when we will discuss
topological terms which appear in systems with $d$-GLSs.

In a similar fashion a multitude of other field theories can be
constructed for systems with  $d$-GLSs which reside in $D \ge d$
dimensions and for which topological defects are characterized by
topological indices which are usually reserved for  systems in $d$
dimensions. For quantum spin systems,  Berry phase terms may be
inserted into the  functional $W$ in Eqs.~(\ref{LWd}), and (\ref{soft})
to enable to the same dimensional reduction. 

As we discuss next, it is precisely these topological defects which 
eradicate local order for systems endowed with  low $d$-GLSs and which
consequently may enable  TQO. 

\subsection{Absence of spontaneous symmetry breaking}
\label{assb}

Next, let us ask ourselves whether we can spontaneously break
$d$-GLSs.  As it turns out \cite{BN}, and reviewed in the Appendix, 
the absolute values of quantities not invariant under ${\cal{G}}_d$ are
bounded from above by the expectation values that they attain in a
$d$-dimensional Hamiltonian $\bar{H}$  (or corresponding action
$\bar{S}$) which is globally invariant under ${\cal{G}}_d$ and
preserves the range of the interactions of the original systems.  [In
this proof, we take (as we must) the perturbing field favoring order to
zero only after the thermodynamic limit is taken.] The physical  origin
of this effect lies in the large entropic contributions associated with
low-dimensional defects [see, e.g., Fig.~\ref{soliton}]. As the
expectation values of local observables vanish in low-$d$ systems, this
bound strictly forbids SSB of  non-${\cal{G}}_d$ invariant local
quantities in systems with interactions of finite range and strength
whenever $d=0$ (Elitzur's theorem)\cite{Elitzur}, $d=1$ for both
discrete and continuous  ${\cal{G}}_d$, and (as a consequence of the
Mermin-Wagner-Coleman theorem \cite{Mermin})  whenever $d=2$ for
continuous symmetries \cite{BN}. Discrete $d=2$ symmetries may be
broken (e.g. the finite-$T$ transition of the $D=2$ Ising model, and
the $d=2$ Ising GLS of $D=3$ orbital compass systems). In the presence
of a finite gap in a system with continuous $d=2$ symmetries, SSB is
forbidden even at $T=0$ \cite{BN}. The engine behind  the absence of
SSB  of non-GLSs invariant quantities in high-dimensional systems are
the same low-dimensional  topological defects which destroy order in $d
\le 2$ systems (e.g. domain walls/solitons in systems with  $d=1$
discrete symmetries, vortices in systems with $d=2$ $U(1)$ symmetries,
hedgehogs for $d=2$ $SU(2)$ symmetries). Transitions and crossovers can
only be discerned by quasi-local symmetry invariant {\em string}-like 
quantities (e.g. Wilson-like loops in pure gauge theories)  or, by
probing global topological  properties (e.g. {\em percolation} in  more
general matter coupled gauge theories \cite{percolation1},
\cite{percolation2}, \cite{percolation3}). Extending the bound of
\cite{BN} to $T=0$, we  now find that if $T=0$ SSB is precluded in the
$d$-dimensional system then it will also be precluded in the
higher-dimensional system for quantities not invariant under exact or
$T=0$ {\em emergent} $d$-GLSs.  Exact symmetries refer to $[U,H]=0$. In
{\em emergent} symmetries \cite{GJW} unitary operators $ U \in
{\cal{G}}_{\sf emergent}$ are not bona fide symmetries ($[U,H] \neq 0$)
yet  become exact at low energies: when applied to any GS, the
resultant state must also  reside in the GS manifold, 
\begin{eqnarray}
U| g_{\alpha} \rangle = \sum_{\beta} u_{\alpha 
\beta} | g_{\beta} \rangle.
\label{emergequation}
\end{eqnarray}  
These operators thus become symmetries when confined to the low-energy
sector \cite{emergexact}. 

Putting all of the pieces together, SSB at $T=0^{+}$  (even for $T \ll
\Delta$ with $\Delta$ the gap size in gapped systems)   of exact or
emergent $d \le 1$ discrete or SSB even at $T=0$ of  $d \le 2$
continuous symmetries in systems of finite  interaction strength and
range cannot occur. 
  


\subsection{Absence of quasiparticle excitations} 

When the bound of \cite{BN} is applied anew  to correlators and
spectral functions, it dictates  the absence of quasiparticle (qp)
excitations in many instances \cite{NBF}. Here we elaborate on this:
The bound of \cite{BN} mandates that the absolute values of
non-symmetry invariant  correlators 
\begin{eqnarray}
|G| \equiv
|\sum_{\Omega_{{\bf j}}} a_{\Omega_{{\bf j}}} \langle \prod_{{\bf i}
\in \Omega_{{\bf j}}}  \phi_{{\bf i}} \rangle|
\end{eqnarray} 
with $\Omega_{{\bf j}} \subset {\cal C}_{{\bf j}}$, and 
$\{a_{\Omega_{{\bf j}}}\}$  $c$-numbers, are bounded from above (and
from below for $G\geq0$ (e.g., that corresponding to $\langle
|\phi({\bf k}, \omega)|^{2} \rangle$)) by absolute values of the  same
correlators $|G|$ in a  $d$-dimensional system defined by ${\cal
C}_\bj$. In particular, $\{a_{\Omega_{{\bf j}}}\}$ can be chosen to
give the Fourier transformed  pair-correlation functions. This leads to
stringent bounds on viable qp weights and establishes the absence of qp
excitations in many cases.   In high-dimensional systems, retarded
correlators $G$ generally exhibit a resonant (qp) contribution (e.g.,
\cite{spectral,vns}).   Here, 
\begin{eqnarray}
G= G_{{\sf res}}({\bf k}, \omega)  + G_{{\sf non-res}}({\bf k},
\omega)
\end{eqnarray} 
with 
\begin{eqnarray}
G_{{\sf res}}({\bf k}, \omega)  = \frac{Z_{{\bf
k}}}{\omega - \epsilon_{{\bf k}} + i 0^{+}}.
\end{eqnarray} 
In low-dimensional systems, the qp weight  $Z_{{\bf k}} \to 0$ and the
poles of $G$ are often replaced by weaker branch cut behavior. If the
momentum ${\bf k}$ lies in a lower $d$-dimensional region
${\cal{C}}_{\bf j}$  and if no qp resonant terms   appear in the
corresponding lower-dimensional spectral functions  in the presence of
non-symmetry breaking fields then the upper bound \cite{BN} on the
correlator $|G|$  (and on related qp weights  given by $\lim_{\omega
\to \epsilon_{{\bf k}}} (\omega - \epsilon_{{\bf k}}) G({\bf k},
\omega)$) of non-symmetry invariant quantities mandates the absence of
normal qps. If quasi-particle fractionalization occurs in the
lower-dimensional system then its higher-dimensional realization
follows.

\subsection{Charge Fractionalization} 

As we noted in Section \ref{Section 3}, whenever the center of the
group linking the various GSs  is (or contains a subgroup) of the
$\Z_{N}$ type, then the basic {\it primitive} charge can be quantized
in integer multiples of $1/N$. Whenever a discrete $d$-GLS symmetry is
present for which such a fractionalization occurs,  then
fractionalization may occur in the $D$-dimensional system.  If the
$d$-GLSs have a $\Z_{N}$ group as their center then  a fractional
primitive charge of size $(1/N)$ is possible. The $d=1$ quasi-particle
quasi-hole insertion operators  in the Quantum Hall problem
\cite{Karl,sdh}  lead to fractionalization in just the same fashion  as
they do in the Peierls chain. $d$-GLSs are obviously  not required for
the appearance  of fractional charge (see our discussion of $N$-ality
in Section \ref{Section 3}). The most prominent of all examples of
charge fractionalization [that of quarks ($SU(N=3)$)] occurs in 3+1
dimensions. What we wish to stress is that the presence  of $d$-GLSs
makes the identification of fractional charge more immediate. 

\subsection{Topological terms}

A related consequence of systems with $d$-GLSs is that topological
terms which appear in $d+1$-dimensional theories also appear in higher
$D+1$-dimensional systems ($D>d$). These can be related to  Chern
number invariants in these theories whose form is similar to that of
general topological indices  which we discussed in Section
\ref{tindex}.  These topological terms appear in actions $\bar{S}$
which bound quantities not invariant under the  $d$-GLSs. For instance,
in the isotropic $D=2$ (or 2+1)-dimensional  orbital-less general spin
version of the $t_{2g}$ KK model [See (b) of Section \ref{section12}
for the complete Hamiltonian.]  \cite{KK}, of exchange constant  $J>0$,
the corresponding continuum Euclidean action is of the 1+1 form
\begin{eqnarray}
\bar{S} = \frac{1}{2g} \int dx d \tau  \Big[ \frac{1}{v_{s}}
(\partial_{\tau} \hat{m})^{2} - v_{s} (\partial_{x} \hat{m})^{2} \Big] 
+ i \theta {\cal{Q}}  + S_{{\sf tr}}, \nonumber
\end{eqnarray}
where
\begin{eqnarray}
{\cal{Q}} = \frac{1}{8 \pi} \int dx d\tau \epsilon_{\mu \nu} \ \hat{m}
\cdot (\partial_{\mu} \hat{m} \times \partial_{\nu} \hat{m}) 
\end{eqnarray}
with  $\mu, \nu \in \{ x, \tau \}$ and $\epsilon_{\mu \nu}$ the
rank-two Levi Civita symbol.  Here, as in the non-linear-$\sigma$ model
of a spin-$S$ chain, $\hat{m}$ a normalized slowly varying  staggered
field, $g=2/S$, $v_{s} = 2 JS$, $\theta = 2 \pi S$, and $S_{{\sf tr}}$
a {\it transverse-field} action term which does  not act on the spin
degrees of freedom along a given chain \cite{BN}. ${\cal{Q}}$ is the
Pontryagin index  corresponding to the mapping between the
(1+1)-dimensional space-time $(x,\tau)$ plane and the two-sphere on
which $\hat{m}$ resides. This (1+1)-dimensional topological term
appears  in the (2+1)-dimensional KK system even for arbitrary large
positive coupling $J$. This, in turn, places bounds on the spin
correlations and implies, for instance, that in $D=2$ integer-spin
$t_{2g}$  KK systems, a finite correlation length exists. 

\section{Finite temperature TQO and Gauge-like symmetries}
\label{Section 5}

Our central contention is that in all systems known to harbor TQO (and
in new examples), $d$-GLSs are present. Old examples include: Quantum
Hall systems, $\Z_{2}$ lattice gauge theories, the Toric-code model
\cite{kitaev} and other systems.  In all cases of TQO, we may cast
known {\it topological} symmetry operators as general low-dimensional
$d \le 2$  GLSs (e.g. in the Toric code model,  there are $d=1$
symmetry operators  spanning toric cycles). The presence of these
symmetries allows for the existence of freely-propagating decoupled
$d$-dimensional topological defects  (or instantons in $(d+1)$
dimensions of Euclidean space time)   which eradicate local order.
These defects enforce TQO. In what follows, we will first investigate
finite-$T$ TQO in general systems and then turn to gapped systems with 
continuous symmetries.

\subsection{Sufficient conditions for finite temperature TQO}
\label{finitetcond.}

We now state a central result:

{\bf{Theorem:}} {\em A system will display  {\rm TQO} for all
temperatures $T \ge 0$  if it satisfies the following three conditions:
(i) It obeys the $T=0$ TQO conditions of Eq.~(\ref{def.}),  (ii) all of
its interactions are of finite range and strength,   and (iii) all of
its GSs may be linked by  discrete $d \le 1$ or by  continuous $d \le
2$ GLSs $U \in {\cal{G}_{\it d}}$.} 

\proof  

Recall the $T>0$ definition of TQO as given the requirement that both
Eqs. (\ref{def.}), and (\ref{vt}) are satisfied. For any ${\cal G}_d$
symmetry, we separate  $V=V_{{\cal{G}},0} + V_{{\cal{G}}; \perp}$ where
$V_{{\cal{G}},0}$ is the component of $V$ which  transforms under the
singlet representation of ${\cal{G}}_{d}$ (and for which
$[V_{{\cal{G}},0},U]=0$) and where $V_{{\cal{G}}; \perp}$ denotes the
components of $V$  which do not transform as scalars under
${\cal{G}}_{d}$ (and for which $\int dU~ U^{\dagger}  V_{{\cal{G}};
\perp} U =0$).  

We will now first discuss the symmetry invariant component 
$V_{{\cal{G}},0}$ and then turn to the symmetry non-invariant component
$V_{{\cal{G}}; \perp}$.  To prove the finite-$T$ relation of  Eq.
(\ref{vt}),  we write the expectation values over a complete set of
orthonormal states $\{ |a \rangle \}$ which are eigenstates of  the
Hamiltonian $H$ endowed with boundary terms favoring a particular GS
$\alpha$,
\begin{eqnarray}
\!\!\!\!\!\!\!\! \langle V_{{\cal{G}},0}  \rangle_{\alpha}  &=& 
\frac{\sum_{a} \langle a| V_{{\cal{G}},0}| a \rangle  e^{-\beta (E_{a}
+ \phi^{a}_{\alpha}
)}}{\sum_{a} e^{-\beta (E_{a} + \phi^{a}_{\alpha}
)}} \nonumber \\ 
&=& \frac{ \sum_{a} \langle a| U^{\dagger} 
V_{{\cal{G}},0} U | a
\rangle  e^{-\beta (E_{a} + \phi^{a}_{U^{\dagger} \beta}
)}}{\sum_{a} e^{-\beta (E_{a}  + \phi^{a}_{U \beta}
)}} \nonumber \\ 
&=&   \frac{\sum_{b} \langle b| V_{{\cal{G}},0}| b \rangle 
e^{-\beta (E_{b} +  \phi^{b}_{\beta}
)}} {\sum_{b}
e^{-\beta (E_{b} + \phi^{b}_{\beta}
)}} = \langle V_{{\cal{G}},0}
\rangle_{\beta}.
\label{long} 
\end{eqnarray}
Here, we invoked $U|a \rangle \equiv |b \rangle,$ and $E_{a} = E_{b}$
(as $[U,H]=0$). The term $\phi^{a}_{\alpha}$ monitors the effect of the
boundary conditions favoring the state $\alpha$. If $H^{\alpha}$ is the
Hamiltonian endowed with boundary terms corresponding to the GS
$\alpha$ then $\langle a | H^{\alpha} | a \rangle = E_{a} +
\phi^{a}_{\alpha}$. In the above derivation,  $\phi^{a}_{\alpha} =
\phi^{Ua}_{U \alpha}  = \phi^{b}_{\beta}$, which is evident by the
application of a simultaneous  unitary transformation in going from $|a
\rangle \to^{U}  |b \rangle$ and  $|g_{\alpha} \rangle \to^{U}
|g_{\beta} \rangle$.

Next, we turn to the symmetry non-invariant component $V_{{\cal{G}};
\perp}$. For the non-symmetry invariant  $V_{{\cal{G}}, \perp}$,  by
the theorem of \cite{BN}, $\langle V_{{\cal{G}}, \perp} \rangle_\alpha
=0$,  i.e. for systems with low-dimensional GLSs, symmetry breaking is
precluded. [In the proof of \cite{BN}, we take the perturbing field
favoring order is taken, as indeed it must, to zero only after the
thermodynamic limit is taken.] Equation  (\ref{long})  is valid
whenever $[U,V] = 0$ for {\it any} symmetry $U$. However,  $\langle
V_{\cal{G}, \perp} \rangle_\alpha =0$ only if $U$ is a low-dimensional
GLS.  In systems in which not all GS pairs can be linked by the
exclusive use of low-dimensional GLSs $U \in {\cal{G}}_{d}$ ($U|
g_{\alpha} \rangle = | g_{\beta} \rangle$), SSB may occur.  

Taken together, our results for both  $V_{{\cal{G}},0}$ and
$V_{{\cal{G}}; \perp}$ conclude our proof. It is important  to note
that our symmetry conditions for $T>0$ TQO once $T=0$ TQO is
established {\em do not rely on the existence of a spectral gap}.
$d$-dimensional symmetries alone, irrespective of the existence or
absence of a spectral gap, mandate the appearance of finite-$T$ TQO. As
we will discuss in the next Section, the existence of a gap facilitates
the proof of $T=0$ TQO in systems harboring a continuous symmetry. 

As stated initially, the physical engine behind TQO in the presence of
$d$-GLSs are  topological defects: For example, domain walls in $d \le
1$ systems, Goldstone modes in continuous $d \le 2$ systems.  These
topological defects proliferate throughout the lattice to enforce TQO
(and destroy local orders). The topological character of the defects is
particularly evident when upon duality, the (dual) disordering fields
have the form of Eq.~(\ref{Td}). Here it is seen how the defects link
the system bulk to its boundaries.

It is worth noting that the homotopy groups  of such defects can, of
course, be either Abelian or non-Abelian.  A non-commutativity of the
defects implies a topological entanglement in the sense that entangled
defects cannot be undone. Two defect loops can become entangled if
their homotopy group representations do not commute. Such a
non-commutativity of dislocations (defects  in elastic solids) is
responsible for the phenomena of work hardening. General defects may
allow for operations which  include and extend on the particular
braiding  group (which appears for FQHE quasiparticles or Majorana
fermions in the vortex cores of $p+ip$ superconductors) - the principal
non-Abelian group to have been discussed to date in the literature. 
This may be extended to general non-Abelian $d$-GLSs.

\subsection{Gapped systems with continuous low-dimensional Gauge-Like
Symmetries}
\label{gapsection}

We will now show that in systems with a spectral  gap between the
ground and all excited states,  and in which continuous low-dimensional
($d \le 2$)-GLSs are present, TQO can appear at low temperatures. 
Unlike the case for discrete $d = 1$ symmetries, the existence of a
spectral gap for $d=1,2$ symmetries ensures the existence of $T=0$ TQO.
This then, in conjunction with the central theorem of Section
\ref{Section 5}, leads to a very strong result:

{\bf{Theorem:}} {\em When in a gapped system of finite interaction
range and strength, the GSs (each of which can be  chosen by the
application of an infinitesimal field) may be linked by  continuous $d
\le 2$ GLSs $U \in {\cal{G}_{\it d}}$, then the system displays both
zero and finite-$T$ TQO.} 

\proof  

Here, we rely on the discussion of Section \ref{assb} - in particular
Corollary IV of Appendix \ref{app2} \cite{BN}.  As Corollary IV is a
$T=0$ extension of  Elitzur's theorem, the proof of the above theorem 
for $T=0$ TQO follows word for word the same proof which we furnished
earlier for $T>0$ TQO in Section \ref{finitetcond.}.  As before, we will
now first discuss the symmetry invariant component  $V_{{\cal{G}},0}$
and then turn to the symmetry non-invariant component $V_{{\cal{G}};
\perp}$.  For $V_{{\cal{G}},0}$, ($\ket{g_\beta}=U\ket{g_\alpha}$),
$\langle g_{\alpha} | V_{{\cal{G}},0} | g_{\alpha} \rangle = \langle
g_{\alpha} | U^{\dagger} V_{{\cal{G}},0} U | g_{\alpha} \rangle =
\langle g_{\beta} | V_{{\cal{G}},0} | g_{\beta} \rangle$. 

Next, we turn to the symmetry non-invariant component $V_{{\cal{G}};
\perp}$. For the non-symmetry invariant  $V_{{\cal{G}}, \perp}$,  by
the theorem of \cite{BN}, and corollary IV in particular,
\begin{eqnarray}
\langle V_{{\cal{G}}, \perp} \rangle_\alpha =0,
\end{eqnarray}  
at both zero and all finite (non-zero) temperatures. That is, for
systems with low-dimensional GLSs, symmetry breaking is precluded. Eq.
(\ref{long})  is valid whenever $[U,V] = 0$ for {\it any} symmetry $U$.
However,  $\langle V_{\cal{G}, \perp} \rangle_\alpha =0$ only if $U$ is
a low-dimensional GLS.  In systems in which not all GS pairs can be
linked ($U| g_{\alpha} \rangle = | g_{\beta} \rangle$) by the exclusive
use of low-dimensional GLSs $U \in {\cal{G}}_{d}$, SSB may occur.  
Taken together, our results for both  $V_{{\cal{G}},0}$ and
$V_{{\cal{G}}; \perp}$ conclude our proof. The existence of $T=0$   TQO
in conjunction with the theorem of Section \ref{Section 5}, then
establishes the existence of TQO for all $T \ge 0$. 

\section{Gauge-Like Symmetry group selection rules and their consequences}
\label{t0tqo}

It is commonly assumed that the existence of a spectral gap may render
$T=0$ TQO (that satisfying Eq.~(\ref{def.})) stable at small positive
temperatures. In Section \ref{gapsection}, we rigorously proved (for
the first time) that this is indeed so for {\em continuous} low 
$d$-GLSs for the appearance of TQO. We now turn to the more general
analysis of $T=0$ conditions of Eq.~(\ref{def.}). We will illustrate
that symmetry alone ({\it without ever invoking the presence of a
spectral gap}) can guarantee that the conditions of Eq.~(\ref{def.})
are satisfied. Unlike finite-$T$ TQO which follows from low-$d$  ($d\le
1$ discrete or $d \le 2$ continuous) GLSs, the $T=0$ rules which we
will  discuss apply for general $d \ge 1$ GLSs. Our approach highlights
how the states themselves encode TQO. The selection rules which we will
employ below are independent  of the specific Hamiltonian which leads
to those GSs.
 
\subsection{Selection rules for general Gauge-Like Symmetries}
\label{selectlong}

Let us first turn to the off-diagonal portion of Eq.~(\ref{def.}). This
relation is often seen to be immediately satisfied in the basis of GSs
which  simultaneously diagonalize not only the Hamiltonian  but also
the $d$-dimensional symmetry operators. Here, we may invoke symmetry
based selection rules (and, when applicable, the Wigner-Eckart
theorem  in particular) to show that the matrix element of any local
operator $V$ between two GSs which are eigenstates of the
$d$-dimensional symmetry generators vanishes in the thermodynamic
limit.  This, along with a set of equations which enable us to satisfy
the diagonal portion  of Eq.~(\ref{def.}) will  assert the existence of
$T=0$ TQO. For concreteness, in what follows,  the notation will refer
to $d=2$ $SU(2)$ symmetries.  Our results hold in their exact form for
symmetry groups in which the Wigner-Eckart theorem holds (e.g.  integer
spin variants of these $SU(2)$ symmetries). When applicable, we will
comment on the occurrence of similar results in systems in which the
Wigner-Eckart theorem  does not apply (e.g. half-integer spin variants
of these $SU(2)$  symmetries). The $d=2$ rotational symmetries, which
we will use for illustration, act on two-dimensional planes
$\{P_{l}\}$. A realization of these symmetries is afforded by the
$t_{2g}$ KK model which we will discussed at length later on [See (b)
of Section \ref{section12}] \cite{t2gKKexample}. The extension
symmetries (of dimension $d \neq 2$)  of groups other than $SU(2)$ is
straightforward. 

Let us assume that a given quasi-local operator $V$ has its support 
inside a sphere of radius $R$. This means that the operator  $V$ spans
no more than $(2R)$ planes and in each plane its spans no more than
$(4R^{2})$ sites. We can expand $V$ as a sum of planar operators, each
of which has its support on a plane $P_{j}$, i.e. 
\begin{eqnarray}
V = \sum_{P_{1} \cdots P_{2R}} B_{P_{1} P_{2} \cdots P_{2R}} 
V_{P_{1}} V_{P_{2}} \cdots V_{P_{2R}} ,
\end{eqnarray}
with bounded coefficients $B_{P_{1} P_{2} \cdots P_{2R}}$.  Let us
furthermore write the GS wavefunctions in the direct product basis of
individual planes,
\begin{eqnarray}
| g_{l} \rangle = 
\sum_{l_{1} \cdots l_{L}} A_{P_{1} \cdots P_{L}} 
| \phi_{P_{1}}^{l_{1}} \rangle \otimes 
| \phi_{P_{2}}^{l_{2}} \rangle \otimes
\cdots | \phi_{P_{L}}^{l_{L}} \rangle.
\label{GSform}
\end{eqnarray}
Here, $L$ is the number of planes. 
Let us denote the number of states $\{|\phi_{P_{j}}^{l} \rangle\}$ in
the sum by $G_{j}$. Each state  $| \phi_{P_{j}}^{l} \rangle$ has its
support on an entire plane $P_{j}$. Now, it is easily verified that if 
\begin{eqnarray}
\langle \phi_{P_{j}}^{l_{j}} | V_{P_{j}} |
\phi_{P_{j}}^{l^{\prime}_{j}} \rangle = v_{j} \delta_{l_{j},
l^{\prime}_{j}} + c_{l_{j}l_{j}^{\prime}}
\label{reduce}
\end{eqnarray}
for any local operator $V_{P_{j}}$ which has  its support exclusively
on the plane $P_{j}$ then any orthonormal GSs of the form of Eq.
(\ref{GSform}) will satisfy Eq. (\ref{general_cond}). Next, let us
examine in detail this matrix element, $\langle \phi_{P_{j}}^{l_{j}} |
V_{P_{j}} | \phi_{P_{j}}^{l^{\prime}_{j}} \rangle$.  We will employ
selection rules on the matrix elements of the irreducible tensors $\{
T^{kq} \}$. Whenever the Wigner-Eckart theorem is applicable,
\begin{eqnarray}
\langle \alpha' j' m' | T^{kq} | \alpha j m \rangle = 
\frac{1}{\sqrt{2j+1}} \langle \alpha' j'| T_{k}| \alpha j \rangle
\nonumber
\\ \times \langle jm; kq| j'm' \rangle,
\end{eqnarray}
we will have that for any local operator $V_{P_{j}}$  (henceforth
denoted by $V$ for brevity), the matrix element between two orthogonal
eigenstates of the $d \ge 1$-dimensional symmetry operator
\begin{eqnarray}
\langle \alpha j_{1} m_{1} | V  | \alpha' j_{2} m_{2} \rangle 
\to 0
\label{same}
\end{eqnarray}
in the thermodynamic limit ($L \to \infty$) if $j_{1} \neq j_{2}$ or
$m_{1} \neq m_{2}$. The proof of this assertion is given by the
asymptotic form of the Clebsch-Gordan coefficients for $SU(2)$ for
systems of finite spin $S$. [For systems of finite spin, products of
more than $S$ on-site spin operators which appear in $V$ (if they
indeed appear) can be reduced to lower order terms.] 

A similar extension of the more familiar variant of the  Wigner-Eckart
theorem holds for general groups other than $SU(2)$ with  similar
conclusions. Even when the Wigner-Eckart theorem is not applicable - as
in e.g. half-integer spin representations of $SU(2)$ - similar
conclusions may be seen to hold.  The proof of this assertion for
half-integer spins follows by an explicit construction of orthogonal
eigenstates of the $d$-dimensional symmetry operator in such cases when
$\alpha = \alpha', j=j'$. For example,  for the  spin $S=1/2$ $d=2$
$SU(2)$ symmetry of pertinence to the KK model,  we have in the sector
of maximal total spin of the  plane $j=N/2$ with $N$ the number of 
sites in the plane, the states 
\begin{eqnarray}
| j=m= \frac{N}{2} \rangle  &=& | \uparrow \uparrow \cdots 
\uparrow \rangle, \nonumber 
\\ | j = \frac{N}{2}, m = \frac{N}{2}-1  \rangle &= &
\frac{1}{\sqrt{N}}  \Big[ |\downarrow \uparrow \uparrow \cdots \uparrow
\rangle + | \uparrow \downarrow \uparrow \cdots \uparrow \rangle \nonumber
\\ &+& | \uparrow \uparrow \downarrow \uparrow \cdots \uparrow \rangle +
\cdots + | \uparrow \uparrow \cdots \uparrow \downarrow \rangle 
\Big], \nonumber
\\ &\vdots & \nonumber \\
| j =\frac{N}{2}, m = \frac{N}{2} - u \rangle &=& 
\sqrt{\frac{(N-u)! \ u!}{N!}} \Big[ |\downarrow \cdots \downarrow
\uparrow \cdots \uparrow \rangle \nonumber
\\ + \mbox{ all other states} &&\!\!\!\!\!\!\!\!\!\mbox{with $u$ down
spins} \Big]. 
\label{maxspin}
\end{eqnarray}
These orthogonal states are related by the exclusive use of the 
$d$-dimensional symmetry operators. The matrix element of any local
operator between  such orthogonal eigenstates of the $d$-dimensional 
symmetry operators tends to zero in the thermodynamic limit ($N \to
\infty$). For instance, the matrix element of the spin lowering 
operator at site $i$,
\begin{eqnarray}
\langle j\!\!=\!\!m=\!\! \frac{N}{2} | S_{i}^{-}| j\!\!=\!\!
\frac{N}{2}, m \!\!= \!\!\frac{N}{2} -1 \rangle 
\!=\! \frac{1}{\sqrt{N}} \underset{N \to \infty}{\longrightarrow} 0. 
\end{eqnarray} 
As in the discussion prior to Eq. (\ref{maxspin}), $N$ denotes the
number of sites in each  $d=2$-dimensional plane. Similarly, the matrix
elements of  all quasi-local operators are bounded from above by
(normalization) factors which scale to zero in the thermodynamic limit.
The  factors for the case of $j = \frac{N}{2}$ are provided in Eq.
(\ref{maxspin}). Here, for large $u$, the matrix element of a local
operator between two orthogonal states becomes exponentially small
(${\cal{O}}(N^{-u/2})$) as can be seen by invoking the Stirling
approximation.  We next construct states for which the  conditions of
Eq. (\ref{def.}) on all quasi-local operators $V$ are the strongest and
then regress to the more general case. 

For states in which $|m-m'|$ is larger than $r$ (the range of the
operator $V$  for an $S=1/2$ system or in $S>1/2$ systems the number 
of independent spin fields which appear in the various product terms
forming a general quasi-local operator $V$), we will have that 
\begin{eqnarray}
\langle \alpha j m| V | \alpha j m' \rangle =0,~~~ |m-m'|>r.
\label{zeromatrixelement}
\end{eqnarray}

Let us next expand the states $\{|\phi_{P_{a}}^{l_{a}} \rangle\}$ in
the complete orthonormal basis of the $d$-dimensional symmetry
generators. In particular, let us focus on those states which may be
written as
\begin{eqnarray}
| \phi_{l} \rangle = \sum_{ m= (r+1) n} a_{l}^{\alpha j m}  |\alpha
j m \rangle,
\label{expand_in_basis}
\end{eqnarray}
with $n$ an integer. In Eq. (\ref{expand_in_basis}), we omitted the
plane index $P_a$. In order to avoid any complications, we  chose the
states in Eq.~(\ref{expand_in_basis}) such that  there is no
off-diagonal matrix element $\langle \alpha j m | V| \alpha j m'
\rangle =0$ for any two states $ | \alpha j m \rangle$ and $| \alpha j
m' \rangle$ which appear in the sum of  Eq.~(\ref{expand_in_basis}).
This vanishing matrix element follows from Eq.
(\ref{zeromatrixelement}). 

As a consequence of Eq. (\ref{same}), the matrix element of a local
operator between two states is
\begin{eqnarray}
\langle \phi_{l} | V | \phi_{l^{\prime}} \rangle 
= \sum_{ m= (r+1) t } (a_{l}^{\alpha j m})^{*} (a_{l^{\prime}}^{\alpha
j m}) \langle \alpha j m| V | \alpha j m \rangle.
\end{eqnarray}
Next, when allowed (e.g. for integer spin renditions of $SU(2)$
systems), we expand $V$ in the irreducible representation of the
$d$-dimensional symmetry,
\begin{eqnarray}
V = \sum_{\alpha k q} T^{\alpha k q}.
\label{dec}
\end{eqnarray}
In the case of the $d=2$ planar $SU(2)$ symmetries of the KK model,
$\{T^{\alpha k q}\}$ behave as spherical tensors under rotations of all
spins in the plane. Here, $k$ denotes the total spin and $q$ the
magnetic quantum number of the spherical tensor $T$. As $V$ is a local
operator, the total angular momentum that it can carry is bounded from
above by a finite number (of the order of its range), $k, q =
{\cal{O}}(1)$. Here, and in what follows we will make extensive use of
Eq.~(\ref{zeromatrixelement}). We now return to the definition of $T=0$
TQO given by Eq.~(\ref{def.}). Obeying Eq. (\ref{reduce}), from which
Eq.~(\ref{def.}) is satisfied, is tantamount to having
\begin{eqnarray}
t^{\alpha k q} \delta_{ll^{\prime}} + c_{ll^{\prime}} = \sum_{ m} 
\langle \alpha j m | T^{\alpha kq} | \alpha j m \rangle
(a_{l}^{\alpha j m})^{*} a_{l^{\prime}}^{\alpha j m}
\label{wec}
\end{eqnarray} 
with arbitrary constants $\{t^{\alpha k q}\}$ for all $l,l^{\prime}$
and for finite $k,q$  which appear as components of the local operator
$V$. 

By the Wigner-Eckart theorem, Eq. (\ref{wec}) reduces to 
\begin{eqnarray}
t^{\alpha k q} \delta_{ll'} + c_{ll'} =  
\frac{\langle \alpha j || 
T^{ \alpha k} || 
\alpha j \rangle}{\sqrt{2j+1}} 
\sum_{m} \langle jm; kq | jm \rangle \nonumber
\\ \times (a_{l}^{\alpha j m})^{*}
(a_{l^{\prime}}^{\alpha j m}),
\label{final_eq_WE}
\end{eqnarray}
for a finite set of $(k,q)$ values with $\langle jm; kq | jm \rangle$
denoting the Clebsch-Gordan coefficients. The double bar in ``$\langle
\alpha j ||  T^{ \alpha k} ||  \alpha j \rangle$'' is used to emphasize
that this coefficient is independent of $m$ and $m'$. Selection rules
mandate that $q=0$. For $m= (r+1)n$ in the sum of
Eq.~(\ref{final_eq_WE}), the corrections $\{c_{ll'}\}$ may be set to
zero. 

We are now faced with the task of satisfying Eq.~(\ref{final_eq_WE})
for a finite set of $(k,q)$ values and for arbitrarily chosen  constants
$\{t^{\alpha k q}\}$. In general,  Eq. (\ref{final_eq_WE})- a compact
shorthand for  a system of quadratic forms in $\{ a_{l;R}^{\alpha j m},
a_{l; I}^{\alpha j m} \}$ with the subscripts $R$ and $I$ denoting the
real and imaginary components of $a_{l}^{\alpha j m}$ and for arbitrary
engineered constants $\{t^{\alpha k q}\}$- admits multiple solutions.
Let us now show this. If we have $y = {\cal{O}}(r)$ possible $(k,q)$
values  in each representation $\alpha$ for the  decomposition of the
finite range $V$ in Eq. (\ref{dec}) then, recalling the  definition
after Eq. (\ref{GSform}), on the left hand side of Eq.
(\ref{final_eq_WE}),  we have ${\cal{O}}(yG^2)$ independent entries.
[We remind the reader that $G$ denotes the number of independent states
$\{| \phi_{P_{j}}^{l} \rangle \}$ on  each plane $P_{j}$ which appear
in Eq.~(\ref{GSform})]. On the right-hand side we have far more entries
as the symmetry indices $m$ also appear:   On its own, for
high-symmetry states (i.e. $j= {\cal{O}}(N)$) the magnetic number $m$
can span ${\cal{O}}(N)$ values. Eq. (\ref{final_eq_WE}) generally 
admits many solutions as $N \to \infty$. 

Let us next avoid the use of Eq. (\ref{zeromatrixelement}) and examine
the most general situation. Repeating the  steps that we undertook
above, we will find that 
\begin{eqnarray}
t^{\alpha k q} \delta_{l l'} + c_{ll'}  =  \frac{\langle \alpha j || T^{\alpha
k}  || \alpha j \rangle}{\sqrt{2j+1}} 
\nonumber
\\ \times \sum_{m m'} \langle j m ; kq | jm' \rangle  (a_{l}^{\alpha j
m})^{*} (a_{l'}^{\alpha j m'}).
\label{generalsumwe}
\end{eqnarray}

In Eq. (\ref{generalsumwe}), locality is manifest in the allowed 
values of the index $k$: i.e. $k \le r = {\cal{O}}(1)$.  As a
consequence of the behavior of the Clebsch-Gordan coefficients,
$\langle j m ; kq |j m ' \rangle$ tends to zero as $|m-m'|>r$. As we
showed earlier, as a consequence of the asymptotic scaling of  these
coefficients, we find that for local operators, the matrix elements
between different eigenstates of the  $d$-GLSs tend, exponentially, to
zero in the thermodynamic limit.  If we do not restrict ourselves to
values of $m,m' = n(r+1)$  as we have in Eq. (\ref{final_eq_WE}), we
will in general need to insert these small matrix elements in the sum
of Eq. (\ref{generalsumwe}). For $SU(2)$, the matrix elements of the
quasi-local operators (for large $j= {\cal{O}}(L^{d})$  and finite $k =
{\cal{O}}(r)$) lead to the appearance of the general Clebsch-Gordan
coefficients
\begin{eqnarray}
\!\!\!\!\!\!\!\!\!\!\!\!\!\!\!\!\!\! 
 \langle j m ; kq | jm' \rangle ~=~~~~~~~~~~~~~~~~ \nonumber
\\ \delta_{m',m+q} \sqrt{\frac{(2j+1)[k!]^{2}(j+m')!(j-m')!}
{(2j+k+1)!(j-m)!(j+m)!(k-q)!(k+q)!}} \nonumber
\\ \times 
\sum_{p=0}^{j+m'} (-1)^{p+k+q} \frac{(k+j+m-p)!(j-m+p)!}{p!(k-p)!
(j+m'-p)!(p+j-k-m')!} \nonumber
\end{eqnarray}
in Eq. (\ref{generalsumwe}).  For large $j = {\cal{O}}(L^{d})$,  these
matrix elements lead to the exponentially small terms discussed
earlier. 
  
For groups other than $SU(2)$ we have  a similar general relation from
the Wigner-Eckart theorem \cite{wk}. This concludes our  construct.
Note that in the $T=0$ case, we do not require that  all GSs be linked
to each other by the exclusive use of the $d$-dimensional symmetry (if
that were  the case only the magnetic number $m$ changes yet $j$ is
held fixed  in going from state to state). The $d$-dimensional symmetry
simply  provides a very convenient  basis for which the construction of
TQO is simple. Our Wigner-Eckart-type construction and  related
variants enable us to construct general states with TQO.  We are
currently working on extensions of this method to  other groups. For
the FQHE, the associated group is that  of magnetic translations (see
Section \ref{SubSection 1}).

It is noteworthy that, in this construct, we do not demand that the
$d$-dimensional symmetry generators commute. In the above construction,
we assumed that we have $d$-dimensional symmetry generators, each
generator acting within a different  plane, and showed how to construct
GSs with TQO.  The presence/absence of any additional symmetries which
do not commute with these is inconsequential. Similarly, the dimension
$d$ of the GLSs need not be small. Any $d \ge 1$ GLSs will allow us to 
follow this construct. All that matters is that all GSs can be written
in the form of Eqs. (\ref{GSform}), and (\ref{expand_in_basis}).

\subsection{A special set of states}
\label{specialselect}

The discrete cyclic symmetry groups $\Z_{n}$ have the simplest
selection rules amongst all GLSs. These discrete symmetry groups,
either appearing on their own right or as homotopy groups for
continuous $d$-GLSs furnish discrete {\it topological}  numbers. 

Nearly all of the prominent examples of TQO to date exhibit a discrete
Abelian $d=1$ GLS. There the analysis simplifies considerably. In
Section \ref{engineer}, we will show in detail how the $T=0$ conditions
are satisfied for generalized Kitaev type models. The Kitaev model
displays a $d=1$ $\Z_{2}$ symmetry. In what follows,  we note how for
these and other general systems a special set of states may be
constructed for which TQO appears. These states encompass the GSs of
the Kitaev model as well as those of the Rokhsar-Kivelson (RK) model
\cite{RK}. In the Kitaev and RK models, the role of the representation
indices of the $d$-GLS groups is taken on by a discrete parity
($\Z_{2}$) index which captures the $d=1$ $\Z_{2}$ symmetries  which
are present in these systems.  Our construct below is more  general
albeit still being quite special. 

The states of general Abelian theories  can also be viewed as
particular substates of a richer non-Abelian theory, since  non-Abelian
symmetries (such as the ones discussed in the  previous Section)
contain certain Abelian symmetries as special subgroups. For instance,
$\Z_{n} \subset U(1) \subset SU(2) \cdots$.  A very special member of
the general states of Eq. (\ref{generalsumwe}) which exhibit TQO is
that of the states
\begin{eqnarray}
|g_{\alpha} \rangle = {\cal{N}} \sum_{| c_{\alpha} \rangle:~~ |
c_{\alpha} \rangle \in \alpha}  \!\!\! f(c) \ |c_{\alpha} \rangle ,
\label{gentop}
\end{eqnarray}
in which we sum over all states in the local spin representation which
lie in the topological sector $\alpha$, in  which the number of states
in each topological sector is the same. The label of the topological
sector is generally none other than the representation index of a
discrete $\Z_{n}$ $d$-dimensional symmetry. The function $f(c)$ in Eq.
(\ref{gentop}) is such that there is a one-to-one correspondence
between individual states in the topological sectors  $|c_{\alpha}
\rangle  \leftrightarrow |c_{\beta} \rangle$ with  $f(c_{\alpha}) =
f(c_{\beta})$. The coefficients $f(c)$ need to be chosen in order to
ensure Eq. (\ref{def.}).  We will return to this form in Section
\ref{engineer}. 

As noted above, an example  is afforded by the GSs of both the Kitaev
(Eqs. (\ref{kitaevmodel}), and (\ref{AB_defn})), and Rokhsar-Kivelson
models which we will discuss in more detail in later Sections.  In the
GSs of these systems, $f(c) =1$ in Eq. (\ref{gentop}), and the GSs of
the  completely symmetric form of Eq. (\ref{soln2Kit}). Here, the
sectors $\alpha$ are determined by the {\it winding} number  in a $d=1$
surface. This winding number - a consequence of the $d=1$ symmetries
that these systems display in their low-energy sector - takes on the
role of the general group representation index which  we employed in
our analysis of the former Section. 

As we will later return to (Section \ref{engineer}), in the Kitaev
model of Eqs. (\ref{kitaevmodel}), and (\ref{AB_defn}), there is a
$d=1$ symmetry selection rule
\begin{eqnarray}
\langle g_{\alpha} | \prod_{a \in \Omega} \sigma_{a}^{\mu} |
g_{\beta} \rangle =0, ~~~ \alpha \neq \beta,
\label{vkm}
\end{eqnarray} 
with $\mu=x,y,$ or $z$  if the volume $\Omega$ is finite (the operator
$V$ is local). This mandates that the off-diagonal condition of Eq.
(\ref{general_cond}) is satisfied. We now prove the diagonal portion of
Eq. (\ref{general_cond}) for the states of Eq. (\ref{gentop}) which
constitute the solution to the GS problem of the Kitaev model. Let us
denote by $T_{\alpha \beta}$ the operator that generates the one-to-one
correspondence between the local spin basis states in the topological
sectors $\alpha$ and $\beta$, i.e.  $T_{\alpha \beta} |c_{\alpha}
\rangle = | c_{\beta} \rangle$. Let the states $\{ |c_{\alpha, \beta} 
\rangle \}$ be eigenstates of $\bigotimes_{i \in \Lambda}
\sigma^{z}_{i}$ with $\Lambda$ all lattice sites. In the Kitaev model,
there is, up to a gauge transformation (an application of $A_{s}$), 
one such operator for a given horizontal/vertical winding number sector
such as that of $\alpha$ or $ \beta$,
\begin{eqnarray}
T = \prod_{ij \in P} \sigma_{ij}^{x},
\end{eqnarray}
where we may always deform the horizontal/vertical contour $P$  by
multiplying it by $A_{s}$ (as $A_{s}=1$ in the GSs) such that it does
not traverse the  region of support $\Omega$ of the local operator $V$.
All that matters is that $P$ winds once in the horizontal and/or
vertical directions. Here, trivially,
\begin{eqnarray}
[T,V] =0.
\end{eqnarray}
Consequently,  the diagonal matrix element of $V$
\begin{eqnarray}
\langle g_{\alpha} | V| g_{\alpha} \rangle  = \langle g_{\alpha} |
T^{\dagger} V T| g_{\alpha}  \rangle =\langle g_{\beta} |  V| g_{\beta}
\rangle. 
\end{eqnarray}

When combined with the vanishing off-diagonal matrix element of  $V$ in
Eq. (\ref{vkm}), this proves the diagonal  portion of Eq.
(\ref{general_cond}). 

When expressed in the original $SU(2)$ spin language,  the maximally
symmetric state of the Kitaev model solution correspond, in the $d=1$
total spin eigenbasis to the 4 states highest spin $(S_{tot}=L/2$)
states
\begin{eqnarray}
| \psi_{\pm \pm} \rangle = \!\!\!\!\sum_{S_{z,h}^{v/h} = L/2-
~{\sf even/odd}} \!\!\!\!\!\!\!\!\!\!\!\!\!\!C_{S_{z}^{v} S_{z}^{h}} \ 
|S_{tot}=L/2, S_{z}^{v}, S_{z}^{h} \rangle.
\label{kitsymbasis}
\end{eqnarray}
The superscript $v$ and $h$ refer to vertical and  horizontal
directions. Here, the symmetry operators along the horizontal and
vertical directions commute with one another and we can employ the
basis states $\{|S_{tot}^{v}, S_{z}^{v}; S_{tot}^{h}, S_{z}^{h}
\rangle\}$ as we have in Eq. (\ref{kitsymbasis}).  In the thermodynamic
limit, where the total spin states do not admit off-diagonal
expectation values of local operators, the constant $C$ in Eq.
(\ref{kitsymbasis}) is akin to $f$ in the general  Eq. (\ref{gentop}).
As we noted earlier, the  $\Z_{2}$ operators (of Eq. (\ref{AB_defn}))
which appear in the Kitaev problem  are a particular subset of the
larger $SU(2)$ group.

\subsection{Implications for the energy spectrum - exact degeneracy of 
states in the thermodynamic limit} 

A corollary of our construct is that for any Hamiltonian which is of
finite range, i.e. for any Hamiltonian which can be expressed as 
\begin{eqnarray}
H = \sum_{i} V_{i},
\label{HV}
\end{eqnarray}
with quasi-local operators $\{V_{i}\}$ which have their support in a
domain including the site $i$, degenerate states can be immediately
constructed. Applied to Eq.~(\ref{HV}), our conditions imply that all
states $|\phi_{l} \rangle$ (and general combinations thereof such as
those appearing in Eq.~(\ref{GSform})) are degenerate in energy up to
the exponential corrections (if all $c_{ll'}$ in
Eqs.~(\ref{final_eq_WE}), and (\ref{generalsumwe}) are bounded from
above by exponentially small quantities). This is so as
\begin{eqnarray}
|\langle g_{l}|H|g_{l} \rangle - 
\langle g_{l'}|H|g_{l'} \rangle| 
= |\sum_{i} (c_{ll}^{i} - c_{l'l'}^{i})|  \nonumber
\\ \le c_{*} L^{D} \exp(-a/L) \to 0,
\label{degeq}
\end{eqnarray}
on a hypercubic lattice of side $L$ as $L \to \infty$. In Eq.
(\ref{degeq}), for all states $\{|g_{a} \rangle\}$ ($a= l, l'$) and for
all quasi-local operators $V_{i}$ which have support at a site $i$, we
have $\langle g_{l} |V_{i} | g_{l'} \rangle = v_{i} \delta_{ll'} +
c^{i}_{ll'}$, $|c_{ll}^{i}| \le |c_{i}|  \exp(-aL^{d})$ with a 
constant $a>0$ [for a $d \ge 1$-GLS], and $c_{*} = \max_{i}
\{|c_{i}|\}$.

\subsection{The failure of this construct for {\it non-topological}
(local) symmetries}

It is important to emphasize that a crucial ingredient  in our
construct, Eq. (\ref{same}), relies on  the eigenbasis of $d \ge 1$
symmetry generators.  In the presence of  local ($d=0$) symmetries
alone [i.e. if no other symmetries are present], this relation fails.
For $d=0$ symmetries, the eigenstates of the  symmetry generator
contain a finite number of sites and the expectation values of local
operators between two  orthogonal symmetry eigenstates can be finite.
An example is afforded, e.g., by the classical $\Z_{2}$ lattice gauge theory
\begin{eqnarray}
H = - K \sum_{\Box} \sigma^{z}_{ij} \sigma^{z}_{jk} \sigma^{z}_{kl}
\sigma^{z}_{li}
\label{Z2t}
\end{eqnarray}
[which is $H_{\Z_{2}}$ of Eq. (\ref{gz2+}) with $h_{x}=0$]
for which the local symmetry operators
\begin{eqnarray}
G_{i} = \prod_{r} \sigma^{x}_{ir},
\label{GI}
\end{eqnarray}
with $r$ all nearest neighbors, the site $i$ link orthogonal GSs.  In
this particular case the local operators $\{G_{i}\}$ are none other
than $\{A_{s}\}$ of Eq. (\ref{AB_defn}) which defines the Kitaev model.
The local symmetry operators $V=G_{i}$ clearly violate Eq.
(\ref{general_cond}) and the  $\Z_{2}$ gauge theory of Eq. (\ref{Z2t}) 
does not exhibit $T=0$ TQO for all of its GSs. A partial  set of GSs of
Eq. (\ref{Z2t}) does  exhibit TQO.  This partial set for the $D=2$
theory is given by  the GSs of the Kitaev Toric code model which we
discuss next. The finite-$T$ TQO for Eq. (\ref{Z2t}) is immediate  and
follows from Elitzur's theorem. 

\section{Engineering TQO: Generalized Kitaev-type models}
\label{engineer}

In this Section, we will show how many models can be systematically
engineered to have  $T=0$ TQO. This will extend and complement some of
the general results introduced in Ref. \cite{kitaev} as well as our
results of Sections \ref{selectlong}, and \ref{specialselect}. 

We start with a general Hamiltonian $H$ which  has GSs which may all be
linked to one another  by the exclusive use of discrete $d \le 1$ 
symmetries or continuous $d \le 2$ symmetries. Note that $H$ may in
general have $d=0$  symmetries. Henceforth, we discuss systems which
have such quasi-local symmetries and we label these symmetries by the
unitary operators $\{ G_{i} \}$.  We (i) discuss cases in which all of
these symmetries commute with one another. As $[H, G_{i}]=0$, we can
simultaneously diagonalize $G_{i}$  and $H$. (ii) In what follows,  we
consider models in which  the GS sector of $H$ contains, amongst
others,   states with a uniform unit  eigenvalue under all $\{G_{i}\}$,
(iii) If a local quantity is invariant under all of the $d=0$
symmetries $\{G_{i}\}$ then it is also invariant under all remaining 
$d>0$ symmetries which are necessary to link orthogonal GSs. Finally,
(iv) we consider systems in which all basic local observables transform
as a singlet under $\{G_{i}\}$,  $G_{i}^{\dagger} V_{i} G_{i} =
\lambda_{V;i} V_{i}$.  If $[V,G_{i}] \neq 0$ then $\lambda_{V;i} \neq
1$.

Let us start by considering  
\begin{eqnarray}
\tilde{H} = H - h \sum_{i} G_{i},
\label{tH}
\end{eqnarray}
with $h>0$. We claim that  under the conditions stated, the GSs of
$\tilde{H}$ display TQO.  The proof of this assertion is simple: From
conditions (ii) and (iv), it follows that for all GSs of $\tilde{H}$,
are the subset of the GSs of $H$ which have $G_{i} |\tilde{g}_{\alpha}
\rangle = | \tilde{g}_{\alpha} \rangle$.  On the one hand, the
expectation value  
\begin{eqnarray}
\langle \tilde{g}_{\alpha} | G_{i}^{\dagger} V
G_{i} | \tilde{g}_{\alpha} \rangle = \lambda_{V;i}  \langle
\tilde{g}_{\alpha} | V | \tilde{g}_{\alpha} \rangle.
\end{eqnarray} 
On the other
hand,  
\begin{eqnarray}
G_{i} | \tilde{g}_{\alpha} \rangle &=& |\tilde{g}_{\alpha}
\rangle, \nonumber
\\ \langle \tilde{g}_{\alpha} | G_{i}^{\dagger} V G_{i} |
\tilde{g}_{\alpha} \rangle &=& \langle \tilde{g}_{\alpha} | V |
\tilde{g}_{\alpha} \rangle.
\end{eqnarray} 
If there is at least one $i$ for which
$[V,G_{i}] \neq 0$ then it follows that 
\begin{eqnarray}
\langle  \tilde{g}_{\alpha} |
V | \tilde{g}_{\alpha} \rangle =0.
\end{eqnarray}
Putting all of the pieces together, this guarantees that the diagonal
portion of  Eq. (\ref{def.}) is satisfied: (a) If $[V,G_{i}] = 0$ for
all $i$ then $\langle \tilde{g}_{\alpha} | V | \tilde{g}_{\alpha}
\rangle = \langle \tilde{g}_{\beta} | V | \tilde{g}_{\beta} \rangle$ 
as $V$ is invariant under symmetries linking $| g_{\alpha} \rangle$ 
and $|g_{\beta} \rangle$ (assumption (iii)).    (b) If $[V,G_{i} ] \neq
0$ then, as we showed, for all GSs,   $\langle \tilde{g}_{\alpha}| V|
\tilde{g}_{\alpha} \rangle =0$. 

It is clear that condition (ii) is a necessary condition for TQO:
For any local symmetry $G_{i}$, there exist many states GSs for which 
$G_{i}|g_{\alpha} \rangle = | g_{\alpha} \rangle$. 

As we showed earlier in the proof of our finite-$T$ result,
condition (ii) suffices to establish that the off-diagonal  portion of
Eq. (\ref{def.}) is obeyed. 

Let us now discuss how the Kitaev model is a special instance of such
Hamiltonians of Eq.~(\ref{tH}).  Here, $H = - \sum_{\Box}
\sigma^{z}_{ij} \sigma^{z}_{jk} \sigma^{z}_{kl} \sigma^{z}_{li}$ is the
Hamiltonian of the $\Z_{2}$ gauge theory which is the sum over all
plaquettes of the operator $B_{p}$ (Eq. (\ref{AB_defn})). Corresponding
to $H$ are the local symmetry operators 
\begin{eqnarray}
G_{i} = \sigma^{x}_{ij} \sigma^{x}_{il} 
\sigma^{x}_{ir} \sigma^{x}_{is}.
\end{eqnarray} 
These are the $A_{s=i}$ symmetries of Eq. (\ref{AB_defn}). With $h=1$,
the corresponding $\tilde{H}$ of Eq. (\ref{tH}) is none other than the
Kitaev model.  The $d=0$ symmetries of $H$ are just $\{A_{s}, B_{p}\}$
(see  Eq. (\ref{AB_defn})). We now verify that conditions (i) - (iv)
are satisfied for the Kitaev model: (i) All of these symmetries commute
with one another, 
\begin{eqnarray}
 [A_{s}, A_{s'}]= [B_{p}, B_{p'}]= [A_{s},
B_{p}]=0.
\end{eqnarray}
Condition (ii) is satisfied for any of the four maximally  symmetric
states \cite{kitaev}: 
\begin{eqnarray}
|\psi_{q} \rangle = 2^{1-N_{s}/2} \sum_{c \in q} | c \rangle,
\label{soln2Kit}
\end{eqnarray}
where the sum is performed over all states $|c \rangle$ in the
$\sigma^{z}$ basis which have $B_{p} =1$ on every plaquette and for
which the  two $\Z_{2}$ operators $\{Z_{1}, Z_{2} \}$ of Eqs.
(\ref{kit}) assume one of the four specific values $q = (\pm 1, \pm
1)$. Note that $|\psi_{q} \rangle$ is the sum of all GSs of the
classical $\Z_{2}$ model (that with  the transverse field $h=0$ in 
Eqs. (\ref{gz2+}), and (\ref{tH})) which belong to the topological
sector $q$. Eq.~(\ref{soln2Kit}) is a special  realization of
Eq.~(\ref{gentop}). 
  
(iii) Here, if any local operator $V$ is invariant under all of these
$d=0$ symmetries then $V$ must be invariant under the $d=1$ symmetries
of Eq. (\ref{kit}). To see this, we note that for any such operator $V=
f(\{A_{s}\}, \{B_{p}\})$. As a functional of $\{A_{s}\}$ and
$\{B_{p}\}$, $V$ is automatically invariant under the $d=1$ symmetries
of Eq. (\ref{kit}).  Lastly, condition (iv) is straightforwardly
satisfied:  
\begin{eqnarray}
A_{s} \sigma^{z}_{ij} A_{s} &=& - \sigma^{z}_{ij},\nonumber 
\\ A_{s}
\sigma^{z}_{kl} A_{s} &=& \sigma^{z}_{kl} ~~~~(s \neq k,l),
\end{eqnarray} 
and 
\begin{eqnarray}
A_{s} \sigma^{x}_{ij} A_{s} = \sigma^{x}_{ij}.
\end{eqnarray}
\noindent [Similarly, $B_{p} \sigma^{x}_{ij} B_{p} = -
\sigma^{x}_{ij}$  if the bond $\langle i j \rangle$ lies in the
plaquette $p$,  $B_{p} \sigma^{x}_{ij} B_{p}  = \sigma^{x}_{ij}$ if the
bond  $\langle ij \rangle$ does  not lie in $p$,  and $B_{p}
\sigma^{z}_{ij} B_{p} = \sigma^{z}_{ij}$.] As a consequence of these
relations, any multinomial $V_{i}$ in these fields which contains,
amongst others, fields at the site $i$ satisfies $G_{i}^{\dagger} V_{i}
G_{i} = \lambda_{V;i} V_{i}$ with $G_{i} = A_{s}$. Putting all of the
pieces together, the reader can convince him/herself that Eq.
(\ref{soln2Kit}) constitutes the solution to the GS problem of the
Kitaev model. It is noteworthy that here TQO (and the GSs of Eqs.
(\ref{soln2Kit})  appears for all $h>0$.  The generalization to other
Hamiltonians $H$ is straightforward. Any GS of the Kitaev model can be
expressed  as a projection of a reference state to the GS sector (that 
in which $A_{s} =1$ for all sites $s$  and $B_{p}=1$ for all plaquettes
$p$):
\begin{eqnarray}
{\cal{N}}\prod_{s} \Big[ \frac{1}{2} (1+ A_{s}) \Big] \prod_{p} \Big[ 
\frac{1}{2} (1+ B_{p}) \Big] | \phi \rangle.
\label{projkit}
\end{eqnarray}
Here, ${\cal{N}}$ is a normalization constant and $| \phi \rangle$ is
a  reference state.  All GSs of the Kitaev model can be linked to one
another by the exclusive use of $d=1$ symmetries. Due to our theorem
[see Section \ref{Section 5}]  TQO follows for all $T>0$.  This is so
as our analysis  above shows that TQO appears at $T=0$. 

It should be noted that other prominent examples of TQO (e.g. the
Quantum Dimer Model \cite{RK}) have  TQO because of the very same
selection rules for the $d$-GLSs which connect their GSs. In the
Appendix \ref{QDMlink}, we discuss certain technical aspects of the
Quantum Dimer model.

\section{Energy Spectra and thermal fragility of some anyonic 
systems}
\label{Energyspectrumsection}

The entire focus of our work is  on the non-local symmetries ($d>0$)
which relate and define different topologically ordered states; we 
showed how the excitation spectrum associated with the restoration of
these  symmetries through the appearance of low-dimensional topological
defects can eradicate local orders and enforce the appearance
of TQO \cite{NO}. In most cases,  this dimensional reduction insofar as
SSB is concerned  does not lend itself to an exact dimensional
reduction in the form of the associated free energies. In this Section,
we dwell on several symmetries harboring low $d$-GLSs for which an
exact dimensional reduction (of the partition function)  also occurs.
For example, Kitaev's and Wen's models of Eqs. (\ref{kitaevmodel}),
(\ref{AB_defn}), and (\ref{hw}) display $d=1$  $\Z_{2}$ symmetries and
fortuitously these $D=2$ systems can indeed also be mapped onto
nearest-neighbor Ising chains (of $D=d=1$). This accidental  exact
reduction of the partition function for these special  models points at
much broader relations - those of the insufficiency of the  spectra in
determining whether  or not TQO appears and in illustrating the thermal
fragility of topological quantities in systems such as these. Apart
from being mathematically interesting, these systems  and extensions
therein might pave the way for TQO computing.   Kitaev's Toric code
model \cite{kitaev} was the  first {\it surface code}  \cite{pachos}
suggested for {\em anyonic quantum computing}.  
 
As we stated in \cite{NO}, dualities illustrate the fact that  the
quantum states themselves and not the energy spectrum  in a particular
(operator language) representation encode TQO. The energy spectrum of
theories with and  without TQO may be the same. If two systems have the
same energy  spectrum,   then they are related by a unitary
transformation that links the two sets of eigenstates. If one system
has TQO while the other does not then this transformation is generally
non-local. 

That the spectrum, on its own, is insufficient to determine TQO
\cite{NO} is established by  counter-examples, e.g. that of the $D=3$
$\Z_{2}$ lattice gauge theory  which is dual to a $D=3$ Ising model
\cite{wegner}. Albeit sharing the same energy spectrum, in the gauge
theory we may find GSs with, in the thermodynamic limit,   rank-$n =8$
(see our definition of rank-$n$ TQO in Section \ref{Section 2})
finite-$T$ TQO while the Ising model harbors a local order.  Similarly,
the $\Z_{2}$ gauge theory on a square lattice [given by
Eq.~(\ref{gz2+}) with $h=0$] is equivalent to the classical $D=1$ 
Ising model. This last, well known, equivalence can be proven by, for
example, replicating the analysis which we will perform in this Section
for both Kitaev's Toric code model and Wen's plaquette model.  The
$D=1$ Ising model has no TQO while the $D=2$ $\Z_{2}$ gauge theory 
(which is dual to it) has rank-$n=4$ TQO. [Four GSs of the  $D=2$
$\Z_{2}$ gauge theory which satisfy Eq. (\ref{def.}) are  given by Eq.
(\ref{soln2Kit}).]

Another example is given by the Kitaev Toric code model of Eqs.
(\ref{kitaevmodel}) and (\ref{AB_defn}) in a system with open boundary
conditions. [In our convention, 
on the square lattice with open  boundary conditions, the
star operator appears only if all of the 4 bonds which make it up
appear in the lattice. We do not allow for star operators near the
boundary of the  system where only three or two operators appear in the
product of Eq. (\ref{AB_defn}) defining the operator $A_{s}$.] As  all
of the Ising operators $A_{s}, B_{p}$ commute with one another, the
spectrum can be easily determined. Here, the density of states
\begin{eqnarray}
g(E) = \sum_{m=0}^{M} \delta_{E,M-2m} \frac{M!}{m!(M-m)!} ,
\label{ges}
\end{eqnarray}
which is identical to that of a $D=1$ Ising model of $M$ sites with
open boundary conditions  or alternatively of $M$ decoupled Ising spins
in an external magnetic field of unit strength. Here, on the square
lattice of $N_s$ sites with open boundary conditions, there are a total
of 
\begin{eqnarray}
M = 2N_{s} - 6 \sqrt{N_{s}} +5
\label{Mnumber}
\end{eqnarray}
 independent
fields $\{A_{s}, B_{p}\}$.  The specific form of  Eq. (\ref{Mnumber})
for a square lattice of open boundary conditions with $N_{s} = (L+1)^2$
sites follows as there are  $L^2$ plaquettes $\{B_{p}\}$  and $(L-1)^2$
{\it stars} $\{A_{s}\}$. The sum of the number of  plaquettes and stars
is given by Eq. (\ref{Mnumber}).  The origin of Eq. (\ref{ges}) and
similar  relations is the mapping that we can perform between each bond
(in a $D=1$ Ising model) or each site (for a system of decoupled Ising
spins in a magnetic field) and the $M$  decoupled commuting Ising
variables $\{A_{s}\}$ and $\{B_{p}\}$.  An identical correspondence
between the two systems [Kitaev's model and the Ising chain] in the
presence of periodic boundary conditions will be detailed in Section
\ref{Tfragile}.

Any system (whether TQO ordered or not) with (for finite $L$) a finite
number of states ($P$), has a spectrum which is identical to that of a
spin system with $T=0$ TQO. The proof of this assertion follows from
our discussion in Section \ref{selectlong}. To pictorially  summarize
this state of the affairs,  we write \newline

\underline{TQO} ~~~~~~~~~~~~~~~~~~~~~~~~~~~~~~~~~~~~~~~~~~~~\underline{No 
TQO} \newline
$D$=3 Ising gauge ~~~~~~~~~~~~~~~~~~~$\leftrightarrow$  ~ $D$=3 Ising
model \newline
$D$=2 Ising gauge  ~~~~~~~~~~~~~~~~~~~$\leftrightarrow$  ~ $D$=1 Ising
model 
\newline
Kitaev's model = Wen's model 
$\leftrightarrow$  ~~$D$=1 Ising model \newline
\newline
A spectral equivalence exists between other pairs of TQO and non-TQO
systems.  Thus, we cannot, as coined by Kac, ``hear the shape of a
drum". The information regarding the existence/absence of TQO is in the
eigenvectors.  We will further show that at any finite temperature,
some of the  toric cycle operators (or in general other $d$-GLSs
operators) may have vanishing expectation values unless they are held
fixed by the application of  an external field. This occurs
notwithstanding the  existence of spectral gaps. Commonly, the
existence of a finite gap between the ground and excited states is
assumed to protect properties associated with $T=0$ TQO up to a small
finite temperature. Our results show that this assumption is, in general,
incorrect. In the Letter of \cite{NO}, we emphasized these properties.
In the following Sections, we will provide details for these claims. 

\subsection{Dimensional Reduction and Thermal fragility in Kitaev's 
Toric code model}
\label{Tfragile}

In this Section, we illustrate that, as a consequence of a dimensional
reduction to a $D=1$ Ising model, the Kitaev Toric code model
exhibits a loss of its toric operators expectation values at any finite
$T$ (i.e. $T>0$, no matter how small the temperature is).  This loss
occurs notwithstanding the existence of a spectral gap in this system.
This does not imply that the Kitaev Toric  code model does not have
finite-$T$ TQO. Indeed, as we showed by our theorem, this model does
exhibit TQO at all  $T$. If the fields on a chosen contour  (a
boundary) are not allowed to fluctuate then the toric operators are
robust due to the absence of correlations between the boundary and all
interior sites. Fixing these boundary fields has no influence on the
expectation  value of any quasi-local operator  $V$: Eq. (\ref{vt}) is
satisfied. If, however, we were to allow the fields along a toric cycle
to fluctuate then although at $T=0$, the GSs have finite values of the
Toric code operators, at finite $T>0$, there is no SSB which leads to 
finite expectation values of these operators. The Kitaev Toric code
model has a thermodynamic phase transition at $T=0$. We similarly
discuss Wen's plaquette model. We further illustrate in this exactly
solvable system the important property of {\em adiabaticity}: exact
GLSs can be lifted while not changing the phase of the system and 
discuss within this particular system in the framework of our general
Ising chain mapping, the dependence of the  GS degeneracy on topology.

\subsubsection{Mapping of the Kitaev model to Ising chains and 
its consequences}   

To prove the assertions above, we compute the exact (generating)
partition function of the Kitaev Toric code model in the presence of
applied infinitesimal fields, 
\begin{eqnarray}
Z&=& \tr_{\{\sigma^{z}_{ij}\}} \exp \Big( -\beta \Big[ H_K \nonumber \\
&-& h_{x,1} X_{1} - h_{x,2} X_{2} - h_{z,1} Z_{1} - h_{z,2} Z_{2} \Big]
\Big).
\label{pfunction}
\end{eqnarray}
The operators appearing in the argument of the trace are those defined
in Eqs. (\ref{kitaevmodel}), (\ref{AB_defn}), and (\ref{kit}).   The
operators $\{X_{1,2}\}$ and $\{Z_{1,2}\}$ [see Eq. (\ref{kit})]  obey
(abelian) {\em anyonic} statistics.  For a square lattice having
$N_{s}$ sites with periodic boundary conditions,
\begin{eqnarray}
\prod_{s} A_{s} = \prod_{p} B_{p} =1.
\label{ABpc}
\end{eqnarray}
The partition function of Eq. (\ref{pfunction}) is
\begin{eqnarray}
Z &=& \tr  \exp \Big( -\beta \Big[
H_K \nonumber \\
&-& h_{x,1} X_{1} - h_{x,2} X_{2} - h_{z,1} Z_{1} - h_{z,2} Z_{2} \Big] \Big).
\label{pfunction1}
\end{eqnarray}

Equation (\ref{pfunction1}) with the constraint of Eq.~(\ref{ABpc})  is
none other than the partition  function of two decoupled circular chains
of non-interacting $S=1/2$ spins of length $N_{s}$ each. The partition
function is given by  
\begin{eqnarray}
Z_{torus}= [ (2 \cosh \beta)^{N_{s}} + (2 \sinh \beta)^{N_{s}}]^{2} 
\nonumber \\ 
\times [\cosh \beta h_{1}] [ \cosh \beta h_{2}],
\label{partfinal}
\end{eqnarray}
with $h_{i} = \sqrt{h_{x,i}^{2} + h_{z,i}^{2}}$.  It is important to
emphasize that Eq. (\ref{partfinal}) constitutes  the partition
function for the Toric code model for {\em arbitrary} size systems
$N_{s}$ on the torus. That this is the partition follows from, for
example, the  identification of a similar algebra in the $D=1$ Ising
model. Let us consider  two decoupled Ising chains, each of length
$N_{s}$ and satisfying periodic boundary conditions and examine the
algebra of their bonds. To this end, let us define the bond variables 
\begin{eqnarray}
a_{s} &\equiv& \sigma^{z}_{s} \sigma^{z}_{s+1}, \nonumber \\ 
b_{p} &\equiv& \sigma^{z}_{p} \sigma^{z}_{p+1}.
\label{abd}
\end{eqnarray} 
in terms of which the Hamiltonian of the decoupled Ising chains is
\begin{eqnarray}
H = - \sum_{s} a_{s} - \sum_{p} b_{p}.
\label{Isingmp}
\end{eqnarray}
In what follows, we drop the polarization index (${z}$). This
polarization index can be forgotten henceforth until the end of this
Section when we write the  second half of the duality relations of
Eq.~(\ref{abd}). 

The Ising bond fields  $\{a_{s}, b_{p}\}$ on each circular chain
satisfy 
\begin{eqnarray}
\prod_{s} a_{s} = \prod_{p} b_{p} =1
\label{abDUAL}
\end{eqnarray}
along with the same algebra as the star and plaquette terms $\{A_{s}\}$
and $\{B_{p}\}$ of the Kitaev Toric code model (see
Eq.~(\ref{AB_defn})),
\begin{eqnarray}
[a_{s},  a_{s'}]= [b_{p}, b_{p'}]= [a_{s}, b_{p}]=0, \nonumber
\\ a_{s}^{2} = b_{p}^{2} =1.
\label{abdual}
\end{eqnarray}
The number of independent fields in each case (the Kitaev Toric code
model vis a vis two decoupled circular Ising chains of length $N_{s}$)
is  exactly the same - there are $2N_{s}$ independent spin variables
$\sigma_{ij}$ versus (for $N_{s}$ Ising variables $\{\sigma_{s}\}$ and
$N_{s}$ Ising variables $\{\sigma_{p}\}$) $2N_{s}$ independent spin
variables $\{a_{s},b_{p}\}$. In both cases,  these variables satisfy
the same set of   constraints [Eq. (\ref{abDUAL}) for the Ising chains
vs Eq. (\ref{ABpc}) for the Kitaev model  variables] and the same
number of Ising variables  which satisfy the same algebra and span the
same Hilbert space of size  $2^{2N_{s}}$. Moreover, they have exactly
the same Hamiltonian (that of Eq.~(\ref{kitaevmodel})).  In the
presence of additional fields (see last terms in Eq.
(\ref{pfunction1})), we may define the two component vector
\begin{eqnarray}
\hat{n}_{i} = \frac{1}{|\vec{h}_{i}|} (h_{x,i}, h_{z,i})
\end{eqnarray}
and set 
\begin{eqnarray}
Q_{1} &=& X_{1} n_{x1} + Z_{1} n_{z1}, \nonumber \\ 
Q_{2} &=& X_{2} n_{x2} + Z_{2} n_{x2}
\end{eqnarray} 
to be the counterparts of two single Ising spins $\sigma^{z}_{sc}$ and
$\sigma^{z}_{pd}$ which  are located at site numbers $c$ and $d$ of
the  two respective Ising chains (that of the $s$ and  that of the $p$
varieties). $c$ and $d$  can be chosen to be any integers such  that $0
\le c, d  \le (N_{s}-1) $. Here, the algebra and set of constraints
satisfied by the two chain Ising  variables $\{a_{s}, b_{p},
\sigma^{z}_{sc},  \sigma^{z}_{pd} \}$ which span a
$2^{2N_{s}}$-dimensional  space is exactly the same as that satisfied
by the Kitaev variables $\{A_{s}, B_{p}, Q_{1}, Q_{2}\}$ which span the
same space.  The terms ($- h_{i} Q_{i}$) of  Eq. (\ref{pfunction1})
maps onto magnetic field terms acting at a single site ($c$) of each of
the two Ising chains. The partition function of a circular Ising chain
with a magnetic field of strength $h$ acting at only one site $c$ and
having no on-site magnetic field on all other sites is $\Big( [(2 \cosh
\beta)^{N_s} + (2 \sinh \beta)^{N_s}]  \cosh \beta h \Big)$. In our
Kitaev model mapping, we have two such Ising chains. This leads to the
partition  function of Eq.~(\ref{partfinal}).  An equivalent derivation
of this  partition function employs a high temperature expansion (see
e.g. \cite{percolation1} for an  application of these expansions in
plaquette actions similar to  those of the Kitaev model). As seen by
inverse Laplace transforms of the two partition functions (that of
Kitaev's model and that of the two decoupled Ising rings), our  mapping
implies that the degeneracy of each level in Kitaev's model is equal to
the degeneracy of two decoupled circular Ising chains  at the same
energy level. In particular, {\em the four GSs of the two decoupled
Ising chains} (two GSs (all spins up or all spins down) for each  of
the two decoupled chains) {\em imply the  existence of four GSs of
Kitaev's model}. Those four GSs of the Kitaev model are indeed realized
in Eq. (\ref{soln2Kit}). 

As we remarked earlier, for a system with open boundary conditions,
each $A_{s}$ or $B_{p}$ may be regarded as a decoupled Ising spin. [On
the square lattice with open  boundary conditions, the star operator
appears only if all of the 4 bonds which make it up appear in the
lattice. We do not allow for star operators near the boundary of the 
system where only three or two operators appear in the product of Eq.
(\ref{AB_defn}) defining the operator $A_{s}$.] The spins do not
interact with one another and the resulting partition function is 
\begin{eqnarray}
Z_{open}= 2^{N_{s}}  [\cosh \beta]^{M} \cosh \beta h_{1}
\cosh \beta h_{2},
\label{partfinal1}
\end{eqnarray}
with $M$ given by Eq. (\ref{Mnumber}).  The only meaningful
thermodynamic properties  are given by the thermodynamic limit $N_{s}
\rightarrow \infty$ where, as they must, the results of Eqs.
(\ref{partfinal}), and (\ref{partfinal1}) for periodic and open
boundary conditions  lead to identical conclusions.

Alternatively, we briefly note, that Kitaev's model can be identified
as that of two decoupled Ising gauge theories on a square lattice  with
$h=0 $ in Eq.~(\ref{gz2+}). One of these gauge theories corresponds to
the plaquettes $B_{p}$ and the other to the sum of the star operators
$A_{s}$.  As the partition function of the $d=2$ Ising gauge theory is
equivalent to the one of an Ising chain, the correspondence  between
Kitaev's Toric code model and the Ising chain immediately follows. We
now comment on a central observation in this Section: As the partition
functions $Z(\beta)$ for $D=1$ Ising model and of the Kitaev Toric code
model are the same, the Laplace transforms, the densities of states,
are also the same in both systems. 

As a particular consequence of the spectral equivalence between the
$D=1$ Ising model and the Kitaev Toric code model, all non-degenerate
states  of the Kitaev Toric code model are separated from one another
by integer multiples of a uniform  constant gap. We have that
\begin{eqnarray}
\langle Z_{i} \rangle \! &=& \!\! \lim_{h_{z,i} \to 0^{+}}
\frac{\partial}{\partial (\beta h_{z,i})} \ln Z \nonumber \\ 
&=& \!\! \lim_{h_{z,i} \to 0^{+}}\! 
\frac{h_{z,i}}{h_{i}} \tanh(\beta h_{i}); \nonumber \\ 
\langle X_{i} \rangle \! &=& \!\! \lim_{h_{x,i} \to 0^{+}}
\frac{\partial}{\partial (\beta h_{x,i})} \ln Z \nonumber \\ 
&=& \!\!  \lim_{h_{x,i} \to 0^{+}}\!  \frac{h_{x,i}}{h_{i}} \tanh(\beta
h_{i}).
\label{expectpart}
\end{eqnarray}
In the above, we compute the finite-$T$ expectation values of  the
Toric code operators  $\langle X_{i} \rangle$, and $\langle Z_{i}
\rangle$ by inserting Eq. (\ref{partfinal}) into Eq.
(\ref{expectpart}). We see that for all temperatures $T>0$, 
\begin{eqnarray} 
\langle Z_{1} \rangle = \langle Z_{2} \rangle = \langle X_{1} \rangle 
= \langle X_{2} \rangle =0. 
\label{0top}
\end{eqnarray}
Thus, {\em the existence of a gap in this system may not protect a
finite expectation value of the Toric code operators $X_{1,2}$ or
$Z_{1,2}$ for any finite temperature}. The expectation value of any
local quantity is independent of the toric cycle operators.

As we see here, by the use of our theorem and the results of \cite{BN},
we can suggest possible symmetry allowed invariants.  However, we
cannot prove that these must be finite at $T>0$. 
The Ising chain mapping  further enables us to compute all symmetry
invariant correlators which contain a finite number
of fields
\begin{eqnarray}
\langle \Big( \prod_{s \in S} A_{s} \Big) \Big( \prod_{p \in P} B_{p}  
\Big) \rangle &=& \langle A_{s} \rangle^{\|S\|} \langle B_{p}
\rangle^{\|P\|} \nonumber \\
&=& (\tanh \beta)^{\|S\| + \|P\|}.
\label{connected}
\end{eqnarray}
Here, $\|S\|$ is the number of star operators in the product of Eq.
(\ref{connected}), $\|P\|$ is the number of plaquette operators in the
product.  For an explicit proof of Eq. (\ref{connected}) see our
discussion in \cite{explain_factorization}. The factorization manifest
in  Eq. (\ref{connected}) illustrates that there are no connected
correlation functions at any finite temperature. Although it is
redundant for this simple case, in other systems with TQO, we may
similarly employ overlap parameters and  more complicated
multi-particle correlators in order to discern a transition.

\begin{figure}
\centerline{\includegraphics[width=0.85\columnwidth]{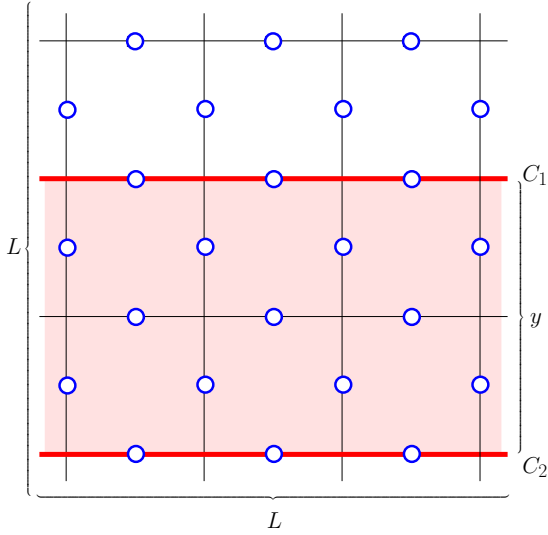}}
\caption{The two contours $C_{i=1,2}$ which define the  toric operators
$Z_{C_{i}}$ of Eq. (\ref{lane}). The two contours $C_{1,2}$ are
separated from each other by a distance $y$. The spins at the center of
each link $(ij)$ are $\sigma_{ij}$. The shaded area denotes the region
$R$ [see Eq. (\ref{c1c2})]. The system is on a torus of horizontal and
vertical cycles of size $L = \sqrt{N_{s}}$. On each link $ij$ there is
an $S=1/2$ field.}
\label{torus_ribbon}
\end{figure}
As the result of Eq. (\ref{0top})  is counter to some common lore, we
elaborate on it in a more general context and derive the same result
from related general considerations. For any finite size system, the
partition function is an  analytic function (as it is the sum of a
finite number of analytic functions). As the magnetizations $\langle
X_{i} \rangle$, $\langle Z_{i} \rangle$ are odd in the external fields
yet must be  analytic (and continuous) in the applied fields we
trivially have that for any  finite size torus at a finite inverse
temperature  $\beta$,  ~$ \lim_{h_{z,i} \to 0^{+}} \langle
Z_{i}(\vec{h}_{i},\beta) \rangle =  \lim_{h_{x,i} \to 0^{+}}\langle
X_{i}(\vec{h}_{i}, \beta) \rangle =0$. As always, if SSB happens, it
must occur in the thermodynamic  limit.  An alternate derivation of Eq.
(\ref{0top}) is as follows. Define on a system of size $\sqrt{N_{s}}
\times \sqrt{N_{s}}$ (with $N_{s} \to \infty$), the two {\it
topological} operators
\begin{eqnarray}
Z_{C_{i}} \equiv \prod_{\langle ij \rangle \in C_{i}} \sigma^{z}_{ij}.
\label{lane}
\end{eqnarray}
Here, we choose the two contours $C_{1,2}$ to be horizontal lines  that
encircle the torus (see Fig. \ref{torus_ribbon}). The two contours are
parallel to each other  and are displaced relative to each other by a
vertical  distance (number of bonds) $y$. Let us denote the ribbon
circumscribed between $C_{1}$ and $C_{2}$ by $R$ (including $C_1$ and
$C_2$). As for any bond $[\sigma_{ij}^{z}]^{2}=1$ we have that 
\begin{eqnarray}
Z_{C_{1}} Z_{C_{2}} = \prod_{p \in  R} B_{p}.
\label{c1c2}
\end{eqnarray}
Now, by our Ising mapping [see \cite{explain_factorization}],
\begin{eqnarray}
\langle \prod_{p \in R} B_{p} \rangle  = \prod_{p \in R} \langle B_{p}
\rangle = [\tanh \beta ]^{\|R\|}.
\label{Ribbon}
\end{eqnarray}
Here, $\|R\|$ denotes the number of plaquettes in $R$. Thus, 
\begin{eqnarray}
\langle Z_{C_{1}} Z_{C_{2}} \rangle = [\tanh \beta]^{y\sqrt{N_{s}}}.
\label{zz12}
\end{eqnarray}
Thus, \cite{explain_limit_period}
\begin{eqnarray}
\lim_{y \to \infty} \langle Z_{C_{1}} Z_{C_{2}} \rangle =0.
\label{yipp}
\end{eqnarray}

This, in turn, suggests that
\begin{eqnarray}
\langle Z_{C_{i}} \rangle =0,
\label{zeroagain}
\end{eqnarray}
which is consistent with the result of Eqs. (\ref{expectpart}), and
(\ref{0top}).  Even if these expectation values did not vanish (as they
do not for general finite $\vec{h}_{i}$), for any  finite $\beta$,
there is no ab initio reason for us to expect $\langle Z_{C_{h}}
\rangle$ to be the same for all contours $\{C_{h}\}$  that encircle one
of the toric cycles (e.g. the horizontal cycle). This is so as for
non-zero temperatures $A_{s}$ and $B_{p}$ are not identical to one. This
implies that, for instance, for two contours $C$ and $C'$ which differ
by the addition of a plaquette $p$, the expectation value  $\langle
Z_{C'} \rangle = \langle Z_{C} B_{p} \rangle$   which is generally
different from $\langle Z_{C} \rangle$.  Thus, at finite temperatures
and fields, not only the  {\it topology} of the contours $C_{i}$ is
important; their precise shape is also important.

Insofar as the operators $\{X_{i}\}$ and  $\{Z_{i}\}$ are concerned,
ergodicity remains unbroken at all finite temperatures.  Generally, we 
can discern transitions by ergodicity breaking at low temperatures
which means that infinitesimal external fields lift ergodicity (states
with positive magnetization have higher weight even for infinitesimal
external $\vec{h}_{i}$).  Phase space is not sampled at low
temperatures - ergodicity is broken and states with positive
magnetization (or negative ones) for infinitesimal positive (negative)
fields are of larger weight. For the case at hand, just as in the $D=1$
Ising model and free spins systems, ergodicity remains unbroken at all
finite temperatures. 

While, obviously, Eq. (\ref{0top}) holds  for all positive
temperatures, at $T=0$ different orders of limits can be taken  in Eq.
(\ref{partfinal}) \cite{order_limits_scroll}.  The only way to have
finite topological $\langle X_{i} \rangle$  or $\langle Z_{i} \rangle$
is to keep them finite by force (apply external perturbations
$h_{i}$).  In the Toric code, we proved that $\langle X_{i} \rangle$, 
$\langle Z_{i} \rangle$ can remain finite at $T>0$ only if they are
kept finite by hand by for example coupling to a non-local external
field. The simplest Toric code devised for pedagogical purposes works
well at $T=0$. It might have additional effects at $T>0$, no matter how
small $T$ is.  We reiterate our main result here: the gap is
inconsequential in the context of keeping $\langle X_i \rangle$ and
$\langle Z_i \rangle$ finite. That these expectation values differ from
their $T=0$ value relates to the non-analyticity of the free energy at
$T=0$. That is, it relates to the $T=0$ transition of the free spin
(Ising chain) system.  

The physical engine for the absence of order in the Ising chains (and
thus also in Kitaev's Toric code model)  are topological defects
(solitons) which proliferate  and propagate freely at any finite
temperature.  The explicit form of these dual operators can be written
to augment half of the  duality transformations written in 
Eq.~(\ref{abd}). The forms in Eq.~(\ref{abd}) appear as a duality  for
the spin for one component (say $S^{z}$) while the other spin component
$(S^{x})$ amounts to a string product (a soliton) on the dual lattice. 
In what follows, we will write the duality relations for the second
half of these dual relations (that for $S^{x}$)  on the two decoupled
Ising chains, e.g. \cite{NF}, \cite{FSU},
\begin{eqnarray}
\overline{a}_{s} &\equiv& \prod_{i \le s} \sigma^{x}_{i} \nonumber \\  
\overline{b}_{p} &\equiv& \prod_{i \le p}  \sigma^{x}_{i}.
\label{abd1}
\end{eqnarray} 
Here, $[\overline{a}_{s}, \overline{a}_{s}'] = [\overline{b}_{p},
\overline{b}_{p}'] = 0$ and  $\{a_{s}, \overline{a}_{s}\} = \{b_{p},
\overline{b}_{p}\}=0$. The duality enables an identification of the
$d=1$ symmetries and makes non-local terms appear local and vice versa.
To make contact with our earlier  discussion of Section \ref{Section 2}
regarding the unimportance of local events on the boundary in
triggering the stabilization of TQO, we remark that  the $d=1$
operators of Eq.~(\ref{Td}) are none other than  the dual operators in
Eq.~(\ref{abd1}).

\subsubsection{Adiabaticity}

 Our finite temperature solution allows us  to demonstrate {\em
adiabaticity} of the Kitaev model. That is, exact GLSs can be lifted
while not changing  the phase of the system. Let us re-examine the
consequences of our solution of Eq. (\ref{pfunction1}). The addition
of, e.g., $h_{x,1} \neq 0$ to the Hamiltonian $H_{K}$ does not lead to
any transition within  the system. Both the Kitaev model and its
counterpart with any additional $h_{x,1} \neq 0$ lead to free energies
that are everywhere analytic apart from $T=0$. Nevertheless, insofar as
GLSs are concerned augmenting $H_{K}$ by an additional $h_{x,1} \neq 0$
lifts the $d=1$ symmetry $Z_{1}$. Thus, the Kitaev model exhibits
finite temperature adiabaticity. 

A similar phenomenon occurs in the adiabatic evolution from a pure Ising
gauge theory [Eq. (\ref{Z2t}) which is equivalent to $H_{\Z_{2}}$ of
Eq. (\ref{gz2+}) with $h=0$]] to another system. Consider the following
change in Hamiltonians
\begin{eqnarray}
\!\!\!\!\!\!\!\!\!\! H_{\Z_{2}}(h_{x}=h_{z}=0) &\to& H_{\Z_{2}}(h_{x}=0, h_{z}) 
\nonumber \\  
&=& H_{\Z_{2}}(h_{x}=0) - h_{z} \sum_{ij} \sigma^{z}_{ij}.
\label{HHH}
\end{eqnarray}
The $\Z_{2}$ gauge theory of Eq. (\ref{Z2t}) maps onto an Ising chain.
By contrast \cite{percolation1}, $ H_{\Z_{2}}(h_{x}=0, h_{z})$ maps
onto a $D=2$ Ising model in a finite magnetic field. Thus,  the systems
described by $H_{\Z_{2}}(h_{x}=h_{z}=0)$ as well as
$H_{\Z_{2}}(h_{x}=0, h_{z} \neq 0)$ exhibit no finite temperature
singularities. Thus,  $H_{\Z_{2}}(h_{x}=h_{z}=0)$ and
$H_{\Z_{2}}(h_{x}=0, h_{z} \neq 0)$ describe systems in the same phase.
However, $H_{\Z_{2}}(h_{x}=h_{z}=0)$ exhibits the local ($d=0$)
symmetries exemplified by  $\{A_{s}\}$ of Eq. (\ref{AB_defn}), as well
as the $d=1$ symmetries $\{X_{1},X_{2}, Z_{1},Z_{2}\}$ of Eq.
(\ref{kit}). On the other hand, $ H_{\Z_{2}}(h_{x}=0, h_{z} \neq 0)$
only has the $d=1$ symmetries of $\{Z_{1},Z_{2}\}$.  In other words, by
setting $h_{z} \neq 0$, the $d=1$ operators $\{X_{1},X_{2}\}$  are no
longer symmetries (and furthermore all local symmetries have been
lifted). Nevertheless, even if a finite $h_{z}$ lifts $d=1$ GLSs, no
transition occurs.  Similar to the Kitaev model, in line with the
lifting of symmetries, the application of a finite longitudinal field
in a gauge theory [$h_{z}$  in Eq. (\ref{HHH})] lifts the GS degeneracy
and leads to a single GS. Nevertheless, at any finite temperature, the
systems  are adiabatically linked to each other.   This is analogous to
the adiabatic continuity seen in matter coupled gauge theories
\cite{FradkinShenker}  as implied by effective magnetic field mappings
\cite{percolation1}.

\subsubsection{Ising mappings of Kitaev models on general manifolds}
\label{mani}

Here, we extend our Ising chain mappings to Kitaev models  on general
manifolds on closed oriented surfaces with genus or number of handles
$g$. Kitaev's Toric code model on the torus (which has been the focus
of our discussion so far) corresponds to the case of $g=1$. 

To investigate what occurs on general manifolds, we apply the
Euler-Lhuillier formula which  states that the genus number $g$ [or
number of handles] satisfies
\begin{eqnarray}
V - E + F = 2(1-g).
\label{Euler}
\end{eqnarray}
Here, $V$ is the number of vertices of the system, $E$ the number of its
edges, and $F$ the number of its faces. In the Kitaev model, $V$
denotes the number of sites of the system and as such counts the number
of fields $\{A_{s}\}$, $E$ is the number of fields which live on the
nearest-neighbor links $(ij)$, and $F$ is the number of plaquettes (or
number of terms $\{B_{p}\}$).  For manifolds of genus number $g>1$, 
the mapping to Ising chains [Eqs. (\ref{Isingmp}),  (\ref{abDUAL}),
and  (\ref{abdual})] will proceed as before. The sole difference is
that the number of edges (the number of fields $\{\sigma_{ij}\}$) is
larger than it was before. Here,  $N_{E} = N_{a} + N_{b} + 2(g-1)$,
with $N_{a,b}$ the number of Ising spins on the two periodic chains
[$N_{a}= V= N_{s}$, $N_{b} = F]$, and $N_{E}$ is the number of fields
$\{\sigma_{ij}\}$. We find that 
\begin{eqnarray}
Z = 2^{2(g-1)} Z_{g=1}.
\label{Kitaevgenus}
\end{eqnarray}
In Eq. (\ref{Kitaevgenus}), $F=V=N_{s}$ and $Z_{g=1}$ the Toric code
partition function which we found earlier [Eq. (\ref{partfinal}) for
$h_{i}=0$]. [For general tessellation of a torus with $F$  plaquettes
and $V$ vertices, we will have $Z_{g=1} = ((2 \cosh \beta)^{V} + (2
\sinh \beta)^{V})  ((2 \cosh \beta)^{F} + (2 \sinh \beta)^{F})$.] Eq.
(\ref{Kitaevgenus}) is the Laplace transform of the density of states
$g(E)$. It implies that  each energy eigenstates of the Kitaev model on
a manifold of genus $g$ is precisely $2^{2(g-1)}$ higher than that of
two decoupled periodic Ising chains.  In particular, the GS degeneracy
is indeed \cite{pachos}
\begin{eqnarray}
2^{2(g-1)} \times 2^{2} = 2^{2g}.
\label{degk}
\end{eqnarray} 
The precise exponential degeneracy of each level [the factor
$2^{2(g-1)}$] holds for each level by comparison to its  multiplicity
on the torus. 

\subsection{Dimensional Reduction and Thermal Fragility in Wen's
 plaquette model}
\label{wentemp}

Although appearing as different models,  Wen's plaquette model of
Eq.~(\ref{hw}) and Kitaev's Toric code model of Eqs.
(\ref{kitaevmodel}) and (\ref{AB_defn})  are one and the same.  The
equivalence between these two systems is demonstrated by a rotation
which  we will explicitly spell out later in this Section. As it must,
according to this equivalence,  Wen's model \cite{wen_plaq} of
Eq.~(\ref{hw}) also exhibits a trivial dimensional reduction just as
Kitaev's Toric code model does. Here, we have that with ${\cal{A}}_{p}
= \sigma^{x}_{i} \sigma^{y}_{j} \sigma^{x}_{k} \sigma^{y}_{l}$ the
Hamiltonian of Eq.~(\ref{hw}),
\begin{eqnarray}
H = -K \sum_{p} {\cal{A}}_{p}, ~~~[{\cal{A}}_{p}, {\cal{A}}_{p'}] =0, 
~~~ {\cal{A}}_{p}^{2}=1.
\end{eqnarray}
When periodic boundary conditions are applied (the  system is on a
torus), the number of plaquettes is equal to the number of vertices. 
As in the Kitaev Toric code model, the partition function of these
decoupled plaquette Ising fields is, for an even by even lattice,
given  by the partition function to two decoupled Ising chains with 
periodic boundary conditions \cite{handong}. In the absence of an
external field that couples to the $d=1$ symmetries of this system, we
have
\begin{eqnarray}
\!\!\!\! \!\!\!Z_{\sf even-even} = [(2 \cosh \beta K)^{N_{s}/2} + (2
\sinh \beta K)^{N_{s}/2}]^{2} .
\end{eqnarray}
For an odd by even lattice, the partition function is that of a single
periodic Ising chain
\begin{eqnarray}
Z_{\sf even-odd}= [(2 \cosh \beta K)^{N_{s}} + (2 \sinh \beta K)^{N_s}].
\end{eqnarray}
The different decoupling for the even by even or even by odd cases
follows from their constraints \cite{wen_plaq}. In both cases, we have
that the product 
\begin{eqnarray}
\prod_{p}  {\cal{A}}_{p} =1
\end{eqnarray} 
with the product over plaquettes $p$ of the lattice. In the even by
even lattice, we have the   additional constraint
\begin{eqnarray}
\prod_{ {\sf even} ~p}  {\cal{A}}_{p} =1.
\end{eqnarray}
``{\sf Even} $p$'' refers here to dual lattice sites (plaquettes) in
which the sum of the $x$ and $y$  coordinates is even. Here, the 
correspondence with Eqs. (\ref{abd}), (\ref{abDUAL}), and
(\ref{abdual}) can be made via 
\begin{eqnarray}
{\cal{A}}_{p'=2s} &\leftrightarrow& a_{s}, \nonumber \\
{\cal{A}}_{p'=2p+1} &\leftrightarrow& b_{p}.
\end{eqnarray}

Similarly, as in Eq.~(\ref{projkit}),  any GS can be expressed as a
projection from a reference state:
\begin{eqnarray}
{\cal{N}} \prod_{p} [1+ {\cal{A}}_{p}] | \phi \rangle
\end{eqnarray}
with $| \phi \rangle$ a reference state.  We will soon rewrite the GS
in  an explicit form similar to that of Eq.~(\ref{soln2Kit}).

\begin{figure}
\centerline{\includegraphics[width=0.85\columnwidth]{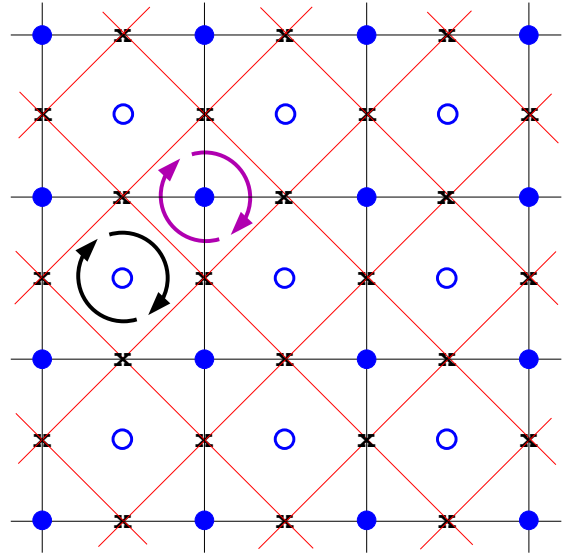}}
\caption{Transformations relating Kitaev's Toric code model of Eqs.
(\ref{kitaevmodel}), and (\ref{AB_defn})  to Wen's plaquette model of
Eq.~(\ref{hw}). The two consecutive  four spin terms correspond to a
plaquette and star terms ($B_{p}$  and $A_{s}$) in Kitaev's Toric code 
model of Eqs. (\ref{kitaevmodel}), and  (\ref{AB_defn}) . In the
rotated lattice formed by the diagonals, Kitaev's Toric code model
becomes Wen's plaquette model of Eq.~(\ref{hw}).  (This occurs upon
applying the spin rotation of Eq.~(\ref{reflWen})  on one sublattice of
the spins.)}
\label{Kitaevdual}
\end{figure}

An explicit correspondence between the Kitaev Toric code model of
Eqs.~(\ref{kitaevmodel}), and (\ref{AB_defn}) and Wen's model of
Eq.~(\ref{hw}) is seen by simple rotations. This is illustrated in
Fig.~\ref{Kitaevdual}.
To this end, we start with Kitaev's model of Eqs.~(\ref{kitaevmodel}),
and  (\ref{AB_defn}). We may draw the diagonal lines as shown  in Fig.
\ref{Kitaevdual}. The spins which formerly lied on the bond of the
lattice are now on the vertices of  the rotated lattice. We may then
identify every other center of the squares formed by these diagonals as
a plaquette center (denoted in Fig. \ref{Kitaevdual} by a hollow
circle) and the centers of the remaining squares by a star center
(denoted by a full circle). Next, we rotate all of the spins (labeled
in Fig. \ref{Kitaevdual}  by the intersection points of the
diagonals) on one sub-lattice by 180 degrees about the symmetric line
in the $x$-$z$ plane in the internal spin space, by
\begin{eqnarray}
\hat{O}_{Wen} = \prod_{i \in {\sf ~even~sublattice}}  \exp[i
\frac{\pi \sqrt{2}}{4} (\sigma_{i}^{x} + \sigma_{i}^{z})].
\label{reflWen}
\end{eqnarray}
In the rotated system formed by the diagonals, what we obtain is Wen's
model of Eq.~(\ref{hw}) [with a relabeling of the spin polarizations].
Due to their exact equivalence, all of the conclusions stated for the
Kitaev model in Section \ref{Tfragile}  will hold for Wen's model as
well. In particular, the GS for a system (of $K>0$ in Eq. (\ref{hw})) 
with periodic boundary
conditions along the diagonals is given by Eq.~(\ref{soln2Kit}) after
undoing the various rotations (in both spin and  real space), 
\begin{eqnarray}
|\psi_{q'} \rangle = 2^{1-N_{s}/2} \sum_{c' \in q'} | c' \rangle.
\label{soln2Wen}
\end{eqnarray}
Here, $q' = (\pm 1, \pm 1)$ denotes the transformed topological
sectors. That is, replacing the topological numbers $\pm 1$ for the
eigenvalues of the $d=1$ symmetries $Z_{i=1,2}$ in Kitaev's model
(Eq.~(\ref{kit})) which now refer to the direction relative to the
diagonals in Fig. \ref{Kitaevdual}, we have 
\begin{eqnarray}
{\cal{Z}}_{i} =  \hat{O}_{Wen}^{\dagger} Z_{i} \hat{O}_{Wen}^{\;}.
\end{eqnarray}
Similarly, the states $|c'\rangle$ denote all states in which
$\sigma^{x}_{a} = \pm 1$  for all sites $a$ of the even sublattice  and
$\sigma^{z}_{b} = \pm 1$ for all sites $b$ of the odd sublattice
for which the product of these quantities around any plaquette is one.
Following the same considerations as for Kitaev's Toric code model
which we outlined in the previous Section, we see that the $d=1$
symmetry operators attain a vanishing expectation value at any finite
temperature unless a finite external  field is applied.


\subsection{Finite-temperature transitions in related
higher-dimensional analogues}

As we proved above, Kitaev's Toric code model as well as Wen's related
plaquette model do not exhibit a finite-$T$ phase transition by virtue
of their mapping onto a system of Ising chains which exhibit  a $T=0$
transition. It is obviously the case that other systems with TQO can
exhibit finite-$T$ transitions. For instance, the $D=3$  Ising gauge
theory exhibits rank-$n=8$ TQO yet, as can be seen by the fact that it
is dual to a $D=3$ Ising model, it exhibits a singularity in the free
energy at a finite critical temperature ($T_{c}$). A related system is
a $D=2$ extension of the Kitaev Toric code model of Eqs.
(\ref{kitaevmodel}), and (\ref{AB_defn}) \cite{bombin}. Here, the
product in the star operator $A_{s}$ of Eq. (\ref{AB_defn}) spans  all
six sites which, on the cubic lattice, are the nearest-neighbors of a
given lattice site.  Similarly, in the product defining the plaquette
operator $B_{p}$ of Eq. (\ref{AB_defn}), we have (as in the $D=2$ 
case), the product of $\sigma^{z}$ on all four bonds which form a
planar  plaquette (parallel to either the $xy$, $xz$, or $yz$ planes). 
It is easy to prove that the Hamiltonian that results is  none other
than that of the $D=3$  Ising gauge theory (given by $- \sum_{p}
B_{p}$) augmented by the sum of all local symmetry generators ($-
\sum_{s} A_{s}$).  It is Eq. (\ref{tH}) with, in the notation of
Section \ref{engineer}, $H$ being the $D=3$ Ising gauge theory
Hamiltonian [i.e. Eq. (\ref{Z2t}) with the sum performed over all
plaquettes that appear in the cubic lattice] and now with $\{G_{i}\}$
the set of all local symmetries ($\{A_{s}\}$) 
\begin{eqnarray}
H = H_{3D~Ising~gauge} + H_{vertex} \nonumber \\ 
H_{3D~Ising~gauge} = -\sum_{p} B_{p} , ~ H_{vertex} = - \sum_{s} A_{s}.
\label{3dK}
\end{eqnarray}
Unlike the $D=2$ case, in three dimensions,  the plaquette terms
$\{B_{p}\}$ which appear in Eq. (\ref{3dK}) satisfy too many
constraints that replace the single constraint of Eq. (\ref{ABpc}). For
instance, if we consider a single cube on the lattice which is of unit
length and label its six faces by {\bf Cube}$_{p=1,2,\cdots,6}$ then we
have that $\prod_{p \in \mbox{\bf{Cube}}} B_{p} =1$.  Precisely all of
these constraints (and none additional)  appear in the $D=3$ Ising
gauge theory. All terms in Eq. (\ref{3dK})  commute and as no
constraints couple the vertex  fields to the plaquette fields, the
partition function is the product of the decoupled plaquette and vertex
fields, 
\begin{eqnarray}
Z = Z_{3D ~Ising~gauge}~ \times~  Z_{1D~Ising~chain}.
\label{z3d}
\end{eqnarray}
The  first term in Eq. (\ref{z3d}) arises from the plaquette terms
whereas the second term is that from the vertex ({\it star}) fields
$\{A_{s}\}$ which, as in the $D=2$ case, adhere to the single
constraint they satisfied in Eq. (\ref{ABpc}). The counting of the
degrees of freedom proceeds as follows: $Z_{3D~Ising~gauge}$ is the
trace of $\exp[-\beta H_{3D~Ising~gauge}]$ over all gauge independent
degrees of freedom whereas $Z_{1D~Ising~chain}$ is the trace of
$\exp[-\beta H_{vertex}]$ over all of the gauge degrees of freedom
$\{G_{i}\}$ (which are here exemplified by the star terms $\{A_{s}\}$).
In the thermodynamic limit, for the system with both periodic or open 
boundary conditions, the free energy per site is given by
\begin{eqnarray}
F = F_{1}(\beta) + F_{3}(\beta^*).
\label{Ffull}
\end{eqnarray}
$F_{1}$ is the free energy per site of an Ising chain; this
contribution is borne by the star fields $\{A_{s}\}$. Similarly,
$F_{3}$ is the free energy  per site of a $D=3$ Ising model at the dual
$\beta^{*}$ \cite{wegner}: 
\begin{eqnarray}
\sinh 2 \beta \sinh 2 \beta^{*} = 1.
\end{eqnarray}
The contribution of $F_{3}$ originates from the  free energy of the
Ising gauge theory ($\{B_{p}\}$)  which is exactly dual to the $D=3$
Ising model. The free energy of Eq. (\ref{Ffull})  exhibits a $T=0$
($\beta_{c1} = \infty$) singularity  (borne by the $T=0$ singularity of
the $D=1$ Ising contribution $F_{1}$) as well as a  finite-$T$
singularity,
\begin{eqnarray}
\beta_{c2} \simeq 0.761423,
\end{eqnarray}
which originates from the {\it $D=3$ Ising} contribution $F_{3}$. 

As in the $D=2$ case, no toric operators can attain a finite
expectation value at any finite temperature, e.g.
\begin{eqnarray}
\langle Z_{P_{\alpha}} \rangle = 
\langle \prod_{ij \in P_{\alpha}} \sigma^{z}_{ij} \rangle =0,
\end{eqnarray}
with $P_\alpha$ an entire {\em plane}.  The steps of  Eqs.
(\ref{lane}), (\ref{c1c2}),  (\ref{yipp}), and (\ref{zeroagain}) may be
reproduced with the former linear contours $\{C_{1,2}\}$  replaced by
planes $P_\alpha$. Due to the decoupling of the plaquette $(\{B_{p}\}$)
and vertex ($\{A_{s}\}$)  operators, the expectation value on the
righthand side of Eq. (\ref{c1c2}), with the contours $\{C_{1,2}\}$ 
replaced by two planes $\{P_{1,2}\}$,  is given by its value for a
classical $D=3$ Ising gauge theory. As seen
by high and low coupling expansions \cite{kogut} 
now applied to {\em surface} operators, 
the correlator
$\langle Z_{P_{1}} Z_{P_{2}} \rangle = \langle \Big( \prod_{ij \in P_{1}} \sigma^{z}_{ij} \Big)
\Big( \prod_{ij \in P_{2}} \sigma^{z}_{ij} \Big) \rangle$ 
vanishes in the
$D=3$ Ising gauge theory as the bounding planes $P_{1,2}$ 
are taken to be infinite. In 
both (i) the {\it confined} phase  [$\beta
< \beta_{c2}$], and in (ii)  the {\it deconfined} phase [$\beta> \beta_{c2}$],
the pair correlator 
$\langle Z_{P_{1}} Z_{P_{2}} \rangle$ vanishes exponentially in the total 
area of the planes $P_{1}$ and $P_{2}$. 

For general dimension $D$, the partition function associated with a
simple extension of Eq. (\ref{3dK}) is given by 
\begin{eqnarray}
Z = Z_{D-dimensional~Ising~gauge} \times  Z_{1D~Ising~chain}.
\end{eqnarray}
In the two dimensional case, which we discussed
earlier,
the first term- the partition
function of the two dimensional Ising gauge theory-
is nothing but the partition function
of an Ising chain. 

\section{Shape and state Dependent Entanglement Entropy: results
and a conjecture}
\label{eent}

With $\rho_{\Gamma}$ the marginal density matrix obtained by  tracing
out all degrees of freedom exterior to a  boundary surface $\Gamma$ (of
size $R$), the entanglement entropy between the 
two regions which are separated by $\Gamma$ is defined as 
\begin{eqnarray}
S_{{\sf ent}} = -\tr [\rho_{\Gamma}
\ln \rho_{\Gamma}].
\end{eqnarray} 
Recent work computed the entanglement entropy for a variety of systems
of pertinence, e.g., AKLT type integer spin chains \cite{hosho}. The
quantity $S_{{\sf ent}}$ measures the entanglement between fields in
the interior of $\Gamma$ to those lying  outside. It was recently 
conjectured \cite{entropy} that a non-local borne deviation $\xi>0$
from an asymptotic  area law scaling,  
\begin{eqnarray}
S_{{\sf ent}} \sim (\eta R-\xi)
\label{sscaling}
\end{eqnarray}
in $D=2$, constitutes a precise measure of TQO.  In \cite{NO} we
conjectured that in some systems, this criteria is either incorrect or,
at best, may require an additional explicit average and/or limit, of a
form yet to be prescribed, over all allowed contours as the size of
the  contour tends to infinity. The central point  of \cite{NO} was
that the entanglement entropy may depend on the shape of $\Gamma$  [or,
equivalently, that for a given shape we may find GSs for which the bare
entanglement entropy criteria  is inconsistent with the definition of
TQO given by Eq.~(\ref{def.})]. These  deviations from an area law are
compounded by other deviations in other arenas (e.g. those for tight 
binding fermions \cite{explainwolf}) which are unrelated to any
appearance of TQO. We now elaborate on this. 

In Klein spin models which host TQO  (by satisfying Eq. (\ref{def.})) -
e.g. arbitrary rank $T>0$ TQO of the Klein model on the pyrochlore
lattice \cite{NBNT} the entanglement entropy criterion is not
satisfied.  In a pure dimer state- a state which is  a product of
individual singlet pairs, all GS correlations are manifest as local
singlets and, at $T=0$, the entanglement entropy  $S_{{\sf ent}}$ is,
up to constants,  the number of dimers partitioned by $\Gamma$:
\begin{eqnarray}
S_{{\sf ent}}^{pure~ dimer} = N_{\partial \Gamma} \ln 2.
\end{eqnarray}
Here,  $N_{\partial \Gamma}$ is the number of dimers which are
partitioned by $\Gamma$:  the number of singlet dimers in which one
spin forming the singlet lies inside $\Gamma$ and the other lies
outside $\Gamma$.  A random surface $\Gamma$  intersects
${\cal{O}}(\|\Gamma\|)$ dimers, with $\|\Gamma\|$ the number of dual lattice 
sites which lie on $\Gamma$.   However, by setting the contour $\Gamma$ 
along the 
dimers present in any given GS, we find arbitrarily large surfaces such
that  no singlet correlations are disrupted and $S_{{\sf ent}}=0$.  We
similarly find many GSs/contours with arbitrary $\xi$ for fixed 
contours $\Gamma$. The arbitrariness of the entanglement entropy and of
the deviation $\xi$  in systems with arbitrary rank - TQO suggests that
the conjecture raised by \cite{entropy}, although very insightful and
correct in some cases,  may be inconclusive. Similar conclusions
also appear for the more sophisticated subtractions 
[termed a {\em topological} entropy] which were suggested
by \cite{entropy} for the extraction of $\xi$. 

In what follows, we elaborate on this conjecture. We start with  (i) a
rigorous demonstration that non-TQO states or states that, at best, can
capture arbitrary rank TQO  (yet do not satisfy Eq.~(\ref{def.}) for
all GSs) in which the entanglement entropy can assume any value between
zero and a quantity which scales linearly with the area $\|\Gamma\|$,
(ii) Show that also the topological entropy can similarly assume
any value (and in particular) be finite in systems which 
do not host TQO, conjecture on (iii) the existence of specific TQO
states in the Klein model whose TQO is manifest not necessarily in
entanglement entropy but rather in topological invariants (line number)
that classify sectors of the GS subspace, and finally construct
states [not necessarily only GSs of the Klein 
Hamiltonian] which have TQO yet display vanishing
entanglement entropy.

\subsection{Pure singlet product states}
\label{pureKlein}

Pure singlet product states $|c \rangle$ - states which are a product
of disjoint singlets -  do not support TQO (i.e. there the local
operators - $[\vec{S}_{i} \cdot \vec{S}_{j}]$ on nearest-neighbor sites
$i j$ which  distinguish the states).

\subsubsection{Short-ranged $SU(2)$ Klein models on a lattice
geometry}

\begin{figure}
\centerline{\includegraphics[width=0.96\columnwidth]{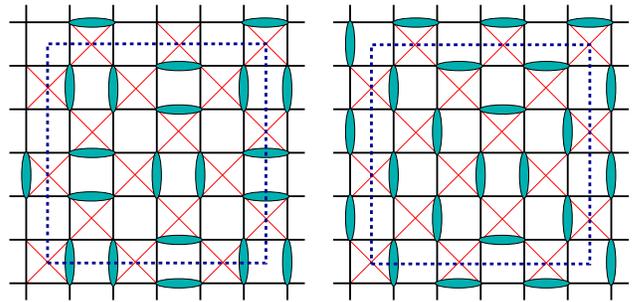}}
\caption{Two singlet product states which are GSs of the Klein model
Hamiltonian of  Eq. (\ref{KleinHamN}) on a checkerboard lattice.  The
ovals denote singlet states formed by nearest-neighbor spins. The
dashed line denotes a contour $\Gamma$. The entanglement entropy
between the two regions separated by $\Gamma$ is different for the two
above states: it is $(5 \ln 2)$ for the singlet product state on the
left while it is $\ln 2$ for the state on the right. Both of these
singlet product states can be differentiated by local 
observables and thus do not satisfy the TQO conditions 
of Eq. (\ref{def.}).}
\label{kleinloop}
\end{figure}

Spin $S=1/2$ [$SU(2)$] Klein models were briefly reviewed in 
Section \ref{section12}. Here, dimer states can  support arbitrary rank TQO 
yet do not support Eq.~(\ref{def.}) for  all GSs of the Klein model.
Here, various  contour shapes $\Gamma$ can be tailored to lead to any
value of the entanglement entropy (from zero up to an upper bound which
scales with the area $\|\Gamma\|$).  In a dimer system, nearly any
value of the entanglement entropy bounded from above by $\|\Gamma\|$
times a constant is possible (see Fig.~\ref{kleinloop}). By setting
$\Gamma$ to lie along dimers, the entanglement  entropy strictly
vanishes. Similarly, in the {\it ferroelectric} GS (see \cite{NBNT}),
if we choose the boundary $\Gamma$ (which circumscribes an inner domain
including all sites on $\Gamma$ itself) then the number of partitioned
dimers for a boundary $\Gamma$ which is of a square form and lying
along the non-diagonal axes of the checkerboard lattice) is exactly
half the perimeter. In fact, for any given dimer state, we can find a
multitude of contours $\Gamma$ of arbitrary size such that 
\begin{eqnarray}
0 \le S_{{\sf ent}}^{pure~ dimer} (\Gamma) \le \|\Gamma\| \ln 2.
\label{eebound}
\end{eqnarray}
[For an elementary derivation of the factor of $\ln 2$,
see Section \ref{app4.5}.]
In particular,  it is not a problem to construct dimer states for a
given contour $\Gamma$ - or equivalently  construct contours $\Gamma$
for a given dimer  state - in which, any desired function of the 
linear scale of $\Gamma$ is possible so long as it satisfies
Eq.~(\ref{eebound}). To conclude, we illustrated that systems exist 
where the entanglement entropy can assume any value between
zero and a quantity which scales linearly with the size 
of the contour $\|\Gamma\|$. In the singlet product states under
consideration, Eq. (\ref{def.}) does not hold and 
the system does not exhibit TQO.
This arbitrary deviation from an area law scaling  does not conform
with either a pure area law scaling anticipated for non-TQO systems nor
to a unique {\it topological} deviation $\xi$.

It should be noted that here the topological entropy $S_{topo}$ as
defined by \cite{entropy} is
\begin{eqnarray}
S_{topo} = \Big[ S_{A} + S_{B} + S_{C} - S_{AB} - S_{AC}  - S_{BC}
\nonumber \\   
+S_{ABC} \Big] 
\label{st}
\end{eqnarray}
for a partition of a large contour $\Gamma$ which spans a  region [ABC]
into three separate regions A, B, and C gives a result which is
identically zero: $S_{topo} =0$. The motivation for introducing the
$S_{topo}$ of Eq. (\ref{st}) in \cite{entropy} was that in many cases,
$S_{topo} = \xi$ of Eq. (\ref{sscaling}).  

\subsubsection{$SU$($N \ge 4$) Klein models on a small world network}

The basic idea underlying the Klein models can be extended to
$SU$($N>2$) systems. This can be done for any lattice/graph in which we
can choose $N$ sites  within each fundamental unit (each ``plaquette'')
to belong to an $SU(N)$ singlet. Obviously, for $N \ge 4$, the 
basic ``plaquette'' spans more 4 sites. 
The covering of the lattice/graph with
these singlets  is that with hard core units (singlets) each of which
spans $N$ sites.   These Klein models are what we will now introduce
for the first time. Here, we will have
\begin{eqnarray}
H = J \sum_{\alpha} P_{\alpha}^{S_{tot} = max}
\end{eqnarray}
with $J>0$, and $P_{\alpha}^{S_{tot}=max}$ the projection operator onto
$SU(N)$ spin states on a given plaquette $\alpha$ which belong to the
highest irreducible representation of $SU(N)$. Now, let us consider the
Klein model on a small worlds network. In a small worlds network 
\cite{its_a_small_world}, most links connect sites which are
geometrically close to each other yet once in a while there is a link
which connects very far separated sites. Such a {\it small worlds}
geometry allows for the existence of singlet states composed of far
separated sites. 

\begin{figure}
\centerline{\includegraphics[width=0.87\columnwidth]{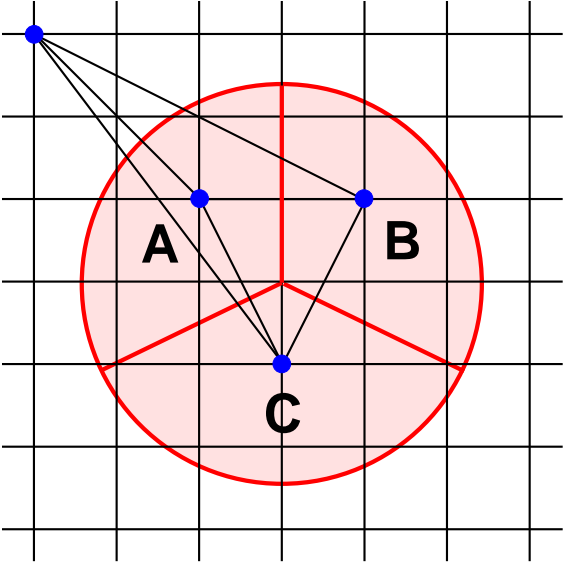}}
\caption{Shown above is a singlet state formed by distant sites in the
regions A, B, C, and the region exterior to these. The lines delineate
the members of this 4 site clique. An $SU(4)$ Klein
model Hamiltonian on a small worlds network will include  a vast number
of short-range singlets yet also allow for singlet states between remote
spins.  The {\em topological} entanglement entropy  of Eqs. (\ref{st}),
and  (\ref{sdetail}) which is associated with the singlet state shown
above is $S_{topo} = \ln  \Big[\frac{32}{27} \Big]$.}
\label{smallworld}
\end{figure}

If we consider a {\it long-ranged} $SU(4)$ singlet (see Fig.
\ref{smallworld}) which is composed of a single site in each of the 4
regions A, B, C, and the region exterior to ABC, then Eq.~(\ref{st})
reads
\begin{eqnarray}
S_{topo}^{long~singlet} = 4 S^{ent}_{1} - 3 S^{ent}_{2}.
\label{sdetail}
\end{eqnarray}
Here, $S^{ent}_{1}$ is the entanglement entropy between one site and
the remaining three sites of an $SU(4)$ singlet. Similarly,
$S^{ent}_{2}$ denotes the entanglement entropy between the two pairs of
sites  which constitute an $SU(4)$ singlet. Here, we find that  for an
$SU(4)$ Klein model on a small-worlds network [where the plaquettes can
contain sites which are arbitrarily close or far apart] the
entanglement entropy is equal to 
\begin{eqnarray}
S_{topo} = N_{long} [4S^{ent}_{1}- 3S^{ent}_{2}].
\label{stop}
\end{eqnarray}
Here, $N_{long}$ denotes the number of singlets in which each of the 
four sites lies in a different region. We have that 
\begin{eqnarray}
S_{1}^{ent} &=& \ln 4, \nonumber \\ 
S_{2}^{ent} &=& \ln 6.
\end{eqnarray}
Inserting this in Eq.~(\ref{stop}) leads to
\begin{eqnarray}
S_{topo} = N_{long} \ln \Big[ \frac{32}{27} \Big].
\label{stopo}
\end{eqnarray}
For general $SU(N)$ systems ($k=1,\cdots$), 
\begin{eqnarray}
S_{k}^{ent} =\ln \left( \begin{array}{c}
N \\ 
k
\end{array} \right). 
\label{SUNK}
\end{eqnarray}
A detailed derivation of these combinatorial forms is  given in
Appendix \ref{app4.5}.  The entanglement entropies associated  with the
completely antisymmetric  $SU(N)$ singlet state are equivalent to those
of an $N$ fermion system in which no interactions are present.

Returning to Eq.~(\ref{stopo}) we see that in general, for $N_{long}
\neq 0$  in an $SU(4)$ Klein model on a  small worlds network we have 
\begin{eqnarray}
S_{topo} \neq 0.
\end{eqnarray}
Nevertheless, whenever singlets are formed between nearest-neighbor
quartets, we have a local operator $[\sum_{i=1}^{4} \vec{S}_{i}]^{2}
=0$.   When no singlet is formed by these 4 points,  $\langle
[\sum_{i=1}^{4} \vec{S}_{i}]^{2} \rangle > 0$. We thus, in general, 
have a local operator ($[\sum_{i=1}^{4} \vec{S}_{i}]^{2}$) which
attains a different expectation value in different GSs - the system
does not possess strict TQO. Nevertheless, here, in general, $S_{topo}
\neq 0$. Furthermore, here, as before, $S_{topo}$ is quite arbitrary. 

In principle, for any scheme for $S_{topo}$ similar to Eq.~(\ref{st})
which requires partitions into ($R$-1) regions, an $SU$($N \ge R)$ model 
on a small worlds network can  generally give rise to an arbitrary
$S_{topo}$. In this system, general local order parameters can
differentiate different {\it pure} [non-superposed] singlet product
states. 

\subsection{TQO ground states in the nearest-neighbor $S=1/2$ Klein model
on the checkerboard lattice}

We can construct a set of states  which satisfy Eq.~(\ref{def.}) in
which we conjecture that a direct topological, rather 
than entanglement entropy, marker might be
the most transparent.  As stated earlier, in  individual dimer states,
local observables can differentiate different GSs. For instance, if a
dimer connects the  two sites $i$ and $j$ then $\langle \vec{S}_{i}
\cdot \vec{S}_{j} \rangle = -3/4$. As such local measurements attain
different values in different GSs, individual dimer states do not
exhibit TQO at $T=0$. Linear combinations of intra-plaquette dimer
states \cite{NBnew} exhaust all of the GSs of the Klein  model on any
lattice and may enable the construction of TQO states. Let us focus on
the Klein model of Eq.~(\ref{KleinHamN}) on a checkerboard lattice
of size $L \times L$.
This Hamiltonian has GSs given by Eq.~(\ref{GSK}). Now, let us focus on
a subset of these GSs. To this end, let us draw a square contour
$\Gamma$ of edge whose length is $R$ and consider the states 
\begin{eqnarray}
| \psi_{l} \rangle = {\cal{N}}  \!\!\!\!\!\!\!
\sum_{\mbox{$|c \rangle$ has $l$ lines, no dimers
partitioned by $\Gamma$}} \!\!\!\!\!\!\!\!\!\!\!\!\!\!|c \rangle. 
\label{symmmmmstate}
\end{eqnarray}    
Here, $| \psi_{l} \rangle$ is defined  for each individual number $l$
of lines in  the line representation of the dimer coverings (see {\bf
d} of  Section \ref{section12} (Figs.~\ref{fig_rules}, and
\ref{fig_local} in particular)). We may focus on the physical  regime
defined by $(L/2 + R)>l > (L/2- R)$  and $L \gg R$.   In
Eq.~(\ref{symmmmmstate}), the  states $|\psi_{l} \rangle$ are sums over
all dimer states $|c \rangle$ which belong to a very specific chosen
sector.  The multiplicative constant ${\cal{N}}$ is a normalization
factor. Here, the last constraint amounts to the requirement that each
dimer configuration summed over exhibits a dimer covering GS within the
square contour of size $R$ defined by $\Gamma$ with no spins unpaired. 
${\cal{N}}$ is a normalization factor. Clearly, there are no local
processes linking the different $\{ | \psi_{l} \rangle \}$. As
quasi-local ($d=0$) symmetries [sums of  local permutations] link all
dimer states within a given sector of $l$ lines to each other, the
expectation value of any quasi-local quantity, $\langle \psi_{l}| V|
\psi_{l} \rangle$, cannot be used to differentiate the different states
which are given by  Eq.~(\ref{symmmmmstate}) and Eq.~(\ref{def.}) may
follow. Here, the different states and their character are clearly
marked by the topologically conserved line number $l$ appropriate to
each sector. As no dimers cross $\Gamma$, the $T=0$ entanglement
entropy across $\Gamma$ may be conjectured to vanish, $S_{ent}^{\Gamma}
= 0$  for each of the states $| \psi_{l} \rangle$. The contour $\Gamma$
may be taken  to be of an arbitrary shape.  The states of  Eq.
(\ref{symmmmmstate}) form the analogue of the states of highest spin in
$SU(2)$  discussed in Section \ref{specialselect}. Regardless of the
form of the entanglement entropy in these and other cases, the
topologically conserved quantities (the line number here) may be the
simplest marker of TQO. 

\subsection{Vanishing topological entanglement entropy
in states with TQO}

The above conjecture can be made rigorous in a somewhat  more
restricted set of states. These states will be constructed in a manner
similar to the TQO GSs of the Klein model of Eq. (\ref{symmmmmstate}).
Nevertheless, these states need not be GSs of the $S=1/2$ $SU(2)$
Klein  Hamiltonian of \ref{section12}. To this end, let  us first note
that Eq. (\ref{def.}) implies that if $T=0$ TQO appears in a set of
states  $\{\phi_{a}^{\alpha}\}_{a=1}^{t}$ for any given $\alpha= 1, 2,
\cdots, N_{g}$ then it appears in their product state. That is, let us 
consider a set of ``fragmented" states $\{\phi_{a}\}_{a=1}^{t}$  each
of which satisfies Eq. (\ref{def.}) and  each of which has its support
in a region of the lattice [or volume]  $\Lambda_{a}$ such that 
\begin{eqnarray}
\Lambda_{a} \cap \Lambda_{b} =0
~~\mbox{for all $a \neq b$, and} \nonumber
\\ \Lambda_{1} \cup \Lambda_{2} \cup \cdots \cup \Lambda_{t} = \Lambda.
\label{decompose}
\end{eqnarray}
In Eq. (\ref{decompose}), $\Lambda$ denotes the entire lattice [or
volume] of the system in question. Let us next construct the product
states
\begin{eqnarray}
| \phi^{\alpha} \rangle =  \bigotimes_{a=1}^{t} | \phi_{a}^{\alpha} \rangle
\label{composeprod}
\end{eqnarray}
which has the entire lattice [or volume] $\Lambda$ as its support.  The
reader can easily convince her/himself that if Eq. (\ref{def.}) holds
for all states $\{\phi_{a}^{\alpha}\}_{a=1}^{t}$ of Eq.
(\ref{decompose}) then the states of Eq. (\ref{composeprod}) satisfy
Eq. (\ref{def.}) as well. In other words, these product states display
$T=0$ TQO. 

Next, let us consider a lattice partitioned into four regions
$\Lambda_{a}$ with $a=1,2,3,4$. In the notation of Fig.
(\ref{kleinloop}) these correspond to the four regions A, B, C, and the
remainder of $\Lambda$ which  we henceforth label by D. Now, let us
construct the following fragmented states
\begin{eqnarray}
|\phi_{a;l_{a}} \rangle = {\cal{N}}_{al_{a}}  \!\!\!\!\!\!\!
\sum_{\mbox{$|c \rangle$ has $l_{a}$ lines in $\Lambda_{a}$}} 
\!\!\!\!\!\!\!\!\!\!\!\!\!\!|c \rangle. 
\label{symmmmmstatefrag}
\end{eqnarray}    
By $l_{a}$ lines we allude to the number of lines piercing a horizontal
plane as it bisects region $a$. Next, we examine the states
\begin{eqnarray}
|\phi_{L_{A}L_{B} L_{C} L_{D}} \rangle = |\phi_{A;L_{A}} \rangle
\otimes |\phi_{B;L_{B}} \rangle \nonumber \\
 \otimes |\phi_{C;L_{C}} \rangle
\otimes |\phi_{D;L_{D}} \rangle .
\end{eqnarray} 

As in Eq. (\ref{composeprod}), these states display TQO.  A calculation
of the topological entanglement entropy of Eq. (\ref{st}) reveals that
in each of these states $S_{topo} =0$. This occurs notwithstanding that
each of these states exhibit TQO.

\subsection{A Summary and a new conjecture} 

In conclusion, the considerations of the singlet product states of
Section \ref{pureKlein}  suggest that if the conjecture of
\cite{entropy} indeed holds universally on lattice systems, an
appropriate averaging and/or limit process needs to be spelled out for
different contours $\Gamma$ such as those detailed here for, at least,
some lattice systems. Otherwise, the entanglement entropy measure of
TQO is inconsistent with  the original definition of TQO [Eq.
(\ref{def.})]  as representing a robustness to local perturbations.  
We raise here the conjecture that if topological entanglement entropy
is to capture TQO then, at the very least, $S_{topo}$ needs to be
contour independent. This leads to an exponentially large - in the
system size - number of conditions (the topological entanglement
entropy for each of the exponentially  large number of partitions must
be set to a fixed value). 

\section{Graph problems: entropy, spectra, and topology}
\label{graphsection}

The insufficiencies of spectra and entropy in determining if TQO is
present have counterparts in the topology of graphs and in the Graph
Equivalence Problem (GEP) in particular. The adjacency matrix of a
graph has elements  $C_{\bi\bj} =1$ if vertices $\bi$ and $\bj$ are
linked by an edge and $C_{\bi\bj}=0$ otherwise. Vertex relabeling $\bi
\to p(\bi)$ leaves a graph invariant but changes the adjacency matrix
$C$ according to  $C \to C^{\prime} =  P^{\dagger} C P$ with $P$ an
orthogonal matrix which represents the permutation: $P =
\delta_{\bj,p(\bi)}$. The GEP is the following \cite{godsil}: ``Given
$C$ and $C^{\prime}$, can we decide if both correspond to the same
topological graph?"  The spectra of $C$ and $C^{\prime}$ are
insufficient criteria. Entropic measures \cite{godsil} are useful but
also do not suffice. Similar to the current ideas in  the study of TQO,
both measures (spectra and entropy) were long ago suggested as a way to
partially flesh out  topological structures in graphs. In order to make
this connection more rigid, we discuss in the Appendix new {\it
Gauge-Graph Wavefunctions} - states defined on a graph in which TQO is
manifest.

\section{String and brane type correlators}
\label{section20}

Hamiltonians potentially capturing a few aspects  of the properties of
novel materials display non-local {\it string} orders
\cite{tJz,BO,Jurij,OS, KMNZ, Nijs, GM, coldatomstring, todo, NR,
Anfuso}. Here, there are non-local  correlators which display enhanced
(or maximal) correlations vis-a-vis standard two-point correlation
functions.  In this way, the presence of off-diagonal long-range order
(ODLRO) in those string correlators does not imply the existence of TQO
as defined by Eq.~(\ref{def.}). However, this ODLRO in the string/brane
correlators signals the presence of a {\it hidden} order of a non-local
character with important physical consequences.

It is important to emphasize that these string correlators are,
generally, different from those in gauge theories or more generally
systems with exact low $d$-GLSs. To make the distinction  clearer, we
reiterate the symmetry considerations of  Section \ref{assb}. At finite
temperatures,  by Elitzur's theorem for $d=0$ \cite{Elitzur} theories
and by its extension  to $d=1$ discrete or $d \le 2$ GLSs \cite{BN}, 
all non GLSs invariant quantities must vanish.  Only GLSs invariant
quantities can be finite. In the case of gauge theories, as well as in
other theories, this only allows for string correlators (closed Wilson
loops or open {\it meson}-like lines) to be finite. At $T=0$, similar
considerations often follow while working in  the eigenbasis in which
the $d$-GLSs are diagonalized: in this basis, only $d$-GLSs invariant
quantities can  attain a finite expectation value. For a $d$-GLSs $U$
we have that  $\langle g_{a}|U|g_{a} \rangle= u_{a}$. If $U$ is a $d=1$
GLS operator then we will also have another trivial example of a {\it
string} correlator. For higher $d>1$ GLSs, we will have {\it brane}
like correlators. These considerations are different  from the maximal
(or enhanced) string correlators  of for example  Refs. 
\cite{tJz,BO,Jurij,OS, KMNZ, Nijs, GM, coldatomstring, todo,  NR,
Anfuso} in which string correlators are not mandated by an exact 
symmetry. We will come back and make contact and comparison with these
considerations towards the end of this Section. We will suggest an
extension to higher-dimensional {\em brane} correlators, detailed
examples of which will be provided in future work.   

\subsection{An algorithm for string/brane correlators which relies 
on ground state selection rules and non-local transformations}

We now outline an algorithm for the construction of such non-local
correlators which are not mandated by symmetries. (In systems with
uniform global order already  present in their GS, the algorithm leads
to the usual two-point correlators.) We seek a unitary transformation
$U_{s}$ which rotates the GSs into  a new set of states which have
greater correlations as measured by a  set of local operators
$\{V_{\bi}\}$. These new states may have  an appropriately defined {\em
polarization} (eigenvalues $\{v_{\bi}\}$) of either (i) more slowly
decaying (algebraic or other) correlations, (ii) a uniform sign ({\em
partial polarization}) or (iii) maximal  expectation values $v_{\bi} =
v_{max}$ for all $\bi$ ({\em maximal polarization}). Cases (i) or (ii)
may lead to an emergent lower-dimensional gauge-like structure for the
enhanced correlator. Generally, in any system with  known (or
engineered) GSs, we may explicitly construct polarizing transformations
$U_{s}$.  In some cases, these operators $U_{s}$ embody a local
gauge-like  structure. Our contention is that these unitary operators
which rotate the (generally) entangled states onto uniformly states of
high {\em polarization} embody an operator $P_{{\sf low}}$ for which 
\begin{eqnarray}
\langle g| P_{{\sf low}}|g  \rangle = 1
\label{Pselectionrule}
\end{eqnarray}
for any state $| g\rangle$ in the GS manifold.  In the AKLT spin chain
which we will focus on shortly, there is associated with each pair  of
sites an operator $P_{ij}$ for which the selection rule of
Eq.~(\ref{Pselectionrule}) is satisfied.

\subsubsection{Selection rules and polarizing transformations in the
AKLT spin chain}

In what follows, we illustrate that in the presence of GS entanglement
characteristic (although not exclusive) to TQO, maximal (or in more
general instances finite) amplitudes of correlation functions evaluated
within a certain GS are seen with a non-local {\it string} correlator 
borne by a non-trivial polarizing operator $U_{s}$.  To provide a
concise known example where these concepts become clear, we focus  on
case (ii) within the  well-studied  $S=1$ AKLT  Hamiltonian
\cite{AKLT}
\begin{eqnarray}
H_{\sf AKLT} &=& \sum_{j=1}^{N_s} \Big[ \vec{S}_{j} \cdot \vec{S}_{j+1}
+ \frac{1}{3} (\vec{S}_{j}  \cdot \vec{S}_{j+1})^{2} \Big] \nonumber \\
&=& \sum_{j=1}^{N_s} \Big[ P_{j,j+1}^{S_{tot}=2} - \frac{2}{3} \Big],
\label{AKLT_Ham}
\end{eqnarray}
for a chain with $N_s$ sites. The argument of the summand is, up  to
constants, nothing but the projection operator of the total spin on
sites $j$ and $(j+1)$ onto a value of $S^{tot} =2$, i.e.
$P_{j,j+1}^{S_{tot}=2}$. This allows for the construction of GSs
\cite{AKLT} which are of the form 
\begin{eqnarray}
\prod_{j} P^{j}_{symm} \bigotimes_{i} [|\uparrow_{i} \downarrow_{i-1}
\rangle - |\downarrow_{i} \uparrow_{i+1} \rangle].
\label{aklts}
\end{eqnarray}
In Eq. (\ref{aklts}), each $S=1$ is replaced by two $S=1/2$ spins and
singlets are formed between of these nearest-neighbor $S=1/2$ spins,
finally the $S=1/2$ basis state is projected onto the $S=1$ basis
(which is attained by the symmetric projection operator $P^{j}_{symm}$)
\cite{AKLT}.  Consequently, the total spin of any nearest-neighbor pair
can at most be 1. Within such a state,
\begin{eqnarray}
\langle g| \prod_{j} (1-P_{j,j+1}^{S_{tot}=2}) | g \rangle =1,
\end{eqnarray}
a realization of Eq. (\ref{Pselectionrule}).  In a system with open
boundary conditions, there are 4 degenerate GSs in which there remain
fractionalized spins at the two endpoints of the chain. 

In what follows,  we discuss gauge-like aspects of this  system. 
(Details underlying the assertions below are given in Appendix
\ref{app5}.) Here, there is a non-trivial unitary operator
\begin{eqnarray}
U_{s} \equiv \prod_{j<k} \exp[i \pi S^{z}_{j} S^{x}_{k}] , \ \mbox{
with } [U_{s},H_{\sf AKLT}] \neq 0 ,
\label{ak}
\end{eqnarray}
which maps the GSs $\{|g_{\alpha}\rangle \}$ into linear superpositions
of states in each of which the local staggered magnetization 
$V_{j}=(-1)^{j}S_{j}^{z}$ is uniformly non-negative (or non-positive)
at every site $j$. As all transformed states 
\begin{eqnarray}
|p_{\alpha} \rangle = U_s | g_{\alpha} \rangle
\label{pusg}
\end{eqnarray}
are superpositions of states with uniform sign  $v_{j}$ (allowing for
two non-negative or non-positive  $v_{j}$ values at every site out of
the three $S=1$ states), the correlator
\begin{eqnarray}
\tilde{G}_{ij}  &\equiv&  \langle p_{\alpha} | V_{i} V_{j} |p_{\alpha}
\rangle = \langle g_{\alpha}| U_{s}^{\dagger} V_{i} V_{j} U_{s} |
g_{\alpha}  \rangle \nonumber \\
&=& (-1)^{|{i}- {j}|}  \langle g_{\alpha} | S_i^{z} Q_{ij}S_{j}^{z}  |
g_{\alpha} \rangle
\label{gijg}
\end{eqnarray}
can be computed to give $|\tilde{G}_{ij}| = (2/3)^{2}$ for arbitrarily 
large separation $|i-j|$, i.e. $\tilde{G}_{ij}$ displays ODLRO.  Here,
$Q_{ij}$ is a shorthand for the string product 
\begin{eqnarray}
Q_{ij}=\prod_{i < k < j} \exp[i \pi S_{k}^{z}].
\end{eqnarray}
Given that the two-point spin correlators
\begin{eqnarray}
G_{ij}=\langle S_{i}^{z} S_{j}^{z} \rangle
\end{eqnarray}
decay algebraically, we have, for $|i-j| \gg 1$, 
\begin{eqnarray}
| \langle S_{i}^{z} Q_{ij} S_{j}^{z} \rangle | \gg  |\langle S_{i}^{z}
S_{j}^{z} \rangle|.
\label{central-string}
\end{eqnarray}

The inequality of Eq. (\ref{central-string}) was shown to hold in a 
large family of $D=1$ $S=1$ Hamiltonians which extends the AKLT point
\cite{KT}. In Appendix \ref{app5} we provide a proof of Eq.
(\ref{central-string}) that is based on classes of states and not
specific Hamiltonians. Similar large amplitude {\it string} or 
non-local correlators appear in other arenas \cite{tJz,BO,Jurij,OS,
KMNZ, AKLT,KT, Nijs, GM}.  Thus far, systems hosting such correlations 
(and the form of the correlations themselves) were found on an example
by example basis. In what follows, we suggest how  exact inequalities
of the form of Eq. (\ref{central-string}) may be derived in a
systematic fashion. To couch our discussion in the simplest possible
terms, the terminology will often follow is that of the spin systems
(and of the AKLT  Hamiltonian, in particular). Nevertheless, the ideas
introduced below allow a construction of non-local string operators in
any instance in which the GSs of a certain system are known and in
which these GSs are entangled. Entanglement implies that  the operator
$U_{s}$ which maps the GSs into a set of uniform ({\it polarized})
states  $\{|p_{\alpha} \rangle\}$ cannot be a uniform product of local
operators (or of sums thereof). That is, 
\begin{eqnarray}
U_{s} \neq \prod_{j \in \Lambda} O_{j},
\end{eqnarray}
with $\{O_{j}\}$ local operators.

\begin{figure}
\centerline{\includegraphics[width=1.0\columnwidth]{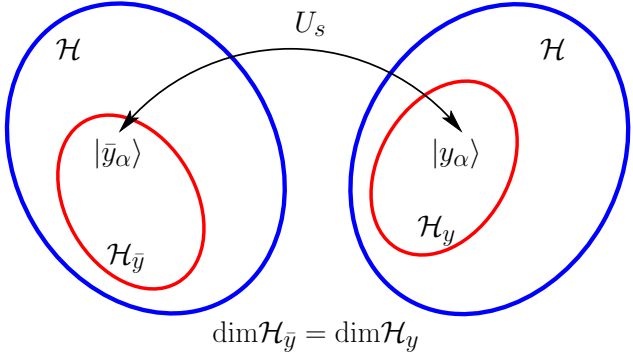}}
\caption{A cartoon of the manifold of states  ($\{|y_{\alpha}\rangle
\}$)  where a local gauge structure exists.  The double arrow above
describes the action of the non-local unitary operator $U_{s}$  (and
$U_{s}^{-1}$). Upon applying $U_{s}$, the states $|y_{\alpha} \rangle$
become partially polarized,  $U_{s} | y_{\alpha} \rangle = |p_{\alpha}
\rangle \equiv |\bar{y}_{\alpha} \rangle$, and have enhanced
correlations relative to a chosen local observable (see text). By its
very definition, the space ${\cal{H}}_{y} = \{ \ket{y_{\alpha}} \}$
includes the GS manifold. Generally, it is far larger. In the AKLT
problem, ${\cal{H}}_{y}$  is of size $2^{N_{s}+1}-1$, while the net
number of GSs is no larger than four. The unitary transformation
$U_{s}$ preserves the dimension $(2^{N_{s}+1}-1)$  of ${\cal{H}}_{y}$.
The total number of possible linearly independent states in the  $S=1$
system is $3^{N_{s}}$. The subspace ${\cal{H}}_{y}$ is an
infinitesimally small fraction of all possible states (${\cal{H}}$). 
States in ${\cal{H}}_{y}$ are linked to each other by local gauge-like
operators. [See Eq. (\ref{piji}) and the description of these 
gauge-like operators in the text.]}
\label{fig2}
\end{figure}

This unitary transformation generalizes the exact symmetry of the GSs
which we discussed in earlier Sections. Of all $3^{N_{s}}$ linearly
independent basis states $\{|S_{1}^{z}, S_{2}^{z}, \cdots,S_{N_{s}}^{z}
\rangle \}$ only $(2^{N_{s}+1}-1)$ might have a finite overlap with a
GS:  In the AKLT example, all GSs are superpositions of N\'eel-like
segments  of the chain (with one of the two sublattice parities:
$S_i^{z} = (-1)^{i}$ or $S_{i}^{z} = (-1)^{i+1}$) \cite{KT} which are
separated by sites at which $S_{i}^{z}=0$, e.g.
\begin{eqnarray}
|+-+-0+-0+ \cdots \rangle.
\label{AB0BA}
\end{eqnarray}
The rule of thumb for determining the  sublattice parity of the N\'eel
orders across the zero sites is as follows: opposite parity N\'eel
orders exist across an odd number of zeros (as in Eq. (\ref{AB0BA})), 
and similar ones exist across even length numbers of zeros.  We find
that this structure can be captured by the following  $d=0$ operators
within the GS basis,
\begin{eqnarray}
P_{ij} = - S_{i}^{z} Q_{ij} S_{j}^{z} + (1- (S_{i}^{z}
S_{j}^{z})^{2}), 
\label{piji}
\end{eqnarray}
for which
\begin{eqnarray}
P_{ij} | g \rangle = |g \rangle 
\label{piji2}
\end{eqnarray}           
for all GSs $|g \rangle$ and for all site pairs $(i,j)$. That Eqs.
(\ref{piji}), and  (\ref{piji2}) are satisfied follows from the form of
Eq. (\ref{aklts}).  The wavefunction of Eq. (\ref{aklts}) - that of
consecutive singlets - does not allow the total $S^{z}$ on any string
to have an absolute value that exceeds 1. That is,
\begin{eqnarray}
|\langle g| \sum_{l=i}^{j} S^{z}_{l}| g \rangle| \le 1.
\end{eqnarray}
It is worth to notice that Eqs. (\ref{piji}), and (\ref{piji2}) imply
\begin{eqnarray}
\langle g| S_{i}^{z} Q_{ij} S_{j}^{z}| g \rangle = - \langle g|
(S_{i}^{z} S_{j}^{z})^{2}| g  \rangle. 
\end{eqnarray}
That is, {\em the hidden non-local order is equal to a local
nematic-type  correlator}. ( As we will see below, this is true for any
state that belongs to the subspace ${\cal H}_y$.) In particular, the
fact that the string correlator $\tilde{G}_{ij}$ of Eq. (\ref{gijg}) is
equal to $(2/3)^{2}$ implies that, within the GS subspace,  $\langle
(S_{i}^{z})^{2} \rangle = 2/3$ at all sites $i$. The operators of
Eq.~(\ref{piji}) act as the identity  operator within the GS basis:
they form a realization of our general rule of thumb of
Eq.~(\ref{Pselectionrule}). They should not be confused with obeying
exact $d=0$ GLSs selection rules which appear throughout the entire
spectrum. In Fig.~\ref{fig2}, we schematically represent  the set of
all states $\{ | y_{\alpha} \rangle \}$ which satisfy $P_{ij} |
y_{\alpha} \rangle = |y_{\alpha} \rangle$.  This set includes the GS
basis as a very special subset.  The set of independent basis vectors
$\{|y_{\alpha} \rangle \}$ may be equivalently  defined by finding all
such basis states  such that the target space $U_{s} |y_{\alpha}
\rangle$ degenerates into {\it partially polarized states} $ U_{s}
|y_{\alpha} \rangle = |p_{\alpha} \rangle$. In the AKLT system, these
partially  polarized states are those with non-negative (non-positive)
entries, e.g. 
\begin{eqnarray}
| ++++0++0+ \cdots \rangle. 
\label{partialp}
\end{eqnarray}
Equation (\ref{partialp}) is the state of Eq. (\ref{AB0BA})  after it
would be acted upon by the partially polarizing transformation operator
of Eq. (\ref{ak}). The polarization transformations are just another
way  of stating the selection rules of Eqs. (\ref{piji}), and
(\ref{piji2}). The operator of Eq. (\ref{ak}) leads to a partially
polarized state [of the form of Eq. (\ref{partialp})] for all states
which satisfy  the selection rules of Eq. (\ref{piji2}) [e.g. the basis
state of Eq. (\ref{AB0BA})].  We denote by $ {\cal{H}}_{y}$ the Hilbert
subspace spanned by $\{\ket{y_{\alpha}} \}$. By unitarity,
${\cal{H}}_{y}$ spans $(2^{N_{s}+1}-1)$ states. This is so as the
number of partially polarized states $\{|p_{\alpha} \rangle\}$ is
$(2^{N_{s}+1}-1)$.  Furthermore, let us denote by $T_{y}$ the set of
operations which have finite matrix elements only within this basis.
These operators (albeit not symmetries of the Hamiltonian) constitute
analogues of the ($d=0$) gauge transformation  symmetry operators of
$\Z_{2}$ gauge theories linking degenerate states. These $d=0$ $\Z_{2}$
transformations correspond (in the $\bigotimes_{j=1}^{N_{s}} S^{z}_{j}$
basis) to the creation (annihilation) of an $S^{z}_{j}=0$ state at any
site ${j}$  followed by a unit displacement of $S^{z}_{k}$ at all sites
${k}>{j}$. 

In the AKLT Hamiltonian, the entire set of basis vectors $\{|y_{\alpha}
\rangle\}$ exhibits an invariance under local Ising gauge-like
operations ($\Z_{2} \otimes \Z_{2} \otimes \cdots  \otimes \Z_{2}$ with
a $\Z_{2}$ appearing for each of the $j=1,2,\cdots,N_{s}$  sites) which
keep the non-negative (non-positive) entries  of $(-1)^{j} S_{j}^{z}$
non-negative (non-positive). Gauge-like operators enumerate the number
of independent rays $\{|y_{\alpha} \rangle \}$ (the size of the Hilbert
space in which the GSs exist in the $\bigotimes_{j=1}^{N_{s}}
|S^{z}_{j} \rangle$ basis): $(2^{N_{s}+1}-1)$ with, asymptotically, a
factor of 2 given  by an Ising  ($\Z_{2}$) degeneracy present at each
of the $N_{s}$ sites.
 
The structure which emerges is not that of the GSs (which is  that of a
global $\Z_{2} \otimes \Z_{2}$ symmetry)  but rather  that of the
$S^{z}$ basis in which the GSs reside.  With $\{U_{\rm GSs}\}$ the set
of  unitary operators which have their support only within  the GS
basis,  we have the obvious  $\{T_{y}\} \supset \{U_{\rm GSs}\}$ with
the subscript referring to the set of symmetries which leave the larger
GSs basis support $|y_{\alpha} \rangle$ ({\it local gauge} in the AKLT
case) or the smaller set of GSs only un-admixed with other states
outside that basis or GS manifold respectively. With generalized
polarized states giving rise to non-local string correlators, we
generalize the notion of unitary transformations associated with the
manifold of the GSs to the larger set of transformations living on the
manifold of states which satisfy the GS selection rule of  Eq.
(\ref{piji2})  (of which the GSs are, by definition, a subset). This is
schematically illustrated in Fig.~\ref{fig2}.
 
The gauge structure that emerges is manifest in the string correlator.
The string correlator is precisely the gauge invariant correlator
involving  {\it charges} ($e_{i}$ and $e_{j}$ which are here portrayed
by the spin components $S^{z}_{i}$ and $S^{z}_{j}$)  in the presence of
a local {\it photon gauge field} $\vec{A}$ associated with the local
gauge structure of the large set of states which become partially
polarized under $U_{s}$.  The group element associated with this gauge
structure in a path linking site $i$ to site $j$ is 
\begin{eqnarray}
Q_{ij} &=& \exp[i  \sum_{k,l \in C} A_{kl}], \nonumber \\ 
A_{t,t+1} &\equiv&   i \pi S_{t}^{z}.
\end{eqnarray}
Enhanced, albeit decaying, string correlators with an underlying
gauge-like structure  also appear in doped spin chains \cite{tJz, OS,
KMNZ}.

We may also generate maximal ($\tilde{G}_{ij} =1$) string correlators
(case (iii) above); here, the number of states with  maximal
polarization is finite and no local gauge-like structure emerges. If we
consider unitary transformations $\{U_{s}^{\max}\}$ of the four GSs of
the AKLT chain to  the two orthogonal states given, in the
$\bigotimes_{j=1}^{N_{s}} S_{j}^{z}$ basis by  
\begin{eqnarray}
|p_{\max}^{1} \rangle &=&  |1,1,1,  \cdots, 1 \rangle, \nonumber \\
|p_{max}^{2} \rangle &=& | -1, -1, -1, \cdots, -1 \rangle,
\end{eqnarray}
and to two other independent states of high polarization, e.g.  
\begin{eqnarray}
|p_{max}^{3} \rangle &= & |1,-1,1 , \cdots, (-1)^{r}, \cdots \nonumber
\rangle, \\ 
| p_{max}^{4} \rangle &= & |-1,1,-1, \cdots, (-1)^{r+1}, \cdots
\rangle,
\end{eqnarray} 
with an inverted staggered spin  at the $r$th site of the chain then,
following the same sequence of steps as undertaken in Eqs.
(\ref{pusg}), and (\ref{gijg}), 
\begin{eqnarray}
|\langle p_{max}|
S_{i}^{z} S_{j}^{z}| p_{max} \rangle| & = &|\langle g_{\alpha}| 
U_{s}^{max ~ \dagger} S_{i}^{z}S_{j}^{z} U_{s}^{max}  | g_{\alpha}
\rangle| \nonumber \\ 
&=& 1 .
\label{MAXSTR}
\end{eqnarray} 
The transformation $U^{s}_{max}$ involves all sites of the lattice.
Unlike the transformation of (\ref{ak}), Eq. (\ref{MAXSTR}) does not
only contain fields between (and including) sites $i$ and $j$.  Eq.
(\ref{MAXSTR}) is a string operator for the two sites $i$ and  $j$
which form the endpoints of the chain.   Here, the spin-spin
correlation is maximal and equal to one (it is not 4/9 as for the
partially polarized case where only all non-zero $S^{z}$ entries
attained a uniform sign). 

Having derived the string-type correlators of the AKLT $S=1$ chains,
we briefly review and notice links with the FQHE problem. Just as
in the spin chain problem, where a polarization operator led to long
range order of string-like objects,  there exists a similar
transformation in the FQHE (see  \cite{explain_longFQHE}). 

Although conceptually clear, we  wish to re-emphasize a simple yet
important point. In the presence of entangled GSs, many transformations
may map the original GS basis to new states in which the correlations, 
albeit not being maximal nor tending to a smaller finite  constant for
arbitrarily large separations between their endpoints, are larger than
in the original basis. In the spin language employed in this Section, 
these transformations do not lead to uniformly polarized  states. Such
transformations are afforded and may be studied by, for example,
Jordan-Wigner and other non-local transformations
\cite{tJz,BO,Jurij,OS, KMNZ, Nijs, GM, NR}. 

\subsubsection{Non-local correlators in general gapped systems}

We next construct non-maximal string and {\it brane} type correlators
in general systems with a spectral gap (of size $\Delta E$).  To this
end, we rely on the work of Ref. \cite{hastingsgap} which showed that
the GSs of a gapped system can be made arbitrarily close to a
generalized matrix product state of finite-size representation. Such a
general product state is similar to what we wish to have in order to
allow for a transformation to a highly {\em polarized} state which we
just discussed.  Let us consider Hamiltonians of the type 
\begin{eqnarray}
H = \sum_{i} H_{i}. 
\end{eqnarray}
For example, in the AKLT Hamiltonian of Eq.~(\ref{AKLT_Ham}),  $H_{i} =
\vec{S}_{i} \cdot \vec{S}_{i+1}  + \frac{1}{3}(\vec{S}_{i} \cdot
\vec{S}_{i+1})^{2}$.  To construct the approximate local projective
states, we set ($q$ is a constant)
\begin{eqnarray}
H_{i}(t) &=& e^{iHt} H_i e^{-iHt}, \nonumber \\ 
\tilde{H}_{i}(t) &=& H_{i}(t) \exp[(-t \Delta E)^{2}/(2q)], \nonumber
\\ 
\tilde{H}_{i}^{0} &=& \frac{\Delta E}{\sqrt{2 \pi q}} 
\int_{-\infty}^{\infty} dt \tilde{H}_{i}(t).
\end{eqnarray}
We next define the quasi-local operators
\begin{eqnarray}
M_{i} &=& \frac{\Delta E}{\sqrt{2 \pi q}} \int_{-\infty}^{\infty}  dt
e^{-(t \Delta E)^{2}/2} H_{i}^{trunc}(t), \nonumber \\ 
H_{i}^{trunc}(t) &=& e^{i H_{loc} t} H_{i} e^{-i H_{loc} t}, \nonumber
\\ 
H_{loc} &=& \sum_{j: |i-j| \le l_{proj} - R} H_{j}.
\label{dec1}
\end{eqnarray}
In Eq.~(\ref{dec1}), the quasi-local Hamiltonian $H_{loc}$ is the sum
of $\{H_{j}\}$ with the distance between sites $i$ and $j$ satisfying
$d(i,j) \le (l_{proj}- R)$. They key point is that the quasi-local
Hamiltonians $M_{i}$ approximate the original $H_{i}$ and  enable the
construction of a product state. Here, if we write all expressions
explicitly, we find that the projection  operators which take us to the
low-energy sector of $\{M_{i}\}$ and their sums and with $| g^{(n)}
\rangle$ low-energy normalized projected states defined iteratively as
in \cite{hastingsgap} {\em lead precisely to size} $n$ {\em string or
{\it brane} operators.} These projection operators play the role of
the selection rules embodied in Eqs.~(\ref{Pselectionrule}), and
(\ref{piji}). These expectation values are those of a string or brane
like operator $Q$ within the GS manifold,
\begin{eqnarray}
\langle g^{(n)} | M^{(n)} | g^{(n)} \rangle =  \langle g| Q^{(n)} | g
\rangle \ge 1 - b_{n},
\label{hb}
\end{eqnarray} 
with bounded $b_{n}$ as defined in \cite{hastingsgap}. 
For general gapped Hamiltonians for
which we do not know the GSs and the unitary transformations which will
maximally {\em polarize} them, the string/brane correlators that we
find in Eq.~(\ref{hb}), are the best that one can do. 

\subsubsection{Low-energy selection rules in general systems}

We now return to make connections between  (i) the {\em
polarization}/projective algorithm which  we outlined above for
generating string correlators in  systems where no GLSs are present to
(ii) systems with exact GLSs for the special case of the AKLT chain in
which all GSs satisfy the $d=0$ selection rules of  Eqs.~(\ref{piji}),
and (\ref{piji2}). In systems with no GLSs, general two-point and 
other correlators may be non-zero. This forces us to think of only
string-like correlators. Obviously, unlike the situation in gauge and
other theories (e.g. Kitaev's Toric code model of
Eqs.~(\ref{kitaevmodel}), and (\ref{AB_defn}) to which we will return
to momentarily), general spin-spin and other correlators may be
non-zero  in the AKLT and many other systems with low-range {\em string
order}: there are no GLSs symmetries which force these correlators to
vanish by an application of Elitzur's theorem and its generalization.  
Nevertheless, the form of the GSs  as adhering to low-$d$ selection 
rules (as in Eqs.~(\ref{piji}), and (\ref{piji2}) for the AKLT chain)
enables maximal correlators of a form similar to that encountered in
systems with exact local symmetries. Such an example is furnished by
Kitaev's  model of Eqs.~(\ref{kitaevmodel}), and (\ref{AB_defn})  for
which the GSs attain the form of Eq.~(\ref{projkit}). Here, the general
$d=0$ projection operator 
\begin{eqnarray}
\!\!\!\!\!\!\!P_{s_{1}\cdots s_{n}; p_{1}\cdots p_{m}} \equiv 
\prod_{i=1}^{n} \Big[ \frac{1}{2} (1+ A_{s_{i}}) \Big]
\prod_{j=1}^{m} \Big[ \frac{1}{2} (1+ B_{p_{j}}) \Big]
\end{eqnarray}
is not only a symmetry but also acts as the identity operator within
the GS basis (similar to Eqs.~(\ref{piji}), and (\ref{piji2}) for the
AKLT chain). The expectation value 
\begin{eqnarray}
\langle g| P_{s_{1} \cdots s_{n}; p_{1}\cdots p_{m}} | g
\rangle =1,
\label{PKitaev}
\end{eqnarray}
for any GS $|g \rangle$.  Here, the set $\{s_{1}, \cdots, s_{n}; p_{1},
\cdots, p_{m}\}$  contains an arbitrary number of sites $(n)$  and
plaquettes ($m$) out of those in the entire lattice. In general, the
expectation value of Eq.~(\ref{PKitaev}) leads to a product of 
string-like objects. At finite temperatures,  this expectation value
will be degraded and we have
\begin{eqnarray}
\langle P_{s_{1} \cdots s_{n}; p_{1}\cdots p_{m}} \rangle =
[\frac{1+\tanh \beta}{2}]^{n+m}.
\label{PKitaevT}
\end{eqnarray}
These brane-type correlators are closely related to (and are sums of)
Wilson loop-type correlators \cite{kogut} which are formed by the
boundaries of connected clusters of  plaquette and vertex terms on the
direct  ($B_{p}$) and dual ($A_{s}$) lattices.  

Similar to the discussion prior to Eq.~(\ref{hb}), for more general
gapped systems, we can construct and evaluate string correlators at
finite temperatures. A more non-trivial example is furnished by {\em
brane}-type correlators recently found \cite{CN} for the Kitaev model
on the hexagonal lattice. In that system, the spin basis GS \cite{CN}
is not of the form of Eq. (\ref{gentop}) with a uniform $f=1$. 

Obviously, Kitaev's Toric model and related systems  are very special: 
most systems cannot have GSs which can be written in the simple form of
Eq.~(\ref{projkit}). Nevertheless, in most entangled systems, there is
a non-local unitary transformation which maps the GSs to states of
maximal correlation. When written out in terms of  local fields, this
unitary transformation, which  will generally involve all sites in the
system,  may lead to string- (or {\it brane}-) type correlators.  In
practical terms, for gapped systems, there is generally an effective
matrix product form which allows to us construct string correlators as
shown in this Section. 

\subsection{The non-equivalence of string/brane orders and TQO} 

Our general polarization (or selection rule) algorithm relies on the
existence of a general non-local transformations which maps the GSs
onto  uniformly factorizable states. The very generality of this
algorithm suggests that the definition of TQO as given by 
Eq.~(\ref{def.}) and the presence of  string (or {\it brane}) orders
are two independent issues. This distinction of the two orders is made
clear by a simple example. The local expectation value $\langle
g_{\alpha} | S_{\bi=1}^{z} | g_{\alpha} \rangle$ depends on
$\ket{g_\alpha}$, violating Eq. (\ref{def.}) and suggesting that the
AKLT chain is not topologically ordered (see Appendix \ref{app5}).

\section{On ground state degeneracy, Topology, and Robustness} 
\label{topology_not_enough}

One of the earliest definitions of TQO was that as classifying systems
in which the GS degeneracy depends on the topology of the manifold on
which the system is embedded \cite{weniu}. In that way, the degeneracy
depends upon a global property. In what follows, we will note that
this  earliest of definitions also  encompasses systems which do not
satisfy the robustness condition of Eq. (\ref{def.}). We will
illustrate this point by three examples.

(i) Let us first consider the Klein model of  Ref. \cite{bt}. By the
theorem proved in \cite{NBnew} all GSs of  this model on the square
lattice are superpositions of singlet dimer coverings. In these dimer
coverings, at least one dimer must appear in each plaquette. An
explicit calculation reveals that all of these special dimer coverings
on the square lattice are  orthogonal to one another. Thus, the
degeneracy is exactly equal to the  number of dimer coverings in which
each plaquette houses at least one dimer. Armed with this result, we
may next compute the GS degeneracy of  this square lattice system when
it is embedded on manifolds of different topology. These are easily
tabulated: a system with open boundary conditions has a degeneracy
equal to
\begin{eqnarray}
{\sf degeneracy}_{\sf Klein; open-bc} = 2^{N_{d}+1} + 3 N_{s}.
\label{opendeg}
\end{eqnarray}
Here, $N_{s}$ denotes the number of sites while $N_{d}$ denotes the
number of diagonals. 

A system with periodic boundary conditions  (a system on a torus) has
fewer states in the GS. The 3$N_{s}$ {\it defect} states appearing in Eq.
(\ref{opendeg}) are incompatible with  periodic boundary conditions.
For different sorts of periodic boundary conditions, the degeneracy can
vary. Systems with periodic boundary conditions along the diagonals
support more GSs than systems with periodic boundary conditions along
the horizontal/vertical axes.

Thus, Klein models of this type have a GS degeneracy which depends on
the topology of the embedding manifold. Nevertheless, as we discussed
in earlier Sections,  the bulk of such Klein model GSs do not obey the
robustness  condition of Eq. (\ref{def.}): In any given dimer state, we
may compute the quasi-local expectation  value $\langle \vec{S}_{\bi}
\cdot \vec{S}_{\bj} \rangle$ to see whether or not the spins at nearest
(or, at worst, next-nearest) sites form a singlet dimer or not.

Although in the above system, the GS degeneracy depends on the
topology, it also depends on other local transformations. We next
construct a system whose degeneracy depends {\em only} on the topology
of the manifold  on which it is embedded.

(ii)  Let us consider a classical $D=2$ Ising model on a closed
oriented manifold of genus $g$. Let us mark  all of the ``+" spins by
vertices of a polyhedra on the manifold. Let us next consider the
Hamiltonian
\begin{eqnarray}
 H= ((\sum_{\langle i j \rangle} s_{i} s_{j})^2 - a[V-E+F]^2)^2.
\end{eqnarray}

Here, $V$ is the number of vertices - the total number of $s_{i}= +1$
spins, $a$ is an arbitrary positive constant, $E$ is the number of
edges (of ``++" pairs on neighboring sites on the polyhedra) on the
polyhedra formed by the ``+" spins, and $F$ is the total number of
faces. As by the Euler-Lhuillier formula for closed oriented surfaces
with genus or number of handles $g$ [Eq. (\ref{Euler})], we will have
that the GSs of $H$ will correspond to all polyhedra (or Ising states)
with a value of $(\sum_{\langle i j \rangle} s_{i} s_{j})^2 $
which is given in terms of the genus number $g$. For a given size
(number of spins) of the Ising system, the degeneracy will depend only
on $g$ and nothing else.

(iii) Let us finally consider the classical $D=2$ Ising gauge theory of
Eq. (\ref{Z2t}).  Similar to Eq. (\ref{degk}), and the discussion of
Section \ref{mani}, the  degeneracy on a manifold of genus $g \ge 1$
is \cite{degexp} $ 2^{N_{s}} 2^{2g-1}$.  The large prefactor of
$2^{N_{s}}$  can be avoided in a legitimate local gauge fix  which
reduces the number of degrees freedom ($\sigma_{ij}^{z}$) yet  does not
introduce additional constraints on the plaquette variables $\{B_{p}\}$
beyond those seen in Eq. (\ref{ABpc}).  To this end, let us  draw a
closed non-intersecting path on the lattice which passes through the
centers of all of the lattice plaquettes. This path is of length $F$ -
the number of plaquettes. Let us set $\sigma_{ij}^z=1$  on all of the
bonds which this path does not intersect \cite{explainwhy}. By Eq.
(\ref{Euler}) with $V=N_{s}$ the number of vertices,  our path leads to
$F=(N_{s}+2g-2)$  uncrossed bonds. There remain two unfixed bonds (the
bonds which are crossed by the path) in each plaquette - each bond being
shared by two plaquettes.  Following this procedure, the classical
Ising gauge Hamiltonian of Eq. (\ref{Z2t}) reduces to that of a
circular Ising chain  of length $F$ which has two GSs.  Such
a procedure sets $[N_{s}+2(g-1)]$  bonds to fixed values whereas  there
are only $[N_{s}-1]$ gauge fix degrees of freedom: $(2g-1)$ additional
Ising degrees of freedom appear in the Ising gauge system by comparison
to that of the Ising chain.  A legitimate gauge fix leads to a few
plaquettes which host more than two unfixed bonds - $(2g-1)$ formerly
pinned bonds by our procedure need to be set active. This leads to a
degeneracy
\begin{eqnarray}
{\sf degeneracy}_{\sf Ising~ gauge} = 2^{2g}
\end{eqnarray}
in any legitimate gauge fix. This result is, as it must, identical to
that of Kitaev's model [Eq. (\ref{degk})] which, in its GS sector
amounts to a particular (non-classical) gauge fix: $A_{s}=1$.
Local measurements can differentiate between
the various GSs (thereby violating Eq. (\ref{def.}))  of this system
whose GS degeneracy depends on topology.  Examples such as this can be
avoided if we only consider rank-$n$ TQO [see the discussion of Section
~ \ref{zerotempthing}]: That is, if we consider $n$ GSs which  satisfy Eq.
(\ref{def.}). However, considering GSs which  satisfy Eq. (\ref{def.})
defeats the purpose of an independent conjectured check on TQO - that
in which the GS degeneracy depends only on the topology. 

Putting all of the pieces together, we note that the link between (1)
the dependence of the GS degeneracy on topology and (2) the robustness
condition of Eq. (\ref{def.}) is tenuous. 

\section{Conclusions}
\label{section21}

To conclude, we analyzed Topological Quantum Order (TQO) from the
perspective of generalized Gauge-Like Symmetries (GLSs) and gauge
structures. This approach  allows us to identify and construct new
physical systems with TQO. This work details, in great length, the
abbreviated results summarized in an earlier Letter \cite{NO}. Below we
list the new results which are provided in our work:

{\bf (1)}  We provide an extension of the zero-temperature definition
of TQO (Kitaev's original definition which relies on robustness to
local perturbations) to finite temperatures.

{\bf (2)} We relate GLSs to general (non-Abelian) Aharonov-Bohm type
transformations. We discuss how low-dimensional GLSs lead (i) to conserved
non-local quantities, (ii) the absence of quasi-particle excitations in
these systems, (iii) how these symmetries enable a natural
identification of charge fractionalization, and to (iv) the appearance,
in high dimensions, of topological terms which are usually only
associated with low-dimensional theories (dimensional reduction).

{\bf (3)} We show how finite-$T$ TQO is mandated  by low-dimensional
GLSs. Our results are particularly strong for gapped systems with $d
\le 2$ continuous GLSs. 

{\bf (4)} We discuss the selection rules associated with these
symmetries and show how these enable the construction of states with 
TQO. In one form or another, these selection rules account for all of
the  known examples of TQO to date. We stress that GLS selection rules
alone can enforce a vanishing effect of local perturbations - no
spectral gaps are invoked. 

{\bf (5)} We further show how GLSs can enforce the appearance of
degenerate states in the large system-size (thermodynamic) limit. This
degeneracy is borne by vanishing of off-diagonal matrix elements (in
the thermodynamic limit) of quasi-local operators in the basis which
is spanned by eigenstates of the GLSs.

{\bf (6)} We identify the GLSs associated with all of the prominent
examples of TQO and identify several  new systems with TQO. It is
important to stress that it is not necessary to have Hamiltonians with
exact GLSs. The same occurs also in systems with standard ({\it local})
order parameters: although no real system displays perfect symmetries,
the fixed points  associated with certain symmetry may have a finite
regime of validity even when perturbations are added. What is important
for TQO, much as in standard  phase transitions, is that  theories may
be {\em adiabatically connected} with systems in which the GLSs are
exact. The adiabatic link between Fractional Quantum Hall systems on a
two-torus and the ($m$-rized)Peierls chain (in which $d=1$ $\Z_{m}$
symmetries are exact) is an example of this principle. 

{\bf (7)} We show that in several well-known  examples of TQO (Kitaev's
model and Wen's model) an exact dimensional reduction occurs. As the
partition functions of these systems are equivalent to those of Ising
chains, no finite-$T$ phase transition occurs. No less important, in
spite of the existence of a spectral  gap, some zero-temperature
topological quantities are eradicated at any finite temperature if all
sites in the system may freely change.  A finite spectral gap might not
protect the robustness of topological quantities even at infinitesimal
temperatures.

{\bf (8)} We show that in certain systems, the entanglement entropy
might not always reflect TQO. To this end, states are constructed in
which local perturbations lead to a null effect in the GS basis (and
thus these states are topologically ordered). It is shown that these
states can have a  vanishing entanglement entropy  in spite of their
TQO character.

{\bf (9)} We provide a general algorithm for the construction of
non-local {\it string} orders in systems where the form of the GSs  (i)
can be shown to adhere to certain selection rules and (ii) specifically
show how {\it string} or higher-dimensional {\em brane}  orders appear
in general gapped systems in arbitrary dimensions. We identify an 
underlying gauge structure associated  with some of these string
orders.

{\bf (10)} We show that, on its own, the spectrum of a theory is not
sufficient to determine whether or not TQO  appears: {\it The
information about TQO is encoded in the state itself}. In order to
prove this, we show that the spectra of several theories with TQO (e.g.
Kitaev's and Wen's models, gauge theories, etc.) are equivalent to
theories with standard ({\it local}) orders (Ising chains, Ising models
in $D$=3 dimensions, etc.). 

{\bf (11)} We demonstrate that the expectation value of the  non-local
string correlator of the AKLT problem (and other $D=1$ models whose GSs
belong to a particular class) is  related to the expectation value of a
{\it local} correlator of the nematic type.

{\bf (12)} We illustrate that systems can have a GS degeneracy which
depends on the topology of the surface on which  they are embedded yet
not display TQO in the sense of robustness to local perturbations.

{\bf (13)} We remark on TQO on graphs and construct {\it Gauge-Graph
Wavefunctions} with TQO.

{\bf (14)} We provide (see Appendix \ref{app3_2}) an  explicit
Dirac-form expression for the symmetry operators of the ($m$-rized)
Peierls chain. 

{\bf (15)} We introduce general $SU(N)$ Klein models on small world
networks and remark  on their {\it topological} entanglement entropy
content.

\noindent {\bf Note added in proof.}
We were delighted to learn that some time after  we reported the
results concerning  the spectra and thermal fragility  in \cite{NO} and
\cite{NOlong},  C. Castelnovo and C. Chamon nicely confirmed our
conclusions. Castelnovo and Chamon \cite{CCe} found an entanglement
entropy signature of the  $T=0$ transition of the Kitaev model that we
first found in \cite{NO} and \cite{NOlong}.

\acknowledgments
We thank C. D. Batista, C. Castelnovo, C. Chamon, H. Chen, E. Fradkin, 
A. Kitaev, H. Katsura, I. Klich,  J. Preskill, K. Shtengel, J.
Slingerland, and C. Wu  for discussions. We especially acknowledge
enlightening discussions with S. Bravyi, and the DOE, the NSA, and the
CMI of WU for support. We are also extremely indebted to C. Castelnovo
for a careful reading of this manuscript.

\appendix

\section{Degenerate Perturbation Theory (DPT) in a Nutshell}
\label{app1}

We now present DPT in the Rayleigh-Schr\"odinger version in a very
simple and pedagogical fashion. This will allow us to establish the
conditions (on the matrix elements of the local perturbation $\hat{V}$,
between states in the degenerate eigenspace) for removal of such
degeneracy in a certain order of DPT. We want to illustrate the
difference between these and the conditions of Eq.~(\ref{def.}).

The equation we want to solve perturbatively is
\begin{equation}
(H_0+\lambda\hat{V})\ket{\Psi}=\epsilon \ket{\Psi} \ , 
\end{equation} 
where the two series
\begin{eqnarray}
\ket{\Psi}= \sum_{p=0}^n \lambda^p \ket{\Psi^{(p)}} \ , \mbox{and} \
\epsilon= \sum_{p=0}^n \lambda^p \epsilon^{(p)} \ , 
\end{eqnarray}
are assumed to be continuous and analytic functions of $\lambda$, for
$\lambda \in [0,1]$.

Let us assume that we want to compute the corrections to a degenerate
eigenspace ${\cal M}_0=\{\ket{g_\alpha}\}$ ($\alpha \in {\cal
S}_0=[1,N_g]$) due to a given perturbation $\hat{V}$. The unperturbed
state ($\langle\Psi^{(0)} | \Psi^{(0)} \rangle=1$)
\begin{equation}
\ket{\Psi^{(0)}}=\sum_{\alpha=1}^{N_g} \gamma_\alpha \ket{g_\alpha}
\end{equation}  
is a linear combination of the degenerate eigenstates $\ket{g_\alpha}$ 
satisfying ($\epsilon_0=\epsilon^{(0)}$)
\begin{equation}
H_0\ket{g_\alpha}=\epsilon_0 \ket{g_\alpha} \ , \ \langle
g_\alpha | g_\beta \rangle = \delta_{\alpha \beta}
\end{equation}  
with coefficients $\gamma_\alpha$ that need to be determined using 
$\hat{V}$ for the perturbation expansion to be analytic. The solution is
subject to the normalization condition $\langle\Psi^{(0)} | \Psi
\rangle=1$, which implies
\begin{equation}
\epsilon=\epsilon_0 + \langle \Psi^{(0)} | \lambda \hat{V}|\Psi \rangle\ .  
\end{equation}  
The corresponding Rayleigh-Schr\"odinger $n$-th order solution can be
obtained  recursively
\begin{equation}
\ket{\Psi^{(n)}}=\ket{\Psi^{(0)}}+G_0
(\one-\ket{\Psi^{(n-1)}}\bra{\Psi^{(0)}})\lambda\hat{V} \ket{\Psi^{(n-1)}}
\end{equation}  
with $G_0=(\epsilon_0-H_0)^{-1}$ and $\one$ the identity operator.

Let us establish the conditions on $\hat{V}$ to remove the degeneracy
of ${\cal M}_0$. Let's start assuming that the degeneracy is removed to
first order in $\hat{V}$. That means that $\langle g_\alpha | 
\hat{V} | \Psi^{(0)} \rangle=\epsilon^{(1)} \langle g_\alpha | \Psi^{(0)}
\rangle=\epsilon^{(1)} \gamma_\alpha$, or equivalently
\begin{equation}
(V^{(1)}_{\alpha\alpha}-\epsilon^{(1)})\gamma_\alpha+\sum_{\beta=1 (\neq
\alpha)}^{N_g}\gamma_\beta V^{(1)}_{\alpha\beta} = 0 \ ,
\end{equation}
$\forall \alpha$, with $V^{(1)}_{\alpha\beta}=\langle g_\alpha | 
\hat{V} | g_\beta \rangle$. We need to determine $\epsilon^{(1)}$ by
solving $\det [ W^{(1)}_{\alpha\beta}]$=0, with
$W^{(1)}_{\alpha\alpha}=V^{(1)}_{\alpha\alpha}-\epsilon^{(1)}$ and
$W^{(1)}_{\alpha\beta}=V^{(1)}_{\alpha\beta}$, otherwise. It is clear
that if 
\begin{eqnarray}
V^{(1)}_{\alpha\alpha}&=&V^{(1)}_{\beta\beta}, \ \forall \alpha,\beta \ , 
\mbox{and} \nonumber \\
V^{(1)}_{\alpha\beta}&=&0 , \ \forall \alpha,\beta \ (\alpha \neq\beta)
\ ,
\label{cond1}
\end{eqnarray}
then  $\epsilon^{(1)}=V^{(1)}_{\alpha\alpha}$, and the degeneracy is
not removed. If any of the conditions above is violated, one can
compute $\epsilon^{(1)}$ and the coefficients $\gamma_\alpha$. At this
point, at least part of, the degeneracy has been removed. 

Let's assume now that Eqs. (\ref{cond1}) are satisfied and investigate
whether degeneracies are removed to second order. That means that 
$\langle g_\alpha | \hat{V} G_0
(\one-\ket{\Psi^{(0)}}\bra{\Psi^{(0)}})\hat{V}| \Psi^{(0)}
\rangle=\epsilon^{(2)} \gamma_\alpha$, or (as can be shown assuming
Eqs. (\ref{cond1})) 
\begin{eqnarray}
\langle g_\alpha | \hat{V} \bar{G}_0 \hat{V}| \Psi^{(0)}
\rangle=\epsilon^{(2)} \gamma_\alpha \ ,
\end{eqnarray}
with $\bar{G}_0 =G_0\hat{P}_\perp$, and $\hat{P}_\perp = \sum_{\eta \in
{\cal S}_0^\perp} \proj{v_\eta}$ a projector onto the subspace
orthogonal to ${\cal M}_0$ ($H_0\ket{v_\eta}=\epsilon_\eta
\ket{v_\eta}$, $\langle v_\eta | v_\xi \rangle=\delta_{\eta\xi}$).
Therefore, the following conditions
\begin{eqnarray}
V^{(2)}_{\alpha\alpha}&=&V^{(2)}_{\beta\beta}, \ \forall \alpha,\beta \ , 
\mbox{and} \nonumber \\
V^{(2)}_{\alpha\beta}&=&0 , \ \forall \alpha,\beta \ (\alpha \neq\beta)
\ ,
\label{cond2}
\end{eqnarray}
with $V^{(2)}_{\alpha\beta}=\langle g_\alpha | \hat{V} \bar{G}_0
\hat{V}| g_\beta\rangle$, guarantee that the degeneracy is not
removed to second order. By induction, if the degeneracy is removed to
order $\ell+1$ in perturbation theory, then the following set of
conditions need to be satisfied ($\forall n \in [1,\ell]$ )
\begin{eqnarray}
V^{(n)}_{\alpha\alpha}&=&V^{(n)}_{\beta\beta}, \ \forall \alpha,\beta \ , 
\mbox{and} \nonumber \\
V^{(n)}_{\alpha\beta}&=&0 , \ \forall \alpha,\beta \ (\alpha \neq\beta)
\label{condl}
\end{eqnarray}
with
\begin{eqnarray}
V^{(n)}_{\alpha\beta}=\langle g_\alpha
|\underbrace{ \hat{V} \bar{G}_0 \hat{V} \dots \bar{G}_0 \hat{V}}_{n
\mbox{ factors } \hat{V}}|
g_\beta\rangle \ ,
\end{eqnarray}
and one of the following conditions
\begin{eqnarray}
V^{(\ell+1)}_{\alpha\alpha}&\neq&V^{(\ell+1)}_{\beta\beta}, \ \mbox{for
some } \alpha,\beta \ , \mbox{or} \nonumber \\
V^{(\ell+1)}_{\alpha\beta}&\neq&0 , \ \mbox{for some } \alpha,\beta
\ (\alpha \neq\beta) \ .
\label{condl1}
\end{eqnarray}
For all these expressions to be analytic we have to assume that there
is no level crossing, i.e. $\epsilon_0\neq \epsilon_\eta$, $\forall
\eta\in {\cal S}_0^\perp=N_g+1,\cdots$.

\section{Generalized Elitzur's theorem}
\label{app2}

We review the results of Ref. \cite{BN} needed for the purposes of this
paper. Reference \cite{BN} showed that ``The absolute mean value of any
local quantity  (involving only a finite number of fields) which is not
invariant under a $d$-dimensional symmetry group  ${\cal G}_d$ of the
$D$-dimensional Hamiltonian $H$ is bounded from above (as well as below
for quantities of fixed sign) by the absolute mean value of the same
quantity computed for a $d$-dimensional Hamiltonian ${\bar H}$ which is
globally invariant under ${\cal G}_d$  and preserves the range of the
interactions''. Non invariant means that the  quantity under 
consideration, $f(\phi_{\bf i})$, has no invariant component:
\begin{equation}
\sum_\bk f[{\bf g}_{{\bf i}\bk}(\phi_{\bf i})] = 0.
\label{nonin}
\end{equation}
For a continuous group, this condition is replaced by $\int f[{\bf
g}_{\bf i}(\phi_{\bf i})] d{\bf g} = 0$. To determine if SSB
occurs, we compute 
\begin{equation}
\langle f(\phi_{\bf i}) \rangle = \lim_{h\rightarrow 0}  \lim_{N_s
\rightarrow \infty}  \langle f(\phi_{\bf i}) \rangle_{h,N_s},
\label{limit}
\end{equation}
where $\langle f(\phi_\bi) \rangle_{h,N_s}$ is the mean value of
$f(\phi_\bi)$ computed on a finite lattice of  $N_s$ sites, and in the
presence of a symmetry breaking field $h$.  Since ${\Lambda}=
\bigcup_{\bl} {\cal C}_\bl$, the site ${\bf i}$ belongs at least to one
set ${\cal C}_\bj$. It is convenient to rename the fields  in the
following way:  $\phi_{\bf i}=\psi_{\bf i}$ if ${\bf i} \notin {\cal
C}_\bj$ and  $\phi_{\bf i}=\eta_{\bf i}$ if ${\bf i} \in {\cal C}_\bj$.
The mean value  $\langle f(\phi_{\bf i}) \rangle_{h,N_s}$ is given by:
\begin{eqnarray}
\label{mast}
&&\langle f(\phi_{\bf i}) \rangle_{h,N_s} =
\frac{\sum_{\{ \phi_{\bf i} \}}  f(\phi_{\bf i}) e^{-\beta
(H(\{\phi_\bi\})+h \sum_{{\bf i}} \phi_{\bf i})}} {\sum_{\{ \phi_{\bf
i} \}} e^{-\beta (H(\{\phi_\bi\})+h \sum_{{\bf i}} \phi_{\bf i})}}=  \\
&&\!\!\!\!\!\!\!\!\frac{\sum_{\{ \psi_{\bf i} \}}  z_{\{ \psi \}}
e^{-\beta h \sum_{{\bf i} \notin {\cal C}_\bj} \psi_{\bf i}}  [\frac
{\sum_{\{ \eta_{\bf i} \} } f(\eta_{\bf i})   e^{-\beta
(H(\{\psi_\bi,\eta_\bi\})+ h \sum_{{\bf i}  \in {\cal C}_\bj} \eta_{\bf
i})}}{z_{{\{ \psi \}}}}]} {\sum_{\{ \psi_{\bf i} \}}  z_{{\{ \psi \}}}
e^{-\beta h \sum_{{\bf i}  \notin {\cal C}_\bj} \psi_{\bf i} }  }
\nonumber 
\end{eqnarray}
where
\begin{eqnarray}
z_{{\{ \psi \}}}= \sum_{\{ \eta_{\bf i} \}}  e^{-\beta 
(H(\{\psi_\bi,\eta_\bi\})+h \sum_{{\bf i}  \in {\cal C}_\bj}\eta_{\bf
i})}.
\end{eqnarray}
From Eq. (\ref{mast}):
\begin{equation}
|\langle f(\phi_{\bf i}) \rangle_{h,N_s}| \leq \left |   
\frac {\sum_{\{ \eta_{\bf i} \} } f(\eta_{\bf i})  
e^{-\beta (H(\{\bar{\psi}_\bi, \eta_\bi\})+ h \sum_{{\bf i} 
\in {\cal C}_\bj}\eta_{\bf i})}}{z_{{\{ {\bar \psi} \}}}} \right |, 
\label{final}
\end{equation}
where $\{ {\bar \psi}_\bi \}$ is the particular configuration  of
fields ${\psi_{\bf i}}$ that maximizes the  expression between brackets
in Eq. (\ref{mast}).  $H(\{\psi_\bi,\eta_\bi\})$ is a $d$-dimensional
Hamiltonian for  the field variables ${\eta_\bi}$ which is invariant
under the  {\it global} symmetry group of transformations
${U}_{\bj\bk}$ over the field $\eta_\bi$.  We can define ${\bar
H}(\{\eta_\bi\}) \equiv H(\{\psi_\bi,\eta_\bi\})$.  The range of the
interactions between the  $\eta$-fields in ${\bar H}(\{\eta_\bi\})$ is
clearly the same  as the range of the interactions between the 
$\phi$-fields in $H(\{\phi_\bi\})$. This completes the demonstration 
of the main theorem in \cite{BN}.  Note that the {\it frozen} variables
${\bar \psi}_{\bf i}$ act like external fields in ${\bar
H}(\{\phi_\bi\})$ which do not break the global symmetry group of
transformations ${U}_{\bj\bk}$. 

{\it Corollary I: Elitzur's theorem} \cite{Elitzur}. Any local
quantity  (i.e. involving only a finite number of fields) which is not
invariant under a local (or $d=0$) symmetry group has a vanishing mean
value at  any finite temperature. This is a direct consequence of  Eq.
(\ref{mast}) and the fact that ${\bar H}(\{\eta_\bi\})$  is a
zero-dimensional Hamiltonian.

{\it Corollary II} \cite{BN}. A local quantity which is not gauge 
invariant under a $d=1$ GLS group has a vanishing mean value at any
finite temperature for  systems with finite range interactions. This 
is a consequence of Eq. (\ref{final}) and the absence of SSB
in one-dimensional Hamiltonians such as ${\bar
H}(\{\eta_\bi\})$ with interactions of finite range and strength. Here,
$f(\eta_{\bf i})$ is a non-invariant quantity under the global symmetry
group ${\cal G}_d$ [see Eq. (\ref{nonin})]. 

{\it Corollary III} \cite{BN}.  In  finite-range systems, local
quantities not invariant  under continuous $d=2$ symmetries  have a
vanishing mean value at any finite temperature. This results  from [Eq.
(\ref{final})] together with the application of the Mermin-Wagner
theorem \cite{Mermin}
\begin{equation}
\!\!\!\!
\lim_{h\rightarrow 0} \lim_{N_s \rightarrow \infty}  \frac {\sum_{\{
\eta_{\bf i} \} } f(\eta_{\bf i})   e^{-\beta
(H(\{\bar{\psi}_\bi,\eta_\bi\})+h \sum_{{\bf i}  \in {\cal C}_\bj} 
\eta_{\bf i})}}{z_{{\{ {\bar \psi} \}}}} = 0.
\end{equation}
We invoked that ${\cal G}_d$ is a continuous symmetry  group of ${\bar
H}(\{\eta_\bi\})$,  $f(\eta_{\bf i})$ is a non-invariant quantity for 
${\cal G}_d$ [see Eq. (\ref{nonin})], and ${\bar H}(\{\eta_\bi\})$ is
a  $d=2$ Hamiltonian that only contains finite range  interactions. 

The generalization of this theorem to the quantum case is 
straightforward if we choose a basis of eigenvectors  of the local
operators linearly coupled to the symmetry  breaking field $h$. Here,
the states can  be written as a direct product  $ |\phi \rangle = |
\psi \rangle \otimes | \eta \rangle$.  Eq. (\ref{final}) is
re-obtained  with the sums replaced by traces over the states $| \eta
\rangle$
\begin{equation}
|\langle f({\phi}_{\bf i}) \rangle_{h,N_s}| \leq \frac {\tr_{\{
\eta_{\bf i} \} } f( \eta_{\bf i})   e^{-\beta
(H(\{\bar{\psi}_\bi,\eta_\bi\})+ h \sum_{{\bf i} \in {\cal C}_\bj}  {
\eta}_{\bf i})}} {\tr_{\{ \eta_{\bf i} \} } e^{-\beta
(H(\{\bar{\psi}_\bi,\eta_\bi\})+h \sum_{{\bf i}  \in {\cal C}_\bj} 
{\eta}_{\bf i})}}, 
\label{quantum}
\end{equation}
In this case, $| {\bar \psi} \rangle$ corresponds to one particular 
state of the basis $| \psi \rangle$ that maximizes the right-hand side
of Eq. (\ref{quantum}). Generalizing standard proofs, e.g. \cite{assa},
one finds a $T=0$ quantum extension of  Corollary III in the presence
of a gap:

{\it Corollary IV} \cite{BN}. If a gap exists in a system possessing a
$d \le 2$-dimensional continuous symmetry in its low energy sector then
the expectation value of  any local quantity not invariant under this
symmetry,  strictly vanishes at $T=0$. As shown in the
present paper, though local order cannot appear, multi-particle
(including TQO) order can exist.  The same holds for emergent discrete
$d \le 1$ symmetries for $T=0^{+}$. (Here, although $T=0$ SSB occurs
wherein in some GSs local observables attain different values,  SSB is
prohibited for $T=0^{+}$.)

{\it Corollary V}  \cite{NBF}.
The absolute values of non-symmetry invariant correlators $|G| \equiv 
|\langle \prod_{\bi \in \Omega_{\bj}} \phi_{\bi} \rangle|$ with 
$\Omega_{\bj} \subset C_{\bj}$ are bounded from above by absolute
values of the same correlators $|G|$  in a $d$-dimensional system
defined by $C_{\bj}$ in the presence of transverse non-symmetry
breaking fields. If no resonant terms appear in the lower-dimensional 
spectral functions (due to fractionalization),  this allows for
fractionalization of non-symmetry  invariant quantities in the 
higher-dimensional system.

{\it Note.}

The existence of a spectral gap in a high-dimensional system may follow
from an exponential bound in gapped low-dimensional  systems in
conjunction with Corollary V  for all non symmetry invariant
correlation functions, e.g. the connected $G^{c}(|{\vec{r}}|) \le  A
\exp(-|{\vec{r}}|/\xi)$. The relation between spectral gaps and
exponential decay of correlations was investigated in \cite{gapproof}
where it was proven that exponentially decaying correlations may imply
a spectral gap.

\section{Peierls problem: The Polyacetylene Story}
\label{app3}

In this Appendix, we first review the trimerized ($m=3$) polyacetylene
chain (Appendix \ref{app3_1}) and then derive new universal symmetry
operators for the general $m$-rized chain  (Appendix \ref{app3_2}). 
These symmetry operators, forming a $\Z_{m}$  group, endow the system
with a {\it primitive} charge quantized in units of $e/m$. Thus, the
Appendix illustrates the relation between symmetry, degeneracy, and
fractionalization. 

\subsection{The trimerized Peierls chain}
\label{app3_1}

We start with the electron-phonon Hamiltonian for a chain of $N_s$
sites introduced by Su, Schrieffer, and Heeger (SSH) \cite{fraction}
\begin{eqnarray}
H = - \!\!\!\!\!\!\sum_{j,\sigma=\uparrow,\downarrow} \!\!\!\!t_{j,j+1}
[c^{\dagger}_{j \sigma} c_{j+1 \sigma}^{\;}  + c^{\dagger}_{j+1
\sigma}  c_{j \sigma}^{\;} ]  +E_{\sl elastic} , 
\label{Peierls}
\end{eqnarray}
where $E_{\sl elastic}=K/2 \sum_j (u_j-u_{j+1})^2$ represents the
(classical) elastic energy with $u_{j}$ denoting the displacement from
equilibrium of the $j$-th atom.  The operators $c_{j\sigma}^{\;}$
($c_{j\sigma}^{\dagger}$) are the electron annihilation (creation)
operators at site $j$ (periodic boundary conditions are assumed, i.e.
$c_{N_s+1\sigma}^{\;}=c_{1\sigma}^{\;}$). By the SSH arguments, the
kinetic hopping term $t_{j,j+1}$ is modulated by the lattice
displacements --- for small separation between the atoms, the hopping
element is enhanced whereas for large separation it decreases. Thus,
$t_{j,j+1} = t_{0} + \alpha (u_{j} - u_{j+1})$ with $\alpha > 0$. Next,
we consider the trimerized case via the Born-Oppenheimer approximation
and set 
\begin{eqnarray}
u_{j} = u \cos (\frac{2 \pi}{3} j - \theta) ,
\label{u.}
\end{eqnarray}
which leads to an elastic energy 
\begin{eqnarray} 
E_{elastic} =  \frac{3Ku^{2}}{4} N_s,
\end{eqnarray}
and to hopping amplitudes
\begin{eqnarray}
t_{1} &=& t_{0} + \sqrt{3} ~\alpha~ u ~\sin(\frac{\pi}{3} - \theta)
\nonumber \\ 
t_{2} &=& t_{0} + \sqrt{3} ~\alpha~ u ~\sin \theta \nonumber \\ 
t_{3} &=& t_{0} + \sqrt{3} ~\alpha~ u~ \sin (\frac{5 \pi}{3} - \theta).
\label{t.}
\end{eqnarray}
With this in hand, note that  $\theta \to \theta + \frac{2 \pi}{3}$
effects the shift
\begin{eqnarray}
t_{1} \to t_{3}, ~ t_{2} \to t_{1}, ~ t_{3} \to t_{2}.
\label{t_permute.}
\end{eqnarray}
Similarly, $\theta \to \theta - \frac{2 \pi}{3}$ leads to 
\begin{eqnarray}
t_{1} \to t_{2}, ~ t_{2} \to t_{3}, ~ t_{3} \to t_{1}.
\end{eqnarray}
In what follows, a shorthand will be employed for the reciprocal
lattice vector $Q \equiv \frac{2 \pi}{3}$ (the lattice constant $a$ is
set to one). Fourier transforming,
\begin{eqnarray}
c_{j\sigma} = \frac{1}{\sqrt{N_s}} \sum_{k} e^{i k x_j}  c_{k\sigma} ,
\end{eqnarray}
the electronic portion of the Hamiltonian reads
\begin{eqnarray}
\!\!H_{kin}\! = \!\!\!\!\!\sum_{k \in {\sf RBZ}, \sigma} \!\!\!\!\left(
\begin{array}{ccc}
c_{k\sigma}^{\dagger}~
c_{k+Q\sigma}^{\dagger}~ 
c_{k+2Q\sigma}^{\dagger} 
\end{array} \right)
{\cal{H}}_{k}  
\left( \begin{array}{c}
c_{k\sigma} \\ 
c_{k+Q\sigma} \\
c_{k+2Q\sigma} 
\end{array} \right) \! ,
\end{eqnarray} 
where ${\sf RBZ}$ is the reduced Brillouin zone with boundaries $\mp
Q/2$, and 
\begin{eqnarray}
{\cal{H}}_{k} =   \left( \begin{array}{ccc}
{\cal E}_0 & V_2 \ e^{- i \theta}& V_1^*\ e^{ i \theta} \\
V_2^*\ e^{ i \theta} &  {\cal E}_1 &  V_0 \ e^{- i \theta} \\
V_1 \ e^{-i \theta} & V_0^*\ e^{ i \theta} &{\cal E}_2
\end{array} \right)\, ,
\label{Hmatrix.}
\end{eqnarray} 
with ${\cal E}_n\!\!=\!-2 t_{0} \cos(k+nQ)$, and $V_n=-i \sqrt{3}
\alpha u \cos (k+ nQ)$.
The eigenenergies are solutions of the cubic equation
\begin{eqnarray}
0&=&E^{3}(k) - [3 t_{0}^{2} + \frac{9}{2} \alpha^{2} u^{2}] E(k) -
\nonumber \\ 
\cos 3k& & \!\!\!\!\!\!\!\!\!\!\![\frac{1}{2} (\sqrt{3}~ \alpha u)^{3}
\sin 3 \theta + \frac{9}{2} \alpha^{2} u^{2} ~ t_{0} - 2t_{0}^{3} ].
\label{energy.}
\end{eqnarray}

If the chain is 2/3 filled (i.e. only two thirds of the $N_s$ sites are
occupied by electrons) then only the lower most band is filled (out of
the three bands) by spin up and down electrons. The net kinetic  energy
per electron is given by 
\begin{eqnarray}
T_{kin} = \frac{1}{2}(E_{1}(k=0) + E_{1}(k=Q/2)) 
\end{eqnarray}
with $E_{1}(k)$ the lowest of the three energy bands given by Eq.
(\ref{energy.}). The net electronic energy is the number of particles
$(2N_s)/3$ times the kinetic energy per electron, i.e. $[E_{kin} \equiv
\frac{2}{3} T_{kin}N_s]$.  The total energy is 
\begin{eqnarray}
E_{tot} = E_{kin} + E_{elastic}.
\end{eqnarray}
Minimizing $E_{tot}$ with respect to $\theta$ we find  the three minima
$\theta_{min}= \frac{\pi}{6},  \frac{5 \pi}{6}, \frac{3 \pi}{2}$.  By
Eq. (\ref{t.}), these correspond to the three possibilities
\begin{eqnarray}
\!\!{\sf I}: t_{1} \!\!&=&\!\! t_{2} = t_{0}+ \frac{\sqrt{3}}{2}~ \alpha  u, ~
t_{3} \!=\! t_{0}  - \sqrt{3}~ \alpha  u  \nonumber \\ 
\!\!{\sf II}: t_{1} \!\!&=&\!\! t_{0} - \sqrt{3}~ \alpha u,~ t_{2} \!=\! t_{3}
\!= \!t_{0}  + \frac{\sqrt{3}}{2} ~ \alpha  u \nonumber \\ 
\!\!{\sf III}: t_{1} \!\!&=& \!\!t_{3} = t_{0}  + \frac{\sqrt{3}}{2}~ \alpha u, ~
t_{2} \!=\! t_{0} - \sqrt{3}~ \alpha u .
\end{eqnarray}
The electronic degrees of freedom ($\{c_{j\sigma}^{\;},
c^{\dagger}_{j\sigma}\}$) are slaved to the distortions $\{u_{j}\}$
caused by the phonons. (This is indeed the essence of the
Born-Oppenheimer  approximation.)

\subsection{New universal symmetry operators in {\it m}-rized Peierls
problems}
\label{app3_2}

To couch our results in a simple familiar setting,  we will in the
spirit of the earlier Appendix focus on the trimerized $(m=3$) chain.
Later on, we will illustrate how to extend these results for arbitrary
$m$-rized chains and derive a universal $m$ independent form of the
symmetry operators. Before  delving into the specifics,  let us make a
few general remarks. First, it is clear that  the electronic only
kinetic portion of each of the Hamiltonians
\begin{eqnarray}
H_{kin} = -\!\!\!\!\!\sum_{j, \sigma=\uparrow,\downarrow} t_{j,j+1}
[c^{\dagger}_{j \sigma} c_{j+1 \sigma}^{\;}  + c^{\dagger}_{j+1
\sigma}  c_{j \sigma}^{\;} ]
\end{eqnarray}
for the three distortion patters ($\alpha = {\sf I}, {\sf II}, {\sf
III}$) is non-degenerate. Henceforth, we will label $H_{kin}$ for the
hoppings $\{t_{j, j+1} \}$ dictated by $\{u_{j}(\theta)\}$ by
$H_\alpha$.  The two thirds full electronic state GS for each of the
corresponding electron-only Hamiltonians is unique and given by
\begin{eqnarray}
| \psi_{\alpha} \rangle = \prod_{k \in {\sf RBZ},\sigma} \beta_{1; k
\sigma; \alpha}^{\dagger}  | 0 \rangle,
\end{eqnarray}
where, in each case, we occupy all of the  states of the lowest band
($E_{1}(k)$).  Note, however, that each single particle  eigenstate
$\beta_{1; k \sigma; \alpha}^{\dagger}  | 0 \rangle$ is different for
each  of the Hamiltonians $H_{\alpha}$. Notwithstanding, the single and
multiple particle  spectra of $H_{\alpha}$ are all the same. This
assertion follows from a relabeling of all indices
(i.e. in going from ${\sf II} \to {\sf I}$ we may define $c_{j +1
\sigma} \to d_{j \sigma}$ with a fermionic problem in $d_{j \sigma}$ 
whose Hamiltonian is given precisely by  $H_{\sf I}[d_{j \sigma}^{\;},
d^{\dagger}_{j \sigma}]$ in lieu of $H_{\sf I}[c_{j \sigma}^{\;},
c^{\dagger}_{j \sigma}]$. As both sets of fermionic operators obey the
same algebra, the spectra  of $H_{\sf II}$ and $H_{\sf I}$ are
identically the same.) The transformations between $H_{\sf I} \to
H_{\sf II} \to H_{\sf III}$ amount to discrete gauge transformations.
The basis vectors are permuted or relabeled  (e.g.
$c^{\dagger}_{j\sigma} \to d^{\dagger}_{j\sigma}$) yet this change of
variable is trivially innocuous in the spectra.  A moment's reflection
reveals that this gauge transformation is not a true symmetry of the
Hamiltonian (e.g. $[H_{\sf I}, H_{\sf II}] \neq 0$). 

Let us consider the full Hamiltonian (Eq. (\ref{Peierls})) containing
both the lattice degrees of freedom $\{u_{j}\}$ and the  electronic
degrees of freedom $\{c_{j\sigma}, c^{\dagger}_{j\sigma}\}$. It is clear
that as the electronic states are slaved to the elastic deformations,
the existence of three GSs of the  elastic deformations leads to three
GSs of the  entire Hamiltonian. 

Next, we construct operators linking the different GSs. Consider the
unitary operator
\begin{eqnarray}
U_{el} &=&  \prod_{k \in {\sf RBZ}, \sigma} U_{k\sigma}  \ , \mbox{
where} \nonumber \\
U_{k\sigma}&=&\exp[i\sum_{n=0}^{2} (k+nQ) \ n_{k+nQ \sigma}] ,
\label{uel}
\end{eqnarray}
with $n_{k \sigma} = c^{\dagger}_{k \sigma} c_{k \sigma}^{\;}$. It is
readily verified that 
\begin{eqnarray}
U_{el} H_{kin}(\theta) U_{el}^{\dagger} = H_{kin}(\theta + Q).
\end{eqnarray}
In particular, it is clear that $U_{el}^{\;}H_{\sf I}U_{el}^{\dagger} =
H_{\sf II}$, ~ $U_{el}^{\;}H_{\sf II}U_{el}^{\dagger} = H_{\sf III}$,
and $U_{el}^{\;}H_{\sf III}U_{el}^{\dagger} = H_{\sf I}$ (whence we
remind the reader that $\theta_{min} = \frac{\pi}{6}, \frac{5 \pi}{6},
\frac{3 \pi}{2}$ correspond to ${\sf I,II, III}$).

The full degeneracy between the different lattice distortions and the
electronic configurations slaved to them is captured by an operator
relating $\theta \to \theta+Q$ for any of the pertinent values of
$\theta$ (and ensuing lattice distortions $u_{j}$ given by Eq.
(\ref{u.})) and at the same time shifting the electronic operators
accordingly. With the hopping amplitudes left explicitly as functions
of $\theta$ ~ (i.e. $t_{j,j+1}= t_{j}(\theta)=t_j$) the unitary
operator
\begin{eqnarray}
U_{\theta} =  \exp[ i Q P_{\theta}]= \exp[Q \frac{d}{d \theta}]  
\label{thetanew}
\end{eqnarray}
shifts $\theta$ by $Q$. [When the 
Baker-Campbell-Hausdorff identity is applied to Eq. (\ref{Hmatrix.}),
we attain 
$U_{\theta}^{\;} H_{kin} U^{\dagger}_{\theta} = H_{kin}(\theta + Q)$].
With Eqs. (\ref{t_permute.}) in mind, the operation 
\begin{eqnarray}
U_{\theta}^{\;} H_{kin}[t_{1}, t_{2}, t_{3}] U_{\theta}^{\dagger} =
H_{kin}[t_{3}, t_{1}, t_{2}].
\end{eqnarray} 
Thus, it is clear (by a similar relabeling as before with the fermionic
$c \to d$) that, for any $\theta$, the spectra of $U_{\theta}^{\;}
H_{kin}[t_{1}, t_{2}, t_{3}]  U_{\theta}^{\dagger}$ is identical to
that of $H_{kin}[t_{1}, t_{2}, t_{3}]$.

By compounding both lattice distortions and  operations on the
electronic operators  we generate the full symmetry of the Hamiltonian
\begin{eqnarray}
U = U_{\theta}^{\dagger}  U_{el}^{\;}= \exp[ - i Q P_{\theta}] \
\prod_{k \in {\sf RBZ},  \sigma}  U_{k\sigma}.
\label{com}
\end{eqnarray}
The operation of $U_{\theta}^{\dagger}$ undoes the  shift $\theta \to
\theta + Q$ effected by $U_{el}$. Trivially, $UH(t_{1},
t_{2},t_{3})U^{\dagger} =H(t_{1}, t_{2},t_{3})$ or, stated otherwise,
$[H,U]=0$.  We immediately see that $\{1, U, U^{2} \}$ form a $\Z_{3}$
group. As a byproduct at the special values of $\theta_{min} =
\frac{\pi}{6}, \frac{5 \pi}{6}, \frac{3 \pi}{2}$  the operator $U$
links the degenerate Hamiltonians $H_{kin}$  between states ${\sf I}
\to {\sf II}$, ${\sf II} \to {\sf III}$, and ${\sf III} \to {\sf I}$.
$U$ is a unitary operator living within each triplet of momentum values
($k, k+Q, k+2Q$) and corresponding $\theta$ values $(\theta, \theta+ Q,
\theta+ 2Q)$. $U$ is a direct product of $SU(3)$ operators for each
value of $0\le k < Q$ and $0 \le \theta< Q$. Following the same
prescription for other periods of the Peierls distortion, we find a
{\em universal} symmetry which holds regardless of the value of $m$
($m=2$ in the dimerized case, $m=3$ in  the trimerized chain and so
on). Very generally, the  $d=1$ connecting operators
\begin{eqnarray}
U_{el} =  \prod_{k \in {\sf BZ}, \sigma} U_{k\sigma} \ , \mbox{ where }
 \ U_{k\sigma}=\exp[ik n_{k \sigma}].
\label{uel_general}
\end{eqnarray}
Here, the product spans the entire Brillouin zone (${\sf BZ}$) with $k
\in (-\pi, \pi]$ regardless of the size of the ${\sf RBZ}$ (determined
by $m$). In Eq. (\ref{uel_general}), the operator  displacing the
electrons by one lattice constant  is $U_{el}= \exp[iP_{tot}]$ with
$P_{tot}$ the total momentum.

We just listed the symmetry operators for the electronic degrees of
freedom via the $k$-space representation of Eq. (\ref{uel_general}). To
make the string correlators slightly more transparent, we now transform
$U_{el}$ to real space. On a chain of length $N_s$ (with $N_{s}$ even),
\begin{eqnarray}
\sum_{k \in {\sf BZ}} k \ n_{k\sigma} = \sum_{i,j} K_{ij} 
c_{i\sigma}^{\dagger} c_{j\sigma}^{\;} ,
\end{eqnarray}
with kernel ($K_{ii}= \frac{\pi}{N_{s}}$, $y = x_{j} - x_{i}$, and $\epsilon
=\frac{2 \pi}{N_s}$) 
\begin{eqnarray}
K_{ij} &=& \frac{1}{N_s} \sum_{k \in {\sf BZ}} k \ e^{iky} =  - i
\partial_{y} \frac{1}{N_s} \sum_{n=-N_s/2+1}^{N_s/2} e^{i n \epsilon
y}  \nonumber \\ 
&=& - i  \partial_{y} \frac{1}{N_s}[ 
\frac{e^{iN_{s} \epsilon y/2} - e^{- i N_{s} \epsilon y/2}}
{e^{i \epsilon y} - 1}] \\ 
&=&\frac{\epsilon}{e^{i \epsilon y} - 1}[\cos \frac{N_{s}\epsilon
y}{2}  - \frac{2i}{N_{s}} \frac{e^{i \epsilon y}} {e^{i \epsilon y} -
1} \sin \frac{N_{s} \epsilon y}{2}] \nonumber ,
\label{finiteLkij1}
\end{eqnarray}
which at $\epsilon N_{s} = 2 \pi$ reduces to
\begin{eqnarray}
K_{ij} = \frac{\epsilon}{e^{i \epsilon y} -1} (-1)^{y} ,
\end{eqnarray}
and, in the thermodynamic limit (for $y \ll N_{s}$), simplifies to
\begin{eqnarray}
K(y)  \to -\frac{i}{y} (-1)^{y}.
\end{eqnarray}

Putting all of the pieces together,
\begin{eqnarray}
U_{el}  = \prod_{\sigma}\exp[i \sum_{i,j} K_{ij} c_{i\sigma}^{\dagger}
c_{j\sigma}^{\;}]
\end{eqnarray}
with (for $y \ll N_{s}$),
\begin{eqnarray} 
\!\!\!\!\!\!\!\!\!\!\!\!\!\!\!\!K_{ij} \!\!\!&=&\!\!\! 
\frac{(-1)^{x_{i}-x_{j}}i}{x_{i} - x_{j}}. 
\label{kijnet}
\end{eqnarray}
Taking into account the overall parity modulation $(-1)^{x_{i}- x_{j}}$
on the two sublattices,  we have a {\em universal} ($m$-independent) 
form for the symmetry operators of the $m$-rized  Peierls chain,
\begin{eqnarray}
U_{el} =&&\prod_{\sigma} \exp \Big[i \sum_{r,s} \left( \begin{array}{cc}
u_{r\sigma}^{\dagger}~
v_{r\sigma}^{\dagger} 
\end{array} \right) \nonumber \\
&&\left( \begin{array}{cc} 
\frac{\epsilon}{e^{i \epsilon y_{1}}-1} &
- \frac{\epsilon}{e^{i \epsilon y_{2}}-1} \\
- \frac{\epsilon}{e^{i \epsilon y_{3}}-1} & 
\frac{\epsilon}{e^{i \epsilon y_{1}} -1}
\end{array} \right) 
\left( \begin{array}{c}
u_{s\sigma}\\ 
v_{s\sigma} 
\end{array} \right) \Big].
\label{mix}
\end{eqnarray}   
Here, $u_{r\sigma} \equiv c_{i=2r\sigma}, v_{r\sigma} \equiv
c_{i=2r+1\sigma}$,  $u_{s\sigma} \equiv c_{j=2s\sigma}$, and
$v_{s\sigma} \equiv  c_{j=2s+1\sigma}$ are the electronic annihilation
operators on the even/odd sites. The coordinates $y_{1} = 2(r-s)$,
$y_{2} = 2(r-s)+1$, and $y_{3}= 2(r-s)-1$  which may further be recast
in terms of the familiar Pauli matrices (or {\it two-dimensional Dirac
matrices}  encountered in field theoretic formulations of the SSH
problem --- similar to those appearing in massless chiral problems). 

The off-diagonal character of Eq. (\ref{mix}) coupling even and odd
sites is reminiscent to the coupling between such two chiral field in
the field theoretical treatment of the SSH problem where the anomalous
behavior and fractional Fermi number are partially brought about by off
diagonal Dirac operators.

\section{A lightning review of Fractional Quantum Hall systems }
\label{app4}

\subsection{Symmetries}

Assume that the electronic system resides on a surface with periodic
boundary conditions, i.e. a {\it flat} torus. The group symmetry
operators  that leave the Hamiltonian of Eq.~(\ref{QHSS}) invariant are
the magnetic translations which can be written as 
\begin{equation}
T({\vec{a}})  = \prod_{i=1}^{N_{e}} t_{i}(\vec{a}) ,
\end{equation}
with $\vec{a}=a_x \hat{e}_{x}+ a_y \hat{e}_{y}$ a vector in the
plane, $t_i({\vec{a}}) = \exp[\frac{i}{\hbar}\vec{k}_i \cdot \vec{a}]$,
and $\vec{k}_i = \vec{\Pi}_i + \frac{e}{c} \vec{B}\wedge \vec{r}_i $
the generators of magnetic translations. With the condition that the
magnetic flux is an integer number of flux quanta $\phi_0$, i.e.
$N_{\phi} \phi_{0}$, 
\begin{equation}
T_{1}^n = (T(L_{x} \hat{e}_{x}/N_{\phi}))^n, \ \ \ T_{2}^m  = (T(L_{y}
\hat{e}_{y}/N_{\phi}))^m ,
\end{equation}
with $m$, and $n$ integers ($L_xL_{y}/N_{\phi} = 2 \pi \ell^{2}$, where
$\ell= \sqrt{\hbar c/(|e|B)}$ is the magnetic length), are translations
which respect the boundary conditions and satisfy
\begin{equation}
T_{1}T_2 = \exp[-i 2\pi \nu] T_2T_1,
\label{magt}
\end{equation}
where the filling fraction $\nu=N_e/N_\phi=p/q$ with $p$ and $q$
mutually prime integers. Most notably, for $\nu<1$, one set of these
operators $\{T_{1}^{n}\}_{n=0,1, \cdots, q-1}$ or $\{T_{2}^{m}\}_{m=0,1,
\cdots, q-1}$ suffices to link all GSs to each other, i.e. the GS subspace
is at least $q$-fold degenerate. Since continuous $d=D=2$ symmetries of
magnetic translations along a given space direction  (which are a
subset) suffice to link all GSs, TQO follows from our theorem.
Therefore, there is no (local) order parameter that characterizes a FQH
liquid state. 


To understand the topological degeneracy of {\it real} FQH liquids
(i.e, including, for example, random potentials) Wen and Niu \cite{weniu}
assumed well-defined quasiparticle-quasihole elementary excitations
with fractional statistics, and finite energy gaps. The {\it symmetry}
operations they considered correspond to $d=1$ quasiparticle-quasihole
propagation  ${\cal{T}}_{x,y}^{n}$ an integer ($n$) number of  cycles
around the toric axis (which are further related to  integer flux
insertions along each of the toric directions). It is assumed that such
processes map GSs to GSs and, by definition, satisfy an equation
similar to Eq. (\ref{magt}). The operators ${\cal{T}}$ linking the
various states lead, for finite $L_{x,y}$, to an exponential splitting
between the various GSs of the infinite $L_{x,y}$ system. The
considerations discussed in \cite{weniu} mirror those which we
elaborated on in the Appendix on the Quantum Dimer Model, although in
the latter case we could write down the linking operators in terms of
the original degrees of freedom. 

Clearly, real FQH systems are not exactly represented by the $H$ of Eq.
(\ref{QHSS}). The presence of impurities, for example, spoils the
invariance of the Hamiltonian under magnetic translations. But this is
not a problem as long as the perturbations added to $H$ do not close
the excitation gap. Of course, those perturbations (which are not
invariant under magnetic translations) will remove the GS degeneracy;
however, as we have shown, the splitting will be exponentially small in
the system size and will eventually vanish in the thermodynamic limit.
This is indeed the essence of TQO, and the $H$ of Eq. (\ref{QHSS}) is a
minimal {\it fixed-point} Hamiltonian capturing the TQO of FQH systems.
We emphasize that the key point to prove existence of TQO in FQH
liquids is the realization of $d=2$ continuous symmetries connecting
{\it all} possible GSs which, by our theorem, cannot be broken. 
. 
\subsection{Reduction to an effective one-particle problem}

In terms of the effective field theory of \cite{zhk},  the Lagrangian
density is
\begin{eqnarray}
{\cal{L}} &=&  \Big[ \phi^{*}(i \partial_{0}- a_{0} - eA_{0}) \phi  -
\frac{1}{2m} \phi^{*}(i \partial_{i} - a_{i} - eA_{i})^{2} \phi
\nonumber \\
&+& \mu |\phi|^{2} - \lambda |\phi|^{4} \Big] + \Big[ - \frac{1}{4 \pi
q}  \epsilon^{\mu \nu \lambda} a_{\mu} \partial_{\nu} a_{\lambda} \Big]
\nonumber \\ 
&\equiv& {\cal{L}}_{\phi} + {\cal{L}}_{a}.
\end{eqnarray}  
Under the decomposition into global and local excitations, $a_{i} +
eA_{i} = \frac{\theta_{i}}{L_{i}} + \delta a_{i}(x,t)$, this leads to 
\begin{eqnarray}
Z= \int D \theta  \exp \Big[ i \int dt \frac{1}{4 \pi q}   (\theta_{1}
\frac{d}{dt} \theta_{2} - \theta_{2} \frac{d}{dt}  \theta_{1})]
\nonumber \\ 
\times \int Da_{0} D \delta a_{i} D \phi  \exp[i \int d^{3}x 
[{\cal{L}}_{\phi}(a_{\mu}, \phi)  + {\cal{L}}_{a}(\delta a_{\mu})]]
\Big]. \nonumber
\end{eqnarray}
This is invariant under the combined transformations \cite{weniu}
\begin{eqnarray}
\phi &\to& \phi' = \exp \Big[ -2 \pi i  \Big( \frac{p_{1} x_{1}}{L_{1}} +
\frac{p_{2} x_{2}}{L_{2}} \Big) \Big] \phi, \nonumber \\ 
(\theta_{1}, \theta_{2}) &\to& (\theta_{1} + 2 \pi p_{1}, \theta_{2} + 2
\pi p_{2}),
\label{qhg}
\end{eqnarray}
with integer $p_{1}$ and $p_{2}$ such that $\phi'$ is single valued. 
The first of Eqs. (\ref{qhg}) is, in the thermodynamic limit,  a
continuous $d=2$ $U(1)$ symmetry while the second is a  discrete $d=0$
transformation.
The two global coordinates ($\theta_{1,2}$) account for non-energetic
excitations. Effectively, all short-range fluctuations $\delta a_{i}$
can be thrown out. The resulting problem has only two fields
$\theta_{1}$ and $\theta_{2}$.  In the low-energy sector the
Hamiltonian $H= \Pi_{x}^{2} + \Pi_{y}^{2}$, where $\Pi_{\mu} =
\partial_{\theta_{\mu}}  - i A_{\theta_{\mu}}$ and the gauge field
corresponds to $m$ flux quanta threading the torus parameterized by
$(\theta_{1}, \theta_{2})$, $B = m/(2 \pi)$. The Hamiltonian is now a
single particle operator which is invariant under the magnetic
translation group symmetry. The magnetic translation group cannot  be
broken for this $d=0$ (single particle) problem. The degeneracy is
clearly invariant under the magnetic translations linking the different
GSs.  This is a central result found by Wen and Niu \cite{weniu}. Local
deformations of the $a$ field cost finite energy and can  therefore be
removed in the low energy sector. We stress that  if we do not restrict
ourselves to the lowest-energy sector then the $d=1$ (or,  more
precisely, the $d=1+1$) quasi-particle evolution  around a toric cycle
is the symmetry operator of the  original  many-body problem.

\subsection{Reduction to a one-dimensional problem with discrete $d=1$
symmetries}
\label{1dQh}

When placed on a thin torus, the $D=2$ Quantum Hall system  reduces, in
the limit in which  the thickness of the torus tends to zero, to a
$D=1$ problem \cite{Karl,sdh}. This infinitely thin limit of the torus
is adiabatically connected to that of  the original $D=2$ systems
where  both of its sides ($L_{x}$ and $L_{y}$) are comparable. As
recently noted in \cite{Karl,sdh}, anyons in Quantum Hall systems may
be regarded as domain walls in $D=1$ systems for which we derived for
the $d=1$ symmetry operators  of the Peierls chain \cite{fraction} in
the previous Appendix.

\section{The Quantum Dimer Model and its relation to Kitaev's Toric code 
model}
\label{QDMlink}

Other systems have GSs identical in form to that of the the Kitaev
model (Eq. (\ref{soln2Kit})) albeit only at isolated points.  
These systems do not have Hamiltonians which conform to 
those of the generalized Kitaev
type of Eq.~(\ref{tH}). The consequence
of TQO generally follows for all states of the form of Eq.~(\ref{gentop}). 
One of these systems is the Quantum dimer model \cite{RK} which 
recently has been looked at anew in the context of
quantum computing in a spirit very similar to
Kitaev's Toric code model, e.g. \cite{mpm}. 

On a triangular lattice this model takes the  form \cite{TQDM},
\begin{eqnarray} 
\tilde{H} &=& -t \tilde{T}+ v \tilde{V}
=
\sum_{i=1}^{N_p}\left\{ -t
\sum_{\alpha=1}^3 \left(|\setlength{\unitlength}{3158sp}%
\begingroup\makeatletter\ifx\SetFigFont\undefined%
\gdef\SetFigFont#1#2#3#4#5{%
  \reset@font\fontsize{#1}{#2pt}%
  \fontfamily{#3}\fontseries{#4}\fontshape{#5}%
  \selectfont}%
\fi\endgroup%
\begin{picture}(319,210)(517,-186)
\thinlines
\multiput(720,-10)(1.54799,-2.57998){58}{\makebox(2.0833,14.5833){\SetFigFont{5}{6}{\rmdefault}{\mddefault}{\updefault}.}}
\multiput(718,-10)(-1.55263,-2.58772){58}{\makebox(2.0833,14.5833){\SetFigFont{5}{6}{\rmdefault}{\mddefault}{\updefault}.}}
\multiput(548,-10)(1.55263,-2.58772){58}{\makebox(2.0833,14.5833){\SetFigFont{5}{6}{\rmdefault}{\mddefault}{\updefault}.}}
\multiput(718,-10)(-3.00000,0.00000){58}{\makebox(2.0833,14.5833){\SetFigFont{5}{6}{\rmdefault}{\mddefault}{\updefault}.}}
\multiput(805,-158)(-3.00000,0.00000){58}{\makebox(2.0833,14.5833){\SetFigFont{5}{6}{\rmdefault}{\mddefault}{\updefault}.}}
\thicklines
\put(720, -9){\circle{28}}
\put(803,-153){\line(-1, 0){171}}
\put(547, -9){\circle{28}}
\put(722, -9){\line(-1, 0){172}}
\put(800,-152){\circle{28}}
\put(627,-152){\circle{28}}
\end{picture}
\rangle
\langle \setlength{\unitlength}{3158sp}%
\begingroup\makeatletter\ifx\SetFigFont\undefined%
\gdef\SetFigFont#1#2#3#4#5{%
  \reset@font\fontsize{#1}{#2pt}%
  \fontfamily{#3}\fontseries{#4}\fontshape{#5}%
  \selectfont}%
\fi\endgroup%
\begin{picture}(317,216)(525,-421)
\thinlines
\multiput(720,-235)(1.54799,-2.57998){58}{\makebox(2.0833,14.5833){\SetFigFont{5}{6}{\rmdefault}{\mddefault}{\updefault}.}}
\multiput(718,-235)(-1.55263,-2.58772){58}{\makebox(2.0833,14.5833){\SetFigFont{5}{6}{\rmdefault}{\mddefault}{\updefault}.}}
\multiput(548,-235)(1.55263,-2.58772){58}{\makebox(2.0833,14.5833){\SetFigFont{5}{6}{\rmdefault}{\mddefault}{\updefault}.}}
\multiput(718,-235)(-3.00000,0.00000){58}{\makebox(2.0833,14.5833){\SetFigFont{5}{6}{\rmdefault}{\mddefault}{\updefault}.}}
\multiput(805,-383)(-3.00000,0.00000){58}{\makebox(2.0833,14.5833){\SetFigFont{5}{6}{\rmdefault}{\mddefault}{\updefault}.}}
\thicklines
\put(643,-386){\circle{28}}
\multiput(809,-386)(-5.48713,9.14522){17}{\makebox(8.3333,12.5000){\SetFigFont{7}{8.4}{\rmdefault}{\mddefault}{\updefault}.}}
\put(556,-237){\circle{28}}
\multiput(644,-388)(-5.53125,9.21875){17}{\makebox(8.3333,12.5000){\SetFigFont{7}{8.4}{\rmdefault}{\mddefault}{\updefault}.}}
\put(807,-385){\circle{28}}
\put(720,-234){\circle{28}}
\end{picture}
|+h.c. \right) \right. \nonumber
\\ 
&+& \left. v
\sum_{\alpha=1}^3  \left( |\setlength{\unitlength}{3158sp}%
\begingroup\makeatletter\ifx\SetFigFont\undefined%
\gdef\SetFigFont#1#2#3#4#5{%
  \reset@font\fontsize{#1}{#2pt}%
  \fontfamily{#3}\fontseries{#4}\fontshape{#5}%
  \selectfont}%
\fi\endgroup%
\begin{picture}(319,210)(517,-186)
\thinlines
\multiput(720,-10)(1.54799,-2.57998){58}{\makebox(2.0833,14.5833){\SetFigFont{5}{6}{\rmdefault}{\mddefault}{\updefault}.}}
\multiput(718,-10)(-1.55263,-2.58772){58}{\makebox(2.0833,14.5833){\SetFigFont{5}{6}{\rmdefault}{\mddefault}{\updefault}.}}
\multiput(548,-10)(1.55263,-2.58772){58}{\makebox(2.0833,14.5833){\SetFigFont{5}{6}{\rmdefault}{\mddefault}{\updefault}.}}
\multiput(718,-10)(-3.00000,0.00000){58}{\makebox(2.0833,14.5833){\SetFigFont{5}{6}{\rmdefault}{\mddefault}{\updefault}.}}
\multiput(805,-158)(-3.00000,0.00000){58}{\makebox(2.0833,14.5833){\SetFigFont{5}{6}{\rmdefault}{\mddefault}{\updefault}.}}
\thicklines
\put(720, -9){\circle{28}}
\put(803,-153){\line(-1, 0){171}}
\put(547, -9){\circle{28}}
\put(722, -9){\line(-1, 0){172}}
\put(800,-152){\circle{28}}
\put(627,-152){\circle{28}}
\end{picture}
\rangle
\langle \setlength{\unitlength}{3158sp}%
\begingroup\makeatletter\ifx\SetFigFont\undefined%
\gdef\SetFigFont#1#2#3#4#5{%
  \reset@font\fontsize{#1}{#2pt}%
  \fontfamily{#3}\fontseries{#4}\fontshape{#5}%
  \selectfont}%
\fi\endgroup%
\begin{picture}(319,210)(517,-186)
\thinlines
\multiput(720,-10)(1.54799,-2.57998){58}{\makebox(2.0833,14.5833){\SetFigFont{5}{6}{\rmdefault}{\mddefault}{\updefault}.}}
\multiput(718,-10)(-1.55263,-2.58772){58}{\makebox(2.0833,14.5833){\SetFigFont{5}{6}{\rmdefault}{\mddefault}{\updefault}.}}
\multiput(548,-10)(1.55263,-2.58772){58}{\makebox(2.0833,14.5833){\SetFigFont{5}{6}{\rmdefault}{\mddefault}{\updefault}.}}
\multiput(718,-10)(-3.00000,0.00000){58}{\makebox(2.0833,14.5833){\SetFigFont{5}{6}{\rmdefault}{\mddefault}{\updefault}.}}
\multiput(805,-158)(-3.00000,0.00000){58}{\makebox(2.0833,14.5833){\SetFigFont{5}{6}{\rmdefault}{\mddefault}{\updefault}.}}
\thicklines
\put(720, -9){\circle{28}}
\put(803,-153){\line(-1, 0){171}}
\put(547, -9){\circle{28}}
\put(722, -9){\line(-1, 0){172}}
\put(800,-152){\circle{28}}
\put(627,-152){\circle{28}}
\end{picture}
|+
|\setlength{\unitlength}{3158sp}%
\begingroup\makeatletter\ifx\SetFigFont\undefined%
\gdef\SetFigFont#1#2#3#4#5{%
  \reset@font\fontsize{#1}{#2pt}%
  \fontfamily{#3}\fontseries{#4}\fontshape{#5}%
  \selectfont}%
\fi\endgroup%
\begin{picture}(317,216)(525,-421)
\thinlines
\multiput(720,-235)(1.54799,-2.57998){58}{\makebox(2.0833,14.5833){\SetFigFont{5}{6}{\rmdefault}{\mddefault}{\updefault}.}}
\multiput(718,-235)(-1.55263,-2.58772){58}{\makebox(2.0833,14.5833){\SetFigFont{5}{6}{\rmdefault}{\mddefault}{\updefault}.}}
\multiput(548,-235)(1.55263,-2.58772){58}{\makebox(2.0833,14.5833){\SetFigFont{5}{6}{\rmdefault}{\mddefault}{\updefault}.}}
\multiput(718,-235)(-3.00000,0.00000){58}{\makebox(2.0833,14.5833){\SetFigFont{5}{6}{\rmdefault}{\mddefault}{\updefault}.}}
\multiput(805,-383)(-3.00000,0.00000){58}{\makebox(2.0833,14.5833){\SetFigFont{5}{6}{\rmdefault}{\mddefault}{\updefault}.}}
\thicklines
\put(643,-386){\circle{28}}
\multiput(809,-386)(-5.48713,9.14522){17}{\makebox(8.3333,12.5000){\SetFigFont{7}{8.4}{\rmdefault}{\mddefault}{\updefault}.}}
\put(556,-237){\circle{28}}
\multiput(644,-388)(-5.53125,9.21875){17}{\makebox(8.3333,12.5000){\SetFigFont{7}{8.4}{\rmdefault}{\mddefault}{\updefault}.}}
\put(807,-385){\circle{28}}
\put(720,-234){\circle{28}}
\end{picture}
\rangle
\langle \setlength{\unitlength}{3158sp}%
\begingroup\makeatletter\ifx\SetFigFont\undefined%
\gdef\SetFigFont#1#2#3#4#5{%
  \reset@font\fontsize{#1}{#2pt}%
  \fontfamily{#3}\fontseries{#4}\fontshape{#5}%
  \selectfont}%
\fi\endgroup%
\begin{picture}(317,216)(525,-421)
\thinlines
\multiput(720,-235)(1.54799,-2.57998){58}{\makebox(2.0833,14.5833){\SetFigFont{5}{6}{\rmdefault}{\mddefault}{\updefault}.}}
\multiput(718,-235)(-1.55263,-2.58772){58}{\makebox(2.0833,14.5833){\SetFigFont{5}{6}{\rmdefault}{\mddefault}{\updefault}.}}
\multiput(548,-235)(1.55263,-2.58772){58}{\makebox(2.0833,14.5833){\SetFigFont{5}{6}{\rmdefault}{\mddefault}{\updefault}.}}
\multiput(718,-235)(-3.00000,0.00000){58}{\makebox(2.0833,14.5833){\SetFigFont{5}{6}{\rmdefault}{\mddefault}{\updefault}.}}
\multiput(805,-383)(-3.00000,0.00000){58}{\makebox(2.0833,14.5833){\SetFigFont{5}{6}{\rmdefault}{\mddefault}{\updefault}.}}
\thicklines
\put(643,-386){\circle{28}}
\multiput(809,-386)(-5.48713,9.14522){17}{\makebox(8.3333,12.5000){\SetFigFont{7}{8.4}{\rmdefault}{\mddefault}{\updefault}.}}
\put(556,-237){\circle{28}}
\multiput(644,-388)(-5.53125,9.21875){17}{\makebox(8.3333,12.5000){\SetFigFont{7}{8.4}{\rmdefault}{\mddefault}{\updefault}.}}
\put(807,-385){\circle{28}}
\put(720,-234){\circle{28}}
\end{picture}
|\right)
\right\} \ .
\label{RKE}
\end{eqnarray} 
Here, the sum on $i$\ runs over all of the $N_p$\ plaquettes, and the
sum on $\alpha$\ over the three different orientations of the dimer
plaquettes, namely \setlength{\unitlength}{3158sp}%
\begingroup\makeatletter\ifx\SetFigFont\undefined%
\gdef\SetFigFont#1#2#3#4#5{%
  \reset@font\fontsize{#1}{#2pt}%
  \fontfamily{#3}\fontseries{#4}\fontshape{#5}%
  \selectfont}%
\fi\endgroup%
\begin{picture}(282,173)(535,-171)
\thinlines
\multiput(720,-10)(1.54799,-2.57998){58}{\makebox(2.0833,14.5833){\SetFigFont{5}{6}{\rmdefault}{\mddefault}{\updefault}.}}
\multiput(805,-158)(-3.00000,0.00000){58}{\makebox(2.0833,14.5833){\SetFigFont{5}{6}{\rmdefault}{\mddefault}{\updefault}.}}
\multiput(718,-10)(-1.55263,-2.58772){58}{\makebox(2.0833,14.5833){\SetFigFont{5}{6}{\rmdefault}{\mddefault}{\updefault}.}}
\multiput(718,-10)(-3.00000,0.00000){58}{\makebox(2.0833,14.5833){\SetFigFont{5}{6}{\rmdefault}{\mddefault}{\updefault}.}}
\multiput(548,-10)(1.55263,-2.58772){58}{\makebox(2.0833,14.5833){\SetFigFont{5}{6}{\rmdefault}{\mddefault}{\updefault}.}}
\end{picture}
 rotated by 0 and $\pm
60^o$.  We refer to the plaquettes with a parallel pair of dimers  as
flippable plaquettes. As a complete orthonormal basis set we use
$\left\{ \left| c\right> | \ c=1, \cdots, N_c \right\}$, where $\left|
c\right>$\ stands for one of the $N_c$\ possible hard-core dimer
coverings of the triangular lattice. $\hat{V}$\ is diagonal in this
basis, with $\hat{V}\ket{c} \equiv n_{fl}(c)\ket{c}$ measuring the
number, $n_{fl}(c)$, of flippable plaquettes in configuration $c$.
In this system, the $d=1$ invariants correspond to  parity conservation in the
following  way \cite{RK,TQDM}:  draw two non-contractible loops
$C_{1,2}$ passing through the bonds and wind the system in the
horizontal and vertical directions respectively.  The parities of the
number of dimer crossings  $(-1)^{N_{b;c_{i=1,2}}}$ reflect two $d=1$
symmetries. On general lattices, whenever $t=V$ (the so-called 
``Rokhsar-Kivelson
point'') \cite{RK}, the GS in each sector of ($d=1$)
topological numbers is a superposition of all dimer configurations-
each appearing with an equal weight- this GS is precisely of the  form
of Eq. (\ref{soln2Kit}) with the sum now  over all dimer coverings of
the lattice belonging to a specific sector $q$ defined by these $d=1$ 
invariants $q$. The GSs of the Kitaev model  are formally the same
as those of the Quantum Dimer Model at the Rokhsar-Kivelson point of
Eq. (\ref{soln2Kit}) with the states $|c \rangle$ one coding for all
dimer coverings which belonging to a given topological sector- the GSs
are all eigenstates of the {\it topological} $d=1$ operators. Unlike
the square lattice sibling of Eq. (\ref{RKE}), on the triangular lattice
there is a finite range of parameters for which no local order appears.
\cite{TQDM} Formally, in all these cases,
the GSs can be expressed (up to an innocuous 
scale factor ${\cal{N}}$) as a projection of reference states 
($| \phi \rangle$)
onto a given topological sector $q$, $|g \rangle = 
{\cal{N}} \times P_{q}| \phi \rangle$.

We should nevertheless emphasize the difference in  the form of the
Hamiltonian giving  rise to these GSs here and in the Kitaev model. We
cannot view the RK model as a realization of  Eq. (\ref{tH}). Formally,
we may set $vV$ to be $H$ and $G_{i}$ to be the local symmetries (in
fact, additional four dimer moves are necessary to  link all dimer
configurations). Here, however,  the maximal eigenvalues of $\{G_{i}\}$
occur when we have  a maximal number of flippable plaquettes. However, 
a maximal number of flippable plaquettes elevates the energy $V$ to a
maximum. Thus, we cannot follow the same procedure as before. 

Ordered valence bond solid crystal GSs cannot be linked to one another
by the exclusive use of $d=1$ symmetries and as a consequence SSB can
occur and TQO need not appear.  The RK model maps onto a fermionic
model \cite{triangleRK}. The different topological sectors
amount to different boundary conditions (symmetric or antisymmetric 
along the two directions) on the free fermion problem leading to the
partition functions $Z_{\pm, \pm}$. The spectrum is gapped in the case
of the triangular lattice and is gapless for the square lattice
\cite{triangleRK}.  The split in energy between the different
topological sectors is bounded from above by the value of the split for
a one-dimensional ray [see, e.g. the ray along $\hat{R} $ where
${\vec{R}}$ maximizes Eq. (18) in Ref. \cite{triangleRK} ($R = {\cal
{O}}(L)$)]. For a finite size system, tunneling between  the various GSs
of the infinite size system lifts the GS degeneracy. The bound is that
corresponding to tunneling events in a $d=1$ system which are mediated
by a soliton. The relations of \cite{BN} (see Appendix \ref{app2}) 
bound this split from above
by the energy split generated by such  events in a $d$-dimensional
system.   In our context, it is this small value of the split $\Delta$
which leads to an  exponential decay (in the system size) of all local
quantities. In the  thermodynamic limit, these tunneling events lead to
zero modes.  This exponentially small split between the system
with the two  different
boundary conditions is precisely the  same as that for a $d=1$ gapped
system. This forms a $d=1$ analogue of  the bounds of \cite{BN} employed
throughout our work. The  event which bounds the expectation  value of
local observables is the propagation of a soliton along a $d=1$
trajectory. On  a torus, such solitons ({\it visons}) must appear in
pairs. Similar to the FQHE problem discussed in 
Appendix \ref{1dQh}, the trajectory of a vison pair 
around a $d=1$ cycle of the torus
is the operation which links different  GSs in the thermodynamic limit.

\section{Von Neumann Entropy of $SU(N)$ singlets}
\label{app4.5}

In what follows, we provide all of the details underlying the  result
of Eq. (\ref{SUNK}) and related combinatorial relations in Section
\ref{eent}. 

To build some of the intuition, let us first consider an $SU(2)$
singlet,
\begin{eqnarray}
\ket{\psi} = \frac{\ket{01}  - \ket{10}}{\sqrt{2}}.
\label{longsu2}
\end{eqnarray}
Here $\ket{ab}$ denotes a product states of two  particles: $\ket{a b}
= \ket{a}_{1} \otimes \ket{b}_{2}$ ($a,b=\{0,1\}$). The reduced density
matrix 
\begin{eqnarray}
\rho_{1} = \tr_2 |\psi \rangle \langle \psi | = \sum_{a =
\{0,1\}_{2}} \langle a | \psi \rangle \langle \psi| a
\rangle.
\label{longtrace}
\end{eqnarray}
Substituting (\ref{longsu2}) into Eq. (\ref{longtrace}) and employing
$_2\langle 0 | \psi \rangle = - \frac{1}{\sqrt{2}}|1 \rangle_1$ and 
$_2\langle 1 | \psi \rangle = \frac{1}{\sqrt{2}}|0 \rangle_1$ we have 
\begin{eqnarray}
\rho_{1} = \frac{1}{2} ( | 0 \rangle \langle 0 |  + | 1 \rangle
\langle 1 | )  = \left( \begin{array}{cc} \frac{1}{2}  & 0 \\ 
0 & \frac{1}{2} \end{array} \right). 
\label{longrho}
\end{eqnarray}
Here, 
\begin{eqnarray}
S_{1}^{ent} = -\tr_1 [\rho_{1} \ln \rho_{1}] = \ln 2.
\end{eqnarray}
This value is what was employed in the $SU(2)$ analysis of  Section
\ref{pureKlein}.

Next, let us turn to $SU(3)$ singlets ($a,b,c=\{0,1,2\}$). Here, 
\begin{eqnarray}
\ket{\psi} \!=\! \frac{\ket{012} + \ket{201} + \ket{120}  - \ket{102} -
\ket{210}  - \ket{021}}{\sqrt{6}}.
\label{su3sing}
\end{eqnarray}
The basis generated by particles two and three spans $3^2=9$
dimensions. The resulting single particle density matrix
\begin{eqnarray} 
\rho_{1} = \frac{1}{3} (|0 \rangle \langle 0 | + |1 \rangle \langle 1 |
+ | 2 \rangle \langle 2 |),
\end{eqnarray}
leads to  $S_{1}^{ent} = - \tr_1 [\rho_{1} \ln \rho_{1}] = \ln 3$. Similar
analysis for general $SU(N)$ singlets results in  $S_{1}^{ent} = \ln N$. 

Let us next turn to the computation of the entanglement entropy between
two particles and the remaining $(N-2)$ particles in a general $SU(N)$
singlet. For $N=2$, we have the obvious  result that $S_{2}^{ent} = 0$. This
is so as $\rho_2 =  | \psi \rangle \langle \psi|$ and consequently
$S_{2}^{ent} = -\tr_{1,2} [\rho_{2} \ln \rho_{2}] =0$ since $\rho_{2}$ is a
density matrix for a pure state. Alternatively, this can also be seen
by the reciprocity relation $-\tr [\rho_{A} \ln \rho_{A}] = -\tr
[\rho_{B} \ln \rho_{B}]$ if the sets $A$ and $B$ are complementary to
each other: $A \cap B = 0$ and $A \cup B = 1$ - the complete space. The
complementary set to the two particle set in an $SU(2)$ singlet is
empty and consequently $S_{2}^{ent} = S_{0}^{ent} =0$.

When applied to $SU(3)$, the reciprocity relation shows that
$S_{2}^{ent} = S_{3-1}^{ent} = S_{1}^{ent} = \ln 3$. For purposes of  a
generalization that will later follow, let us now  forgo this shortcut
in this particular case and compute  $S_{2}^{ent}$ longhand. Here, we
have from Eq. (\ref{su3sing}) that 
\begin{eqnarray}
\rho_{2} = \tr_{3} | \psi \rangle \langle \psi |,
\end{eqnarray}
with 
\begin{eqnarray}
_3\langle 0 \ket{\psi}  &=& \frac{\ket{12} - \ket{21}}{\sqrt{6}} \\ \nonumber
_3\langle 1 \ket{\psi}  &=& \frac{\ket{20} - \ket{02}}{\sqrt{6}} \\ \nonumber
_3\langle 2 \ket{\psi}  &=& \frac{\ket{01} - \ket{10}}{\sqrt{6}} .
\end{eqnarray}
This then leads to 
\begin{eqnarray}
\rho_{2}  = \frac{1}{6} [ | 0 1 \rangle \langle 01 | +  |1 0 \rangle
\langle 10 | + |12 \rangle \langle 12 | \nonumber \\ 
+ |21 \rangle \langle 21 | + |02 \rangle \langle 02 | + |20 \rangle
\langle 2 0 | ] \nonumber \\ 
- \frac{1}{6} [|0 1 \rangle \langle 1 0 |
+ | 1 0 \rangle \langle 01 | + | 12 \rangle \langle 21 | \nonumber \\ 
+ | 21 \rangle \langle 12 | + |02 \rangle \langle 20 | + |2 0 \rangle
\langle 02 | ].
\end{eqnarray}
In the basis spanned by the states $(|01 \rangle, | 1 0 \rangle, |02
\rangle, |2 0 \rangle, |12 \rangle, | 21 \rangle)$ (in that order) this
leads to the following block diagonal reduced density matrix,
\begin{eqnarray}
\rho_{2} = \frac{1}{6}
\begin{pmatrix}
1  & -1 & 0 & 0 & 0 & 0 \cr 
-1 & 1 & 0 & 0 & 0 & 0 \cr
0 & 0 & 1 & -1 & 0 & 0 \cr
0 & 0 & -1 & 1 & 0 & 0 \cr
0 & 0 & 0 & 0 & 1 & -1 \cr
0 & 0 & 0 & 0 & -1 & 1
\end{pmatrix} . 
\label{longrho}
\end{eqnarray}
The basic block diagonal structure is that of the  matrix 
\begin{eqnarray}
\begin{pmatrix}
1  & -1 \cr 
-1 & 1 
\end{pmatrix} . 
\end{eqnarray}
which has the two eigenvalues $\epsilon_{+} =0$ and $\epsilon_{-} =
2$.  The non-zero eigenvalues of $\rho_{2}$ are thus $\epsilon_{1} =
\epsilon_{2} = \epsilon_{3} = 1/3$. This leads to $S_{2}^{ent} = -
\sum_{i} \epsilon_{i} \ln \epsilon_{i} =  \ln 3$.

In general, an $SU(N)$ singlet $| \psi \rangle$ will belong to the
totally antisymmetric representation of the permutation group $S_{N}$.
Thus, it will contain $N!$ terms in its expansion. Of these, $N!/2$
terms will appears with a $(+)$ sign and $N!/2$ terms will appear with
a $(-)$ sign. All of the coefficients are equal to $1/\sqrt{N!}$. The
matrix $\rho_{2}$ has $\left( \begin{array}{cc} N \\ 2 \end{array}
\right)$ eigenvalues $\{\epsilon_{i}\}$ different from zero. All of
these eigenvalues are equal to one another  and 
\begin{eqnarray}
S_{2}^{ent} = - \sum_{i=1}^{\left( \begin{array}{cc} N \\ 2 \end{array}
\right) } \epsilon_{i} \ln \epsilon_{i}  =  - \left( \begin{array}{cc}
N \\
2 \end{array} \right) \epsilon \ln \epsilon.
\end{eqnarray}
Here, $\epsilon = m \epsilon_{-}/N! = (2m)/N!$ where  $m=(N-2)!$.
Putting all of the pieces together, we have  $\epsilon = 1/\left(
\begin{array}{cc} N \\ 2 \end{array} \right)$ which leads to
$S_{2}^{ent} = \ln \left( \begin{array}{cc} N \\ 2 \end{array}
\right)$. The factor of $m$ originates from the following
considerations. Suppose that we isolate the first  two particles (1 and
2) and trace over the rest $(3,4, \cdots, N)$. For each pair of states,
e.g. (0,1), we will have the two kets  $|01 abc \cdots \rangle$ and
$|10abc \cdots \rangle$. Here,  $a, b, c, \cdots = \{ 2,3,4, \cdots,
(N-1)\}$ with $a \neq b \neq c \cdots$ The number of distinct states
$|01abc \cdots \rangle$ that we have   is $(N-2)! = m$. 

Let us next consider the evaluation of $S_{k}^{ent}$ in an  $SU(N)$
singlet with general $k$. $S_{k}^{ent}$ is given by
\begin{eqnarray}
S_{k}^{ent} = -\tr_{1,2,\cdots,k} [\rho_{k} \ln \rho_{k} ]
\end{eqnarray}
where
\begin{eqnarray}
\rho_{k} =\tr_{k+1,k+2, \cdots, N} | \psi \rangle \langle \psi |.
\end{eqnarray}
We partition each state in the $N$ particle basis, e.g. 
\begin{eqnarray}
|\underbrace{012 \cdots k-1}_{k!} \underbrace{k k+1 \cdots
N-1}_{(N-k)!} \rangle .
\end{eqnarray}
This leads to the following structure for $\rho_{k}$
\begin{eqnarray}
\rho_{k} = \frac{(N-k)!}{k!} \Big( \underbrace{[k! \mbox{ terms}]  [k!
\mbox{ terms}] + \cdots}_{\left( \begin{array}{cc} N \\ k \end{array}
\right) \mbox{ terms}} \Big).
\end{eqnarray}
Here, we employed the shorthand
\begin{eqnarray}
[k! ~\mbox{terms}] &=&|012 \cdots k-1 \rangle + \mbox{all other even
perm.} \nonumber \\
&-& |102 \cdots k-1 \rangle - \mbox{all other odd perm.}\nonumber 
\end{eqnarray}
The submatrix $[k! \mbox{ terms}]  [k! \mbox{ terms}]$ is of the form
\begin{eqnarray}
A = \left( \begin{array}{cc} {\sf One} & -{\sf One}\\ -{\sf One}&
{\sf One} \end{array} \right),
\end{eqnarray}
where the square matrix ${\sf One}$ is given by 
\begin{eqnarray}
{\sf One} = \underbrace{\left( \begin{array}{ccccc}
1 & 1 & 1 & \cdots & 1\\
1 & 1 & 1 & \cdots & 1 \\
. & . & . & \cdots & 1 \\
. & . & . & \cdots & 1 \\
1 & 1 & 1 & \cdots & 1
\end{array} \right)}_{\frac{k!}{2}}.
\end{eqnarray} 
The spectrum of the matrix $A$ is given by 
\begin{eqnarray}
{\sf Spec} \{A\} = \{ \underbrace{0,0,0, \cdots, 0}_{k!-1}, k!\}.
\end{eqnarray}
On the other hand, $\rho_{k}$ is block diagonal with  $\left(
\begin{array}{cc} N \\ k \end{array} \right)$ blocks of the type $A$.
Therefore, the number of non-zero eigenvalues of $\rho_{k}$ is $\left(
\begin{array}{cc} N \\ k \end{array} \right)$.  All of these
eigenvalues are equal to one another: $\epsilon_{1}  = \epsilon_{2} =
\cdots = \epsilon_{\left( \begin{array}{cc} N \\ k \end{array} \right)
} = \epsilon = \frac{1}{ \left( \begin{array}{cc} N \\ k \end{array}
\right) }$. This leads to 
\begin{eqnarray}
\!\!\!\!\!\!S_{k}^{ent} \!=\! - \!\!\!\sum_{i=1}^{\left(
\begin{array}{cc} N \\ k \end{array} \right) } \!\! \epsilon_{i} \ln
\epsilon_{i} = - \left( \begin{array}{cc} N \\ k \end{array} \right)
\epsilon \ln \epsilon = \ln \left( \begin{array}{cc} N \\ k \end{array}
\right),
\end{eqnarray}
as stated in Eq. (\ref{SUNK}).

\section{Non-local string correlators: The case of $S=1$ spin chains}
\label{app5}

\subsection{Symmetries}
\label{app5_1}

Given an $S=1$ chain of length $N_s$, we define the following global
$su(2)$ spin operators
\begin{equation}
S^\mu=\sum_{j=1}^{N_s} S^{\mu}_{j} \ , \ \mbox{ with } \mu=x,y,z . 
\end{equation}

The global symmetry operators, $\exp[i \pi S^{x}]$ and $\exp[i \pi 
S^{z}]$ can be related to global symmetries of non-local order
parameters.  The latter non-local order parameters have longer range
correlations (the string correlators of the Haldane chain). 

With $U_{s} \equiv \prod_{j<k} \exp[i \pi S^{z}_{j} S^{x}_{k}]$, 
($U_s=U_s^\dagger=U_s^{-1}$) 
\begin{eqnarray}
\tilde{S}_{j}^{x} &=& U_{s}S_{j}^{x}U_{s}^{-1} 
= S_{j}^{x}  \exp[i \pi
\!\! \sum_{k=j+1}^{N_s} \!\! S_{k}^{x}], \nonumber \\ 
\tilde{S}_{j}^{y} &=& U_{s}S_{j}^{y}U_{s}^{-1} = \exp[i \pi \sum_{k=1}^{j-1}
S_{k}^{z}] \ S_{j}^{y} \exp[i \pi \!\! \sum_{k=j+1}^{N_s}\!\! 
S_{k}^{x}], \nonumber \\ 
\tilde{S}_{j}^{z} &=& U_{s}S_{j}^{z} U_{s}^{-1} = \exp[i \pi \sum_{k=1}^{j-1} 
S_{k}^{z}] \ S_{j}^{z} ,
\label{transfS}
\end{eqnarray}
with global transformed operators $\tilde{S}^\mu=\sum_{j=1}^{N_s}
\tilde{S}^{\mu}_{j}$

Let us now examine the operators  $P_{z}= \exp[i \pi  \tilde{S}^{z}]$
and $ P_{x}= \exp[i \pi  \tilde{S}^{x}]$. We claim that these operators
are none other than $\exp[ i \pi  S^{z}]$ and $\exp[i \pi  S^{x}]$,
respectively. The calculation is elementary. First consider 
\begin{eqnarray}
\exp[i \alpha S^{z}_j] = 1 +  S^{z}_j( i \sin \alpha) + (S^{z}_j)^{2}
(\cos \alpha -1).
\label{angle}
\end{eqnarray}
For $\alpha = \pi$,
\begin{eqnarray}
\exp[i \pi S^{z}_j] = 1 - 2(S^{z}_j)^{2}.
\end{eqnarray}
By rotational symmetry ($\vec{S}_j=(S^x_j,S^y_j,S^z_j)$)
\begin{eqnarray}
\exp[i \pi \vec{S}_j \cdot \vec{n}] = 1 - 2 (\vec{S}_j \cdot
\vec{n})^{2}.
\end{eqnarray}

Let us turn to the evaluation of the various terms in $P_{x,z}$. From
Eqs. (\ref{transfS}), the exponential transforms as
\begin{eqnarray}
\!\!\exp[ i \pi \tilde{S}_{j}^{x}] \!=\! U_{s} \exp[i \pi S_{j}^{x}]
U_{s}^{-1} \!=\! U_{s}(1- 2 (S_{j}^{x})^{2}) U_{s}^{-1}.
\end{eqnarray}
The only non-trivial term is
\begin{eqnarray}
U_{s} (S_{j}^{x})^{2} U_{s}^{-1}\!\!&=&\!\!  U_{s} S_{j}^{x}
U_{s}^{-1} U_{s} S_{j}^{x} U_{s}^{-1} = (\tilde{S}_{j}^{x})^{2}\nonumber \\
\!\!&=& \!\!(S_{j}^{x} \exp[ i \pi \!\! \sum_{k=j+1}^{N_s} \!\! S_{k}^{x}])
(S_{j}^{x} \exp[ i \pi \!\! \sum_{k=j+1}^{N_s}\!\!  S_{k}^{x}]) \nonumber \\ 
\!\!&=& \!\!(S_{j}^{x})^{2} \exp[i2 \pi \!\! \sum_{k=j+1}^{N_s} \!\! S_{k}^{x}]=
(S_{j}^{x})^{2} ,
\end{eqnarray}
where we invoked that for any integer spin  (including the current
$S=1$ of interest, see also Eq. (\ref{angle})),
\begin{eqnarray}
\exp[ i 2 \pi S_{j}^{x}] = 1.
\end{eqnarray}
Thus, 
\begin{eqnarray}
P_{x} = \exp[ i \pi  \tilde{S}^{x}] = \exp[ i \pi {S}^{x}].
\label{pxd}
\end{eqnarray}
By the same token, $P_{z} = \exp[i \pi  S^{z}]$. It is obvious that
$\{\one_x,P_{x},\one_z, P_{z} \}$ satisfy a $\Z_{2} \times \Z_{2}$
group algebra.  First, note that $P_{x,z}^{2} =\one_{x,z}$. Next, we
immediately see that $[P_{x}, P_{z}]=0$. More explicitly, 
\begin{eqnarray}
\!\!\!\!\!\![P_{x}, P_{z}] &=& [ \prod_{k=1}^{N_s} (1-2(S_{k}^{x})^{2}), 
\prod_{k=1}^{N_s} (1-2(S_{k}^{z})^{2})]. 
\end{eqnarray}
As
\begin{eqnarray}
[(S_{j}^{x})^{2},  (S_{k}^{z})^{2})] = 0 \ \mbox{ for any $j,k$},  
\end{eqnarray}
$P_{x}$ and $P_{z}$ commute and the set $\{\one_x,P_{x},\one_z, P_{z}
\}$ form a $\Z_{2} \times \Z_{2}$ group. The physical meaning of the
operators $P_{x,z}$ is very simple --- a rotation by $\pi$ about the
$S^{x,z}$ axis. It is evident that this is a symmetry for any global
rotation-invariant spin chain. In particular all Hamiltonians
discussed in Ref. \cite{KT}, which includes $H_{\sf AKLT}$, display a
global  $\Z_{2} \times \Z_{2}$ symmetry.  A rotation by $\pi$ about the
internal $S^{y}$ axis may be viewed as the product of a rotation by
$\pi$ about $S^{x}$ (the operator $P_{x}$) and a rotation by $\pi$
about $S^{z}$ ($P_{z})$. Thus, the rotation by $\pi$ about $S^{y}$ does
not constitute a new group element and we merely have a global $\Z_{2}
\times \Z_{2}$ group. Furthermore, $P_{x} P_{z} = P_{z} P_{x}  = P_{y}
\equiv \exp[i \pi  S^{y}]$, with commutators
\begin{eqnarray}
[U_{s}, P_{x}] = [U_{s}, P_{z}] = 0.
\end{eqnarray}
Moreover, each element in the product forming the non-local
transformation $U_{s}$ commutes with the GS connecting operators
$P_{x,z}$, i.e.
\begin{eqnarray}
[\exp[i \pi S_{j}^{z} S_{k}^{x}], P_{x}] = [\exp[i \pi S_{j}^{z}
S_{k}^{x}], P_{z}] =0.
\end{eqnarray}
For  the AKLT Hamiltonian  (Eq. (\ref{AKLT_Ham})) reveals that $[H_{\sf
AKLT},U_{s}] \neq 0$.

\subsection{Local operators distinguishing the AKLT ground states and
connecting operators}
\label{app5_2}

In general, the degeneracy of the GS subspace depends on boundary
conditions. For $H_{\sf AKLT}$, the GS is four-fold degenerate in the
case of open, and non-degenerate in the case of periodic, boundary
conditions.  We now construct local operators which lift the four-fold
degeneracy of the AKLT GSs. We further comment on the relation between
the role of the non-trivial string operator $U_{s}$ and the operators
which links pairs of GSs.

In the string transformed basis, the Hamiltonian
\begin{eqnarray}
\tilde{H}_{\sf AKLT} = U_{s} H_{\sf AKLT} U_{s}^{-1}
\end{eqnarray}
is given by
\begin{eqnarray}
\tilde{H}_{\sf AKLT} =  \sum_{j} [h_{j} + \frac{1}{3} h_{j}^{2}],
\label{tHAKLT}
\end{eqnarray}
where 
\begin{eqnarray}
h_{j}=-S^x_jS^x_{j+1}+S^y_j e^{i\pi(S^z_j+S^x_{j+1})}
S^y_{j+1}-S^z_jS^z_{j+1}  .
\end{eqnarray}
The two-sites Hamiltonian  $[h_{j} + \frac{1}{3} h_{j}^{2 }]$  is
easily diagonalized \cite{KT}. The GS subspace is 4-dimensional and
spanned by the states $\ket{\phi_{\alpha}}_j \otimes
\ket{\phi_{\alpha}}_{j+1}$, $\alpha=1,2,3,4$, where
\begin{eqnarray}
| \phi_{1} \rangle_j  &=& \frac{1}{\sqrt{3}}[ |0 \rangle_j + \sqrt{2} \
| + \rangle_j], \nonumber \\
| \phi_{2} \rangle_j  &=& \frac{1}{\sqrt{3}}[ |0 \rangle_j - \sqrt{2} \
| + \rangle_j] , \nonumber  \\ 
| \phi_{3} \rangle_j  &=& \frac{1}{\sqrt{3}}[ |0 \rangle_j + \sqrt{2} \
| - \rangle_j] , \nonumber \\ 
| \phi_{4} \rangle_j  &=& \frac{1}{\sqrt{3}}[ |0 \rangle_j - \sqrt{2} \
| - \rangle_j],
\end{eqnarray}
and ($_j\langle \phi_\alpha | \phi_\alpha \rangle_j=1$)
\begin{eqnarray}
\!\!\!\![h_{j} + \frac{1}{3} h_{j}^{2}]\ket{\phi_{\alpha}}_j \otimes
\ket{\phi_{\alpha}}_{j+1}=-\frac{2}{3}\ket{\phi_{\alpha}}_j \otimes
\ket{\phi_{\alpha}}_{j+1} .
\end{eqnarray}
In this way the (non-orthogonal) AKLT GSs can be compactly written as
\begin{eqnarray}
\ket{\bar{g}_\alpha} = U_{s} \bigotimes_{j=1}^{N_s} \ket{\phi_{\alpha}}_j =
U_{s} \ket{\tilde{g}_\alpha} .
\label{GSAKLT}
\end{eqnarray}

We may ask the question whether a local (or quasi-local) operator {\it
distinguishes} two orthonormal AKLT GSs. To this end we need to consider,
for example, the two orthonormal GSs
\begin{eqnarray}
\ket{{g}_1} &=&U_{s} \ket{\tilde{g}_1} , \nonumber \\
\ket{{g}_2} &=& U_{s} \frac{3^{N_s}}{\sqrt{9^{N_s}-1}} \left
[\ket{\tilde{g}_3}-\frac{1}{3^{N_s}}\ket{\tilde{g}_1} \right ] ,
\end{eqnarray}
and {\it measure} the $z$-component of the spin in the first lattice
site with the result
\begin{eqnarray}
\frac{2}{3} = \langle g_1 | S^z_1 | g_1 \rangle \neq 
\langle g_2 | S^z_1 | g_2 \rangle =-\frac{2}{3},
\label{localAKLT}
\end{eqnarray}
meaning that we can certainly distinguish these 2 states through {\it
local measurements}. This suggests that the GSs of this $S=1$ spin chain
are not topologically ordered according to the definition of Eq.
(\ref{def.}). 

To see how these GSs are connected, we apply the symmetry generators to
them. In the (tilde) basis, the global $\Z_{2}$ symmetry generator is
$\exp[i \pi \tilde{S}^{x}] = \prod_{j} (1- 2 (\tilde{S}^{x}_{j})^{2})$.
An explicit evaluation reveals that 
\begin{eqnarray}
 (1- 2 (\tilde{S}^{x}_{j})^{2}) | \phi_{1} \rangle_j &=& - |\phi_{3} 
\rangle_j, \nonumber \\ 
(1- 2 (\tilde{S}^{x}_{j})^{2}) | \phi_{2} \rangle_j &=& - |\phi_{4} 
\rangle_j,  \nonumber \\ 
(1- 2 (\tilde{S}^{x}_{j})^{2}) | \phi_{3} \rangle_j &=& - | \phi_{1} 
\rangle_j, \nonumber \\ 
(1- 2 (\tilde{S}^{x}_{j})^{2}) | \phi_{4} \rangle_j &=& - | \phi_{2}
\rangle_j.
\end{eqnarray}
Similarly, 
\begin{eqnarray}
 (1- 2 (\tilde{S}^{z}_{j})^{2}) | \phi_{1} \rangle_j &=&  |\phi_{2} 
\rangle_j, \nonumber \\ 
(1- 2 (\tilde{S}^{z}_{j})^{2}) | \phi_{2} \rangle_j &=&  |\phi_{1} 
\rangle_j,  \nonumber \\ 
(1- 2 (\tilde{S}^{z}_{j})^{2}) | \phi_{3} \rangle_j &=&  | \phi_{4} 
\rangle_j, \nonumber \\ 
(1- 2 (\tilde{S}^{z}_{j})^{2}) | \phi_{4} \rangle_j &=&  | \phi_{3}
\rangle_j.
\end{eqnarray}
 Thus, the global $\Z_{2} \times \Z_{2}$ symmetry operators $\{P_{x},
P_{z}\}$ link the GSs to each other and are thus named {\it connecting
operators}.

\subsection{String correlators and string operators as {\it partial
polarizers}}
\label{app5_3}

The objective of this Appendix is to flesh out the details of  the
reason why the string correlator is always larger or equal to the usual
N\'eel correlator.  We use general classes of states to prove it
instead of Hamiltonians; we find this procedure more transparent and
also more easily generalizable to other problems. In this way, we
extend and highlight physical aspects of a previous theorem by Kennedy
and Tasaki \cite{KT}.

Consider the parent states represented by a string of length $(N_s-M)$
of $\pm$ followed by a string of length $M$ of zeros,
\begin{eqnarray}
|\mu_{M}^{1} \rangle &=& |\underbrace{+-+-+ \cdots -}_{N_s-M} \underbrace{00
\cdots0}_{M} \rangle , \nonumber \\ 
|\nu_{M}^{1} \rangle &=& |\underbrace{-+-+- \cdots +}_{N_s-M} \underbrace{00
\cdots0}_{M} \rangle.
\end{eqnarray}
In what follows, the appearance of the zeros in these states allows
for {\it partial polarization} (vis a vis {\it full polarization} if
the zeros were absent). Similarly, consider states with all possible
reshuffling of the zeros. For example, the states
\begin{eqnarray}
| \mu_{M}^{2} \rangle &=&|+0-+-+- \cdots - 0 \cdots 0 \rangle, \nonumber
\\
| \nu_{M}^{2} \rangle &=&|-0+-+-+ \cdots + 0 \cdots 0 \rangle,
\end{eqnarray}
and numerous others. Thus, for a fixed  number $M < N_s$ of zeros there
are  $\begin{pmatrix} N_s \\ M \end{pmatrix}=N_M$ states
$|\mu_{M}^{\alpha} \rangle$ and $N_M$ states $|\nu_{M}^{\alpha}
\rangle$ with $\alpha = 1,2, \cdots, N_M$, with the property $\langle
\eta_M^\alpha | \xi_M^{\beta} \rangle=\delta_{\alpha
\beta}\delta{\eta\xi}$, and $\hat{M} \ket{\eta_M^\alpha}=M
\ket{\eta_M^\alpha}$ ($\eta(\xi)=\mu,\nu$, and the operator
$\hat{M}=\sum_{j=1}^{N_s}(1-(S^z_j)^2)$ counts the number of zeros).
Moreover, they satisfy 
\begin{eqnarray}
P_z \ket{\eta_M^\alpha} &=& (-1)^{N_s-M} \ket{\eta_M^\alpha}, \nonumber
\\
P_x \ket{\mu_M^\alpha} &=& (-1)^{N_s} \ket{\nu_M^\alpha}, 
\end{eqnarray}
with, much as in earlier Sections, $P_{z} = \exp[i \pi S^{z}]$, and
$P_{x} = \exp[i \pi S^{x}]$ ($\exp[i \pi S^{x}_j] |m_j \rangle = 
-|(-m_j)\rangle$ for $z$-axis polarization $m_j= 0, \pm 1$). Thus, the
effect of $P_{x}$ is to conserve the number of zeros and to flip  $+$
to $-$ and vice versa. In this way, one can construct states which are
also eigenstates of $P_x$ and $P_z$
\begin{eqnarray}
\ket{\eta_{M\pm}^\alpha} &=& \frac{1}{\sqrt{2}} (\ket{\mu_M^\alpha}\pm
\ket{\nu_M^\alpha}), 
\end{eqnarray}

Any state with a well-defined number $M$ of zeros and $S^z$ can, most
generally, be written as ($\langle \phi_M | \phi_{M'} \rangle =
\delta_{MM'}$)
\begin{eqnarray}
|\phi_{M} \rangle = \sum_{\alpha=1}^{N_M} a_{\alpha} 
|\eta_{M}^{\alpha} \rangle  .
\end{eqnarray}
This state satisfies ($\sum_{\alpha=1}^{N_M} |a_{\alpha}|^2=1$)
\begin{eqnarray}
\hat{M}\ket{\phi_M}&=&M \ket{\phi_M} , \nonumber \\
P_{z} |\phi_{M} \rangle &=& (-1)^{N_s-M} | \phi_{M} \rangle, \nonumber \\
S^{z} | \phi_{M} \rangle &=& c_{M} |\phi_{M}  \rangle  , \ \ \mbox{ with } 
\end{eqnarray}
\begin{eqnarray}
 c_M \! = \! \begin{cases} 
      +(-) 1 & \text{if $N_s-M$ is odd and $\eta=\mu(\nu)$} \\
   \ \ \  \ \ \ \   0 & \text{if $N_s-M$ is even} 
              \end{cases} .
\end{eqnarray}

{\bf Lemma}: For a general state $|\phi_{M} \rangle$,
\begin{eqnarray}
| \langle \phi_{M} | S_{i}^{z} Q_{ij} S_{j}^{z} | \phi_{M} \rangle| 
&\ge &| \langle \phi_{M} | S_{i}^{z} S_{j}^{z} | \phi_{M} \rangle|,
\ \mbox{ where} \nonumber  \\
Q_{ij} &=& \prod_{i<k<j} \exp[i \pi S_{k}^{z}] .
\end{eqnarray}
The proof of this assertion is immediate. By explicit evaluation, 
\begin{eqnarray}
\langle \phi_{M} | S_{i}^{z} Q_{ij} S_{j}^{z} | \phi_{M} \rangle \! =
\! \sum_{\alpha=1}^{N_M} |a_{\alpha}|^{2} {\cal M}^\alpha_{ij} \ ,
\end{eqnarray}
where ${\cal M}^\alpha_{ij}=\langle \eta_{M}^{\alpha} |S_{i}^{z} 
Q_{ij} S_{j}^{z} | \eta_{M}^{\alpha} \rangle$. Here, $i$ and $j$ are
arbitrary site labels, $i,j = 1, \cdots , N_s$. If either
$S_{i}^{z}\ket{\eta^\alpha_M}$ or $S_{j}^{z}\ket{\eta^\alpha_M}$ vanish
the expectation value above vanishes (as does $\langle \eta^\alpha_M | 
S_{i}^{z} S_{j}^{z} | \eta^\alpha_{M} \rangle$). There are four
remaining cases:
\begin{eqnarray}
S_{i}^{z}\ket{\eta^\alpha_M} &=& \ \ \, S_{j}^{z}\ket{\eta^\alpha_M} = \pm 1
\ket{\eta^\alpha_M} \Rightarrow Q_{ij}=-1 \ , \nonumber \\
S_{i}^{z}\ket{\eta^\alpha_M} &=&  -S_{j}^{z}\ket{\eta^\alpha_M} = \pm 1
\ket{\eta^\alpha_M} \Rightarrow Q_{ij}=+1 \ . \nonumber 
\end{eqnarray}
Therefore, 
\begin{eqnarray}
{\cal M}^\alpha_{ij}=      \begin{cases}
\ \ 0 & \text{if $S_{i}^{z}\ket{\eta^\alpha_M}=0$, or
$S_{j}^{z}\ket{\eta^\alpha_M}=0$} \\
-1 & \text{otherwise}
                            \end{cases} ,
\label{uniform_sign} 
\end{eqnarray}
while (${\cal S}^\alpha_{ij}=\langle \eta_{M}^{\alpha}|S_{i}^{z}
S_{j}^{z}|\eta_{M}^{\alpha}  \rangle$)
\begin{eqnarray}
{\cal S}^\alpha_{ij}=             \begin{cases}
\ \ 0 & \text{if $S_{i}^{z}\ket{\eta^\alpha_M}=0$, or
$S_{j}^{z}\ket{\eta^\alpha_M}=0$} \\
+ 1 & \text{if $S_{i}^{z}\ket{\eta^\alpha_M}=S_{j}^{z}\ket{\eta^\alpha_M}$} \\
- 1 & \text{if $S_{i}^{z}\ket{\eta^\alpha_M}=-S_{j}^{z}\ket{\eta^\alpha_M}$}
                            \end{cases} .
\label{uniform_sign1} 
\end{eqnarray}
As the string correlator has a  uniform sign,
\begin{eqnarray}
|\langle \phi_{M}| S_{i}^{z} Q_{ij} S_{j}^{z} | \phi_{M} \rangle| 
\ge | \langle \phi_{M} | S_{i}^{z} S_{j}^{z} | \phi_{M} \rangle|  .
\end{eqnarray}
\qed
Notice that this proof is easily extended to the case of states which 
are also eigenstates of $P_x$ ($|\phi_{M\pm} \rangle =
\sum_{\alpha=1}^{N_M} a_{\alpha}  |\eta_{M\pm}^{\alpha} \rangle$; they
are not eigenstates of $S^z$ unless $N_s-M$ is even). 

Let us now generalize these statements and examine what happens when 
the number of zeros ($M$) is not a good quantum number but $S^z$
remains being a good one. We define a class of states via ($\langle
\Psi | \Psi \rangle =1$)
\begin{eqnarray}
| \Psi \rangle = \sum_{M} \beta_{M} |\phi_{M} \rangle.
\label{classM}
\end{eqnarray}
with the property ($\sum_M |\beta_M |^2=1$)
\begin{eqnarray}
P_z | \Psi \rangle = \sum_{M} \beta_{M} (-1)^{N_s-M} |\phi_{M} \rangle.
\end{eqnarray}

We have the simple
 
{\bf Corollary:}
For the class of states defined in Eq. (\ref{classM})
\begin{eqnarray}
| \langle \Psi | S_{i}^{z} Q_{ij} S_{j}^{z} | \Psi \rangle | \ge  |
\langle \Psi| S_{i}^{z} S_{j}^{z} | \Psi \rangle |.
\label{stringM}
\end{eqnarray}
The proof of this assertion proceeds as for the Lemma,
\begin{eqnarray}
\langle \Psi| S_{i}^{z} Q_{ij} S_{j}^{z}|\Psi \rangle = \sum_{M}
|\beta_M|^2
\langle \phi_{M} | S_{i}^{z} Q_{ij} S_{j}^{z} | \phi_{M} \rangle ,
\end{eqnarray}
and we apply the  Lemma, leading to the inequality (\ref{stringM}).

If we seek states $| \Psi \rangle$ which are also eigenstates of
$P_{z}$ (see text just after Eq. (\ref{pxd})) 
\begin{eqnarray}
P_{z} | \Psi \rangle = \pm 1 | \Psi \rangle,
\end{eqnarray}
we then have the following constraints
\begin{eqnarray}
P_{z} \ket{\Psi}& =& +1 \ket{\Psi}  \Rightarrow \ket{\Psi} =
\!\!\!\!\!\!\! \sum_{M, (N_s-M)\in \ {\rm even}}\!\!\!\!\!\!\!\!\!\!\!\!\!\! 
 \beta_{M} | \phi_{M} \rangle , \\
P_{z} \ket{\Psi}& =& -1 \ket{\Psi}  \Rightarrow \ket{\Psi} =
\!\!\!\!\!\!\! \sum_{M, (N_s-M)\in \ {\rm odd}} \!\!\!\!\!\!\!\!\!\!\!\!\!\! 
\beta_{M} | \phi_{M} \rangle .
\end{eqnarray}

In order to understand the reason why one gets larger string
(non-local) correlations than local ones we will investigate  the
effect of $U_{s} = \prod_{j <k}  \exp[i \pi
S_{j}^{z} S_{k}^{x}] = \prod_{j<k} (1 - 2 (S_{j}^{z})^{2} 
(S_{k}^{x})^{2})$ on e.g. $|\mu_{M}^{1} \rangle$. In what follows, we
will use a shorthand notation for the states $\ket{\eta^\alpha_M}$ and
consider only those modes  involved in the action of the operators. We
find that ($\ket{m_jm_k}$)
\begin{eqnarray}
\Big[ 1-2(S_{j}^{z} S_{k}^{x})^{2} \Big] \ket{++}  &=& - 
| +- \rangle ,\nonumber \\  
\Big[ 1-2(S_{j}^{z} S_{k}^{x})^{2} \Big] |- - \rangle &=& -
|-+ \rangle, \nonumber \\ 
\Big[ 1- 2(S_{j}^{z} S_{k}^{x})^{2} \Big] |-+ \rangle &=& -
|-- \rangle, \nonumber \\ 
\Big[ 1-2(S_{j}^{z} S_{k}^{x})^{2} \Big] |+- \rangle &=& -
|++ \rangle, \nonumber \\ 
\Big[ 1-2(S_{j}^{z} S_{k}^{x})^{2} \Big] |0 m_k \rangle &=& \ 
|0 m_k \rangle, \nonumber \\ 
\Big[ 1-2(S_{j}^{z} S_{k}^{x})^{2} \Big] | m_j 0 \rangle &=& 
\!\! - |m_j 0 \rangle ~~( m_j \neq 0).
\end{eqnarray}

Let us next consider three sites ($\ket{m_im_jm_k}$) with
$U_{ijk}=\exp[i \pi S_{i}^{z} S_{j}^{x}]\exp[i \pi S_{i}^{z}
S_{k}^{x}]\exp[i \pi S_{j}^{z} S_{k}^{x}]$
\begin{eqnarray}
U_{ijk}\ket{+-+} &=& - \ket{+++}, \nonumber \\  
U_{ijk}\ket{-+-} &=& - \ket{---}, \nonumber    \\ 
U_{ijk}\ket{+-0} &=& - \ket{++0}, \nonumber \\  
U_{ijk}\ket{-+0} &=& - \ket{--0}, \nonumber \\ 
U_{ijk}\ket{0+-} &=& - \ket{0++}, \nonumber \\ 
U_{ijk}\ket{+0-} &=& \ \ \ket{+0+} .
\end{eqnarray}

For a generic chain of length $N_s$, 
\begin{eqnarray}
U_{s} |\mu_{M}^{1} \rangle
&=& (-1)^{\varphi_{1}} |\underbrace{+++++ \cdots +}_{N_s-M} \underbrace{00
\cdots0}_{M} \rangle , \nonumber \\
U_{s} |\nu_{M}^{1} \rangle 
&=& (-1)^{\varphi_{1}} |\underbrace{----- \cdots -}_{N_s-M} \underbrace{00
\cdots0}_{M} \rangle ,
\label{final-polar}
\end{eqnarray}
where the phase is $\varphi_1=N_s(N_s-1)/2-M(M-1)/2$. Thus,  the string
operator $U_s$ acts as a {\it polarizer}.

For generic states $\{\ket{\eta^\alpha_M}\}$,
\begin{eqnarray}
U_{s} |\mu_{M}^{\alpha} \rangle
&=& (-1)^{\varphi_{\alpha}} \ket{\bar{\mu}^\alpha_M} , \nonumber \\
U_{s} |\nu_{M}^{\alpha} \rangle 
&=& (-1)^{\varphi_{\alpha}} \ket{\bar{\nu}^\alpha_M} ,
\label{final-polar1}
\end{eqnarray}
where the state $\ket{\bar{\mu}^\alpha_M}$ ($\ket{\bar{\nu}^\alpha_M}$)
denotes the fully polarized state of $(N_s-M)$ spins having $m=+1$
($m=-1$) and $M$ intercalated zeros. For example, 
\begin{eqnarray}
\!\!\!\!\!\!\!\!\!\! U_{s} |\mu_{M}^{2} \rangle
&=& (-1)^{\varphi_{2}} |+0+++++ \cdots + 0 \cdots 0 \rangle , \nonumber \\
\!\!\!\!\!\!\!\!\!\! U_{s} |\nu_{M}^{2} \rangle 
&=& (-1)^{\varphi_{2}} |-0----- \cdots - 0 \cdots 0 \rangle ,
\label{final-polar11}
\end{eqnarray}
and, in general, the phase is given by
\begin{eqnarray}
\varphi_{\alpha} = \frac{N_s(N_s-1)}{2} - \sum_{i_{0}} (N_s-i_{0}),
\end{eqnarray}
where $\{i_{0}\}$ spans the position of the zeros.

GSs of known Hamiltonians represent particular examples of the classes
$\ket{\phi_M}$ or $\ket{\Psi}$. For example, the GS of the $t$-$J_z$
model \cite{tJz,BO} is a particular $\ket{\phi_M}$ state,  while the
GSs of the Hamiltonians considered by Kennedy and Tasaki \cite{KT}
belong to the class $\ket{\Psi}$. For example, consider the AKLT GSs of
$\tilde{H}_{\sf AKLT}$, Eq. (\ref{tHAKLT}). It is quite clear that
they represent linear combinations of {\it polarized} states with
different number of zeros $M$
\begin{eqnarray}
\ket{\tilde{g}_\alpha}&=&\sum_{M}\beta_M^{\alpha}
\ket{\bar{\phi}_M^\alpha}, \ \mbox{ with} , \nonumber \\
\!\!\!\!\!\!\!\!\!\!\!\! \ket{\bar{\phi}_M^{(1,2)}} &=&
\sum_{\gamma=1}^{N_M}  \ket{\bar{\mu}_M^\gamma} , \ \mbox{ and }
\ket{\bar{\phi}_M^{(3,4)}}  = \sum_{\gamma=1}^{N_M} 
\ket{\bar{\nu}_M^\gamma} . 
\end{eqnarray}
From what we have just shown
\begin{eqnarray}
\ket{\bar{g}_\alpha}&=&\sum_{M}\beta_M^{\alpha}
U_{s} \ket{\bar{\phi}_M^\alpha}= \sum_{M}\beta_M^{\alpha}
\ket{{\phi}_M^\alpha}, \ \mbox{ with} , \nonumber \\
\!\!\!\!\!\!\!\!\!\!\!\! \ket{{\phi}_M^{(1,2)}} &=&
\sum_{\gamma=1}^{N_M} (-1)^{\varphi_\gamma} \ket{{\mu}_M^\gamma} , \
\mbox{ and } \nonumber \\
\!\!\!\!\!\!\!\!\!\!\!\! \ket{{\phi}_M^{(3,4)}}  &=&
\sum_{\gamma=1}^{N_M} (-1)^{\varphi_\gamma} \ket{{\nu}_M^\gamma} , 
\end{eqnarray}
which unambiguously shows that they belong to the class of states
$\ket{\Psi}$.

Similar constructs for operators $U_{s}$ polarizing given GSs and
bringing them to uniform reference  states having high correlations of
one sort or another appear in doped Hubbard chains \cite{tJz,BO,OS,
KMNZ}.  In its $t$-$J_z$ approximant the system may be mapped onto a
$S=1$ chain similar to the one above by a Jordan-Wigner 
transformations \cite{tJz}; there the local state $S^{z}_{i}=0$
corresponds  to the presence of a hole. Similar correlations also
appear in $D=2$  $t$-$J_{z}$ models \cite{Jurij, NR},  and in
Quantum Hall systems \cite{GM}.

\section{Gauge-Graph Wavefunctions}
\label{app6}

In the text we made some analogies between systems with TQO and
graphs. To  complete this circle of ideas, we now illustrate how simple
wavefunctions on a graph may be concocted to have TQO.

Let us consider a graph which on any link has a gauge field. In what
follows, we will focus  on a graph composed of $(2N_s)$ points --- we
label $N_s$ of these site as ``+'' sites  and the other $N_s$ sites as
``$-$'' sites.  We now consider a state $|\phi \rangle$ which has $M=
N_s z$  dimers ($\sigma^{z}_{i_{+}, i_{-}} = -1$ if $i_{+}$ and $i_{-}$
are linked by a dimer; $\sigma^{z}_{i_{+}, i_{-}} = 1$ otherwise) 
connecting ``+'' sites to ``$-$'' sites. Here, $z$ is the coordination
number of the graph. In what follows, we examine the case of $z =
{\cal{O}}(N_s)$. Given any such state $|\phi \rangle$ we define the graph
wavefunctions
\begin{eqnarray}
|\psi_{1,2} \rangle = \frac{1}{N_s!} \sum_{P_{+}, P_{-}} 
(-1)^{0, P_{+}P_{-}} 
P_{+} P_{-}|\phi \rangle.
\end{eqnarray} 
The permutations $P_{+,-}$ permute only the $N_s$ ``+'' or $N_s$
``$-$'' sites amongst themselves --- they do not inter-permute ``+''
sites with ``$-$'' sites. The phase factor $(-1)^{P_{+} P_{-}}$ denotes
the  parity of the combined permutations. Inspection reveals that
$\langle \psi_{1} | \psi_{2} \rangle = 0$.  We claim that $| \psi_{1,2}
\rangle$ exhibit rank-$n=2$ TQO as we now show. It is apparent that
$\langle \psi_{1} |  \sigma_{pq}^{x,z} | \psi_{1} \rangle = \langle
\psi_{2} |  \sigma_{pq}^{x,z} | \psi_{2} \rangle$ where $(p,q)$ are any
two points in the graph ($p \in +, q \in -$). It is also clear that 
for any product of $\{\sigma^{x,z}_{pq}\}$ we will have equivalent
expectation values in $|\psi_{+} \rangle$ and $|\psi_{-} \rangle$. If
$z = {\cal{O}}(N_s)$ then  no {\it local} operation links $|\psi_{1}
\rangle$ to $|\psi_{2} \rangle$. Any operation which links the two
states involves ${\cal{O}}(N_s)$ graph links. With $Q_{+,-}$ being being any
permutation  of the ``+'' or ``$-$'' sites respectively, we have
$Q_{\pm} | \psi_{1} \rangle = | \psi_{1} \rangle$  and $Q_{\pm}
|\psi_{2} \rangle = (-1)^{Q_{+}Q_{-}} |\psi_{2} \rangle$. Any such
permutation involves, at least, $2z$  gauge links. As $z=
{\cal{O}}(N_s)$, the viable order parameters $Q_{\pm}$ are not
quasi-local and involve
${\cal{O}}(N_s)$ fields.


\begin{thebibliography}{99}
\bibitem{emergexact} 
It is worth emphasizing that {\em emergent symmetries} refer to the Eq.
(\ref{emergequation}): they are merely symmetries of the GS
manifold. Within the GS sector, any operator $U$ is a symmetry
\begin{eqnarray}
\langle g_{\alpha}|
[U,H]|g_{\beta} \rangle =0,
\end{eqnarray}
as can be proven by applying  $H$ to the left or right and recalling
that $|g_{\alpha} \rangle$ and $|g_{\beta} \rangle$ are eigenstates of
$H$  with the same eigenvalue.
\bibitem{bounding_KK} 
In the $\alpha = x,z$ plane on which a $D=2$ $t_{2g}$ KK Hamiltonian is
defined (Eqs. (\ref{1}), (\ref{orbHam}), and (\ref{compass1})),  the
general bounding action $\overline{S}$ is the exact action of an
orbital state in which all electrons are in the $|d_{yz} \rangle$
state. In such an orbital state,  spin-exchange processes are the
largest.
\bibitem{t2gKKexample} 
In this system, rotations of each individual flavor in orthogonal
planes commute with one another.  Here, the planar $SU(2)$ rotations
can be parameterized by the total spin $\otimes_{P} |S, S_{z}
\rangle_{p}$ in given $d=2$ planes $P$. As these operators commute with
$H$,  the GSs can be chosen to be simultaneous eigenstates of the
symmetry operators and  can labeled by their respective eigenvalues. 
The matrix elements of $V$ between any two  orthogonal eigenstates of
the $d=2$ symmetry generators  are,  by explicit writing of the
symmetry eigenstates and the use of the Wigner-Eckart theorem,
vanishingly  small in the thermodynamic limit. The matrix element
between the maximally symmetric state, $S_{z}=S=S_{\max}$ and states
which differ by a finite number of spin flips is algebraically small in
the system size. The general matrix element of a quasi-local operator 
between different (orthogonal) symmetry eigenstates can be computed by
the Clebsch-Gordan coefficients and  is seen to be exponentially small
in the system size.
\bibitem{explain_factorization}
The Ising mapping of Eqs. (\ref{abd}), (\ref{Isingmp}), 
(\ref{abDUAL}), and (\ref{abdual}) shows that 
\begin{eqnarray}
\langle \Big( \prod_{s \in S} A_{s} \Big) \Big( \prod_{p \in P} B_{p}  
\Big) 
\rangle
\end{eqnarray}
evaluated with the Gibbs weights that appears in the Kitaev's Toric
code model (whose sum appears in  Eq. (\ref{pfunction})) is the same as
\begin{eqnarray}
\langle  \Big( \prod_{s \in S} a_{s}\Big) \Big(\prod_{p \in P}
b_{p}\Big)  \rangle 
\end{eqnarray}
evaluated with the Gibbs weights given by the Hamiltonian of Eq.
(\ref{Isingmp}). Now,  the Hamiltonian of Eq. (\ref{Isingmp}) as well
as all of the conditions of  Eqs. (\ref{abDUAL}), and (\ref{abdual}) 
are invariant under arbitrary relabeling of the indices $s$ and $p$.
All that matters is that the number of terms $a_{s}$ and $b_{p}$ remain
invariant. Thus, we can choose the simplest labeling for the sets $S$
and $P$ for the average of 
\begin{eqnarray}
\langle \prod_{s \in S} [\sigma_{s}^{z} \sigma_{s+1}^{z}] 
\prod_{p \in P} [\sigma_{p}^{z} \sigma_{p+1}^{z}]  \rangle
\end{eqnarray}
with the Hamiltonian of Eq. (\ref{Isingmp}). If we choose the set $S$
to span $s=0,1,2, \cdots, \|S\|-1$ and the index $P$ to span $p=0,1,2,
\cdots, \|P\|-1$ then  we will have to evaluate
\begin{eqnarray}
\langle \prod_{s=0}^{\|S\|-1} [\sigma_{s}^{z} \sigma_{s+1}^{z}] 
\prod_{p=0}^{\|P\|-1}  [\sigma_{p}^{z} \sigma_{p+1}^{z}] \rangle \nonumber \\ 
= \langle \sigma_{0}^{z} \sigma_{\|S\|}^{z} \rangle_{\mbox{a~ Ising
~chain}} \nonumber \\ 
\times \langle \sigma_{0}^{z} \sigma_{\|P\|}^{z} \rangle_{\mbox{b~
Ising~chain}}.
\end{eqnarray}
Now, for the Ising Hamiltonian of Eq. (\ref{Isingmp}) (in the
thermodynamic limit of the periodic chain  or for a finite chain with
open boundary conditions), $\langle \sigma_{0}^{z} \sigma_{\|S\|}^{z}
\rangle = [\tanh (\beta)]^{\|S\|}$ and similarly $\langle
\sigma_{0}^{z} \sigma_{\|P\|}^{z} \rangle = [\tanh(\beta)]^{\|P\|}$.
From these relations, Eq. (\ref{connected}) follows.
\bibitem{explain_limit_period} 
Strictly speaking, Eqs. (\ref{Ribbon}), and (\ref{zz12}) hold for the
periodic system so  long as $N_{s} \gg \|R\|$ or equivalently $y \ll
\sqrt{N_{s}}$. This is so as only in the  thermodynamic limit ($N_{s}
\to \infty$) of a periodic chain  does Eq. (\ref{connected}) hold. In
the mapping of the Kitaev model to Ising chains, the size of each of
the periodic chains ($N_{s}$) needs to be much larger than the
separation of the spins $\|R\|$.
\bibitem{degexp}
To prove this, we recall that the number of  fields $E$
is given by 
\begin{eqnarray}
E = F+ V+ 2(g-1),
\end{eqnarray}
with $F$ the number of faces (plaquettes) and $V$ the number of
vertices (sites of which there are $N_{s}$). There are $(F-1)$
independent conditions $\{B_{p} =1\}$. [There are $F$ plaquettes
$\{B_{p}\}$; the additive contribution of (-1)  originates from the
single constraint $\prod_{p} B_{p} =1$.] This leads to 
\begin{eqnarray}
E-(F-1) = N_{s} + 2(g-1) + 1
\end{eqnarray}
remaining degrees of freedom and thus to a degeneracy of  $2^{N_{s} +
2g-1}$.
\bibitem{explainwhy} 
When applied to smaller closed paths, this procedure  would lead to
additional unwanted constraints  (supplanting Eq. (\ref{ABpc})) of the
form $\prod_{p \in R} B_{p} =1$ with $R$ a region which does not span the
entire lattice.
\end{thebibliography}
\end{document}